%% file: light_matter.tex
\documentclass[11pt,a4paper]{article}
\usepackage[T1]{fontenc}
\usepackage[utf8]{inputenc}
\usepackage{lmodern}
\usepackage[english]{babel}
\usepackage[a4paper,margin=2.6cm]{geometry}
\usepackage{amsmath,amssymb}
\usepackage{graphicx}
\usepackage{xcolor}
\usepackage{framed}
\usepackage{caption}
\usepackage{placeins}
\usepackage{needspace}
\usepackage[hidelinks]{hyperref}
\hypersetup{pdftitle={How light interacts with matter, revisited},pdfauthor={Giancarlo Reali},
  pdfsubject={A story told by oscillating dipoles and retarded fields},
  pdfkeywords={light-matter interaction, refractive index, extinction theorem, Feynman arrows, Brewster angle, scattering}}

\newif\ifmarked \markedfalse
\newif\ifparked \parkedtrue
\newcommand{\rev}[1]{\ifmarked{\color{red}#1}\color{black}{}\else#1\fi}
\newcommand{\revon}{\ifmarked\color{red}\fi}

\graphicspath{{fig/}}
\newcommand{\ud}{\mathrm{d}}
\newcommand{\bE}{\mathbf E}
\newcommand{\bp}{\mathbf p}
\newcommand{\bn}{\mathbf n}

\definecolor{shadecolor}{RGB}{238,242,248}
\usepackage[breakable,skins]{tcolorbox}
\newtcolorbox{panelbox}{breakable,enhanced,colback=shadecolor,colframe=shadecolor,boxrule=0pt,
  arc=0pt,left=6pt,right=6pt,top=6pt,bottom=6pt,before skip=10pt,after skip=10pt,fontupper=\small}
\newenvironment{panel}[1]{\begin{panelbox}\noindent\textbf{#1}\par\smallskip\noindent\ignorespaces}%
  {\end{panelbox}}

\newcommand{\anim}[2]{\textsf{Animation~#1} (file \texttt{\detokenize{#2}})}

\title{\textbf{How light interacts with matter, revisited}\\[0.4em]
\large A story told by oscillating dipoles and retarded fields}
\author{Giancarlo Reali\\[0.3em]
\small \rev{Bright Solutions S.r.l., Via degli Artigiani 27, 27010 Cura Carpignano (PV), Italy}\\
\small \rev{\texttt{g.reali@brightsolutions.it} \quad -- \quad ORCID 0000-0001-9943-5079}}
\date{\small October 2026; \rev{revised December 2026}}

\begin{document}
\maketitle

\begin{abstract}
\noindent In 1968 Victor Weisskopf explained in \emph{Scientific American} how light interacts with
matter, with a few formulas and many words. We tell the same story again with the tools that have
since become available: the fields are computed and drawn, and their evolution in time is shown in
animations. Light is described as a classical electromagnetic field and matter as a collection of
charged oscillators; every phenomenon follows from adding up the retarded waves radiated by the
oscillators, each with its proper delay and phase. From this single picture we obtain the blue of
the sky and the white of the clouds, transparency and the refractive index, the Ewald--Oseen
extinction theorem, reflection and the colours of thin films, why metals reflect and
what gives metals and semiconductors their colours, polarizers, Brewster's angle and birefringence, the working of
mirrors, lenses, microscopes and the eye, the propagation of a few-cycle pulse through a slab,
\v{C}erenkov radiation, amplification and second-harmonic generation. Feynman's arrows are made quantitative with the
sheet method, and coupled-dipole simulations show the incident wave being cancelled and replaced
inside the medium. The connection with quantum theory is discussed where the classical picture
stops: the photon, spontaneous emission and the forbidden lines of the aurora, vision and
photosynthesis, and the sum over paths. The
text is addressed to undergraduate students and their teachers, and to any reader curious about the
physics behind everyday optical phenomena; derivations are collected in separate panels, and the
animations and the Python codes used for the figures and simulations are provided as ancillary files.
\end{abstract}

{\small\makeatletter\renewcommand\@pnumwidth{1.8em}\makeatother\setlength{\parskip}{0pt}\tableofcontents}
\bigskip

\noindent\emph{About this text.} Formulas are in SI units, with the time dependence $e^{-i\omega t}$
for harmonic fields. Derivations that can be skipped at first reading are collected in shaded
panels. The animations quoted in the text are provided as ancillary files, together with the Python
codes that produced the figures.

\clearpage
\input{ch0_prologue}
\input{ch0b_light}
\input{ch1_one_atom}
\input{ch2_sheet}
\input{ch3_index}
\input{ch4_extinction}
\input{ch5_reflection}
\input{ch6_brewster}
\input{ch6b_mirrors}
\input{ch7_pulse}
\input{ch8_nonlinear}

\input{ch9_nature}
\input{ch10_epilogue}

\section*{Acknowledgements}
\addcontentsline{toc}{section}{Acknowledgements}
The text, the calculations and the figures were developed with the assistance of an AI model
(Claude, Anthropic), used for drafting, coding and checking. The author has verified all the
results and is solely responsible for the content.

\bibliographystyle{unsrt}

\end{document}

%% file: ch0_prologue.tex
\section*{Prologue: wonder}
\addcontentsline{toc}{section}{Prologue: wonder}
\label{sec:prologue}

\begin{figure}[!t]
\centering
\IfFileExists{fig/prologue_forest.jpg}{%
  \includegraphics[width=\textwidth]{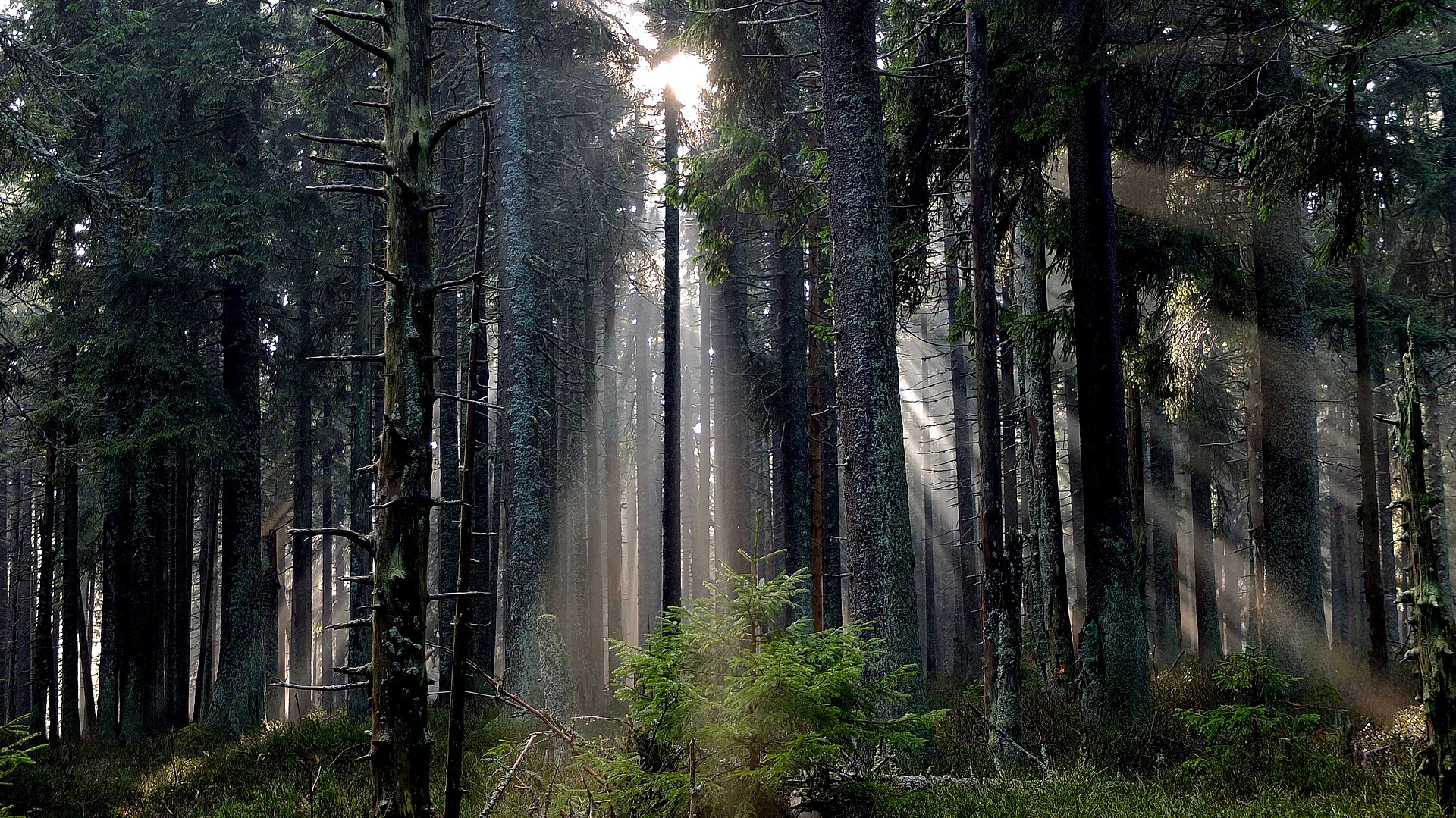}}{%
  \fbox{\parbox[c][0.32\textheight][c]{0.7\textwidth}{\centering\small
  [Photograph to be inserted:\\ \emph{Sun's rays in a dense forest} (Unsplash),\\ via Wikimedia Commons]}}}
\caption{Sunbeams in a misty forest. \rev{Photo: Filip Varga (2017), \emph{Sun's rays in a dense forest}, Unsplash, via Wikimedia Commons; public domain (CC0 1.0).}}
\label{fig:forest}
\end{figure}

Walk into a wood on a misty morning, just as the sun clears the trees. All at once the air is
crossed by shafts of light, sharp-edged, motionless, as if they were solid
(Fig.~\ref{fig:forest}). We stop. Before any thought, there is wonder---a feeling much older than
science, and one that no explanation should take away from us.

A physicist feels it too, and then, almost at once, something else: questions. Why do we see the
beams at all? Light that does not enter our eyes is invisible; we see these shafts only because
countless tiny droplets in the mist send a small part of the sunlight sideways, towards us. Why
are they white, while the sky above is blue? Why are they so straight? Every molecule of the air
they cross sends light in all directions, yet the beams travel on through the wood without
spreading, as if nothing were there. And why do they seem to fan out from the sun, when in fact
they are parallel? (This last one is the easy one: it is perspective, the same illusion that makes
railway tracks meet at the horizon.)

Nature offers such spectacles every day, and each of them hides a question of the same kind.
The blue of the sky and the red of the setting sun. The light that crosses a window pane without
noticing it, and the faint image of ourselves that the same pane reflects. The reflection of the
sky on a lake, white like the light that produced it, and the glare that disappears when we look
at it through polarizing sunglasses. The colours of a soap bubble. The green spot of a laser
pointer, born from an infrared beam inside a tiny crystal. The following chapters try to answer
these questions one by one, and they will find that the answers are all the same answer, told in
different ways.

In 1968 Victor Weisskopf told this story in a celebrated article in \emph{Scientific
American}~\cite{Weisskopf}, with a few formulas and many words. Here we tell it again with what
fifty years have added to the physicist's toolbox: the fields can now be computed and drawn, and
we can watch them evolve in time, wave by wave. The plot is very simple. Matter is made of
electric charges bound to atoms; a light wave sets them oscillating; every oscillating charge
radiates a wave of its own, which travels at the speed of light and reaches the other charges a
little later. Everything we see---scattering, transparency, refraction, reflection, polarization,
colour---comes from adding up these waves, each with its proper delay and its proper phase.

Does understanding diminish the wonder? Keats feared so: in his \emph{Lamia} (1820) he complained
that ``cold philosophy'' would ``clip an Angel's wings'' and ``unweave a rainbow''. Feynman, who
loved to talk about the beauty of a flower, answered that science only adds to it: the scientist
sees the same beauty as the artist, and in addition the beauty of what happens inside, at scales
no eye can reach. We hope the reader will finish these pages, and then walk again into that wood,
feeling the same.

%% file: ch0b_light.tex
\section*{\rev{What is light? Waves, fields and photons}}
\addcontentsline{toc}{section}{What is light? Waves, fields and photons}
\label{sec:whatislight}

{\revon
This text describes light as a wave of the electromagnetic field and matter as a collection of
charged oscillators. Before starting, the reader deserves a word on why this is legitimate, and
on the other word that will appear here and there: the photon.

\paragraph{Waves or particles?} The question is as old as optics (Fig.~\ref{fig:timeline}).
Newton imagined light as a stream of corpuscles; Huygens, a little earlier, as a wave. A century
later Young's interference fringes and Fresnel's theory of diffraction settled the matter in
favour of waves, and Maxwell, in 1865, discovered what was waving: electric and magnetic fields,
travelling at the speed of light. Hertz produced such waves with an electric circuit in 1888.
Then, at the turn of the century, the quanta appeared. Planck, in 1900, explained the spectrum of
thermal radiation by assuming that matter exchanges energy with light only in packets $h\nu$;
he quantized the oscillators of matter, not the light. Einstein, in 1905, went further and
proposed that light itself carries its energy in quanta, an idea supported by the photoelectric
effect and, in 1923, by Compton's discovery that X-rays scattered by electrons behave as in a
collision of particles. In 1926 the chemist G.N.~Lewis named the light quantum ``photon'', and
in 1927 Dirac quantized the electromagnetic field, the first step towards quantum electrodynamics.

\begin{figure}[!htbp]
\centering
\includegraphics[width=0.95\textwidth]{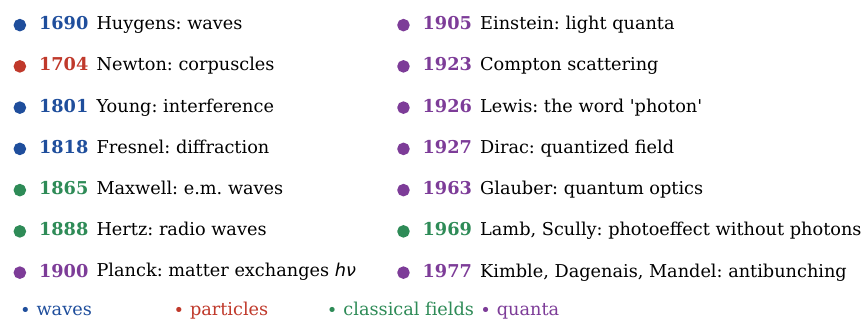}
\caption{\rev{Some steps in the understanding of light.}}
\label{fig:timeline}
\end{figure}

\paragraph{What is a photon?} ``All these fifty years of conscious brooding have brought me no
nearer to the answer to the question: what are light quanta?'', wrote Einstein to his friend
Michele Besso in 1951. The difficulty is not technical. What we actually observe are
\emph{clicks}: a photodetector illuminated by very weak light responds with discrete, localized
events, and when the light is made weaker the clicks become rarer, not fainter. As
Marburger has pointed out in a lucid essay~\cite{Marburger}, physicists use the word ``photon''
in at least three senses: the click itself, the event recorded by the detector; the imaginary
particle that, in Feynman's formulation, explores all the paths from the source to the detector,
each contributing an arrow to the probability amplitude; and the quantized excitation of a mode of
the electromagnetic field, an abstract object of the theory. None of them is a small ball of
light travelling along a definite path. Quantum theory predicts the statistics of the clicks; it
does not provide a persistent, localizable object between one click and the next. In Marburger's
words, ``photon'' is a quantum word, for which no classical word is adequate, whereas ``light'',
as we use it in everyday language, is a classical word: in the classical world we live in, light
behaves, to an excellent approximation, as an electromagnetic wave.

\paragraph{How much of light needs photons?} Less than one might think. Many effects attributed to
photons, the photoelectric effect included, can be explained with a classical field acting on
quantized atoms: this ``semiclassical'' analysis, already explored in the days of the old quantum
theory, was carried out in detail by Lamb and Scully~\cite{LambScully}. The unambiguous
fingerprints of the quantized field came later, with the quantum theory of optical coherence of
Glauber~\cite{Glauber} and with experiments in which a single atom was seen never to produce two
clicks at once (``antibunching'')~\cite{KDM}, or a single photon was seen to interfere with itself
while never being split by a beam splitter~\cite{GRA}: no classical wave can do that. And
spontaneous emission, the light of an excited atom left alone, requires the quantized field from the
start.\ifparked\ (For the reader who wants to see why everyday light looks so classical, the panel below makes
it quantitative.)\fi

\ifparked\begin{panel}{The click and the sparseness of light}
In a beautiful and somewhat forgotten textbook, Kroemer describes the click with care~\cite{Kroemer}.
A photographic emulsion is a dense array of tiny grains, any one of which may intercept a photon. The
blackened grain tells us where the light \emph{interacted}, not where a photon \emph{was}: the
localization is a property of the interception event itself. Kroemer prefers to say that the photon
``materializes'' in the volume of the grain, rather than that it is ``found'' there, because the
everyday word suggests an object already localized before the event; the measurement creates the
position, rather than revealing a pre-existing one.

He then asks how many photons a light beam contains in a volume $\lambda^3$. For a beam of intensity $S$
(energy per unit area and time) and wavelength $\lambda$ the energy density is $S/c$, each photon
carries $\hbar\omega=2\pi\hbar c/\lambda$, and
\begin{equation*}
N_\lambda=\frac{S\lambda^4}{2\pi\hbar c^2}.
\end{equation*}
\begin{center}\small
\begin{tabular}{lccc}
source & $S$ (W/m$^2$) & $\lambda$ & $N_\lambda$\\ \hline
a star seen with the naked eye & $2\times10^{-8}$ & 550 nm & $3\times10^{-17}$\\
sunlight & $10^{3}$ & 600 nm & $2\times10^{-6}$\\
laser pointer, 1 mW on 1 mm$^2$ & $10^{3}$ & 650 nm & $3\times10^{-6}$\\
laser, 1 W focused on 10 $\mu$m & $10^{10}$ & 1064 nm & $2\times10^{2}$\\
mobile-phone signal & $10^{-6}$ & 33 cm & $2\times10^{8}$\\
near an FM radio antenna & $10^{-3}$ & 3 m & $10^{15}$
\end{tabular}
\end{center}
Ordinary light is a \emph{sparse} stream: a volume $\lambda^3$ almost never contains a photon. Radio waves,
because of the factor $\lambda^4$, are extremely dense, and a radio engineer never meets a photon. In
the optical domain, therefore, the interference of light cannot come from photons interacting with one
another: each quantum propagates as a wave, and interferes with itself. The numbers should be read as the
average probability of finding a quantum in $\lambda^3$, not literally as separate wave packets that never
overlap: for laser light the number of photons in a volume is not even defined, and the random clicks it
produces can be reproduced by a classical field acting on quantized detectors (Lamb and Scully, above).
What the estimate shows beyond doubt is that, in optics, what propagates is described by a field, and
discreteness appears only at the moment of the click.
\end{panel}\fi

\paragraph{The agreement with the reader.} In this text, light is the classical electromagnetic
field, and matter is made of classical oscillators whose strengths and frequencies are given by
quantum mechanics (Sec.~\ref{sec:one-atom}). This is enough for everything that follows:
scattering, refraction, reflection, polarization, the propagation of pulses, even amplification and
nonlinear optics. The word ``photon'' will appear only where the classical description stops:
spontaneous emission, the first step of vision, photosynthesis. The arrows we shall add in
Secs.~\ref{sec:sheet}--\ref{sec:reflection} are the classical counterpart of the arrows of
Feynman's quantum electrodynamics~\cite{FeynmanQED-book}: the same phasors, applied to waves
instead of probability amplitudes---Marburger's second kind of photon, used here as what it is, a
wonderful device for adding up waves.
}

%% file: ch1_one_atom.tex
\section{One atom}
\label{sec:one-atom}

Everything in this story is built from a single object: an electric charge bound to an atom, set
into oscillation by a passing wave, and radiating in turn. Before we let many such charges act
together---which is where transparency, refraction and reflection come from---we must understand
very well what one of them does. This chapter is about that one atom.

\subsection{An electron on a spring}

\begin{figure}[!htbp]
\centering
\includegraphics[width=0.92\textwidth]{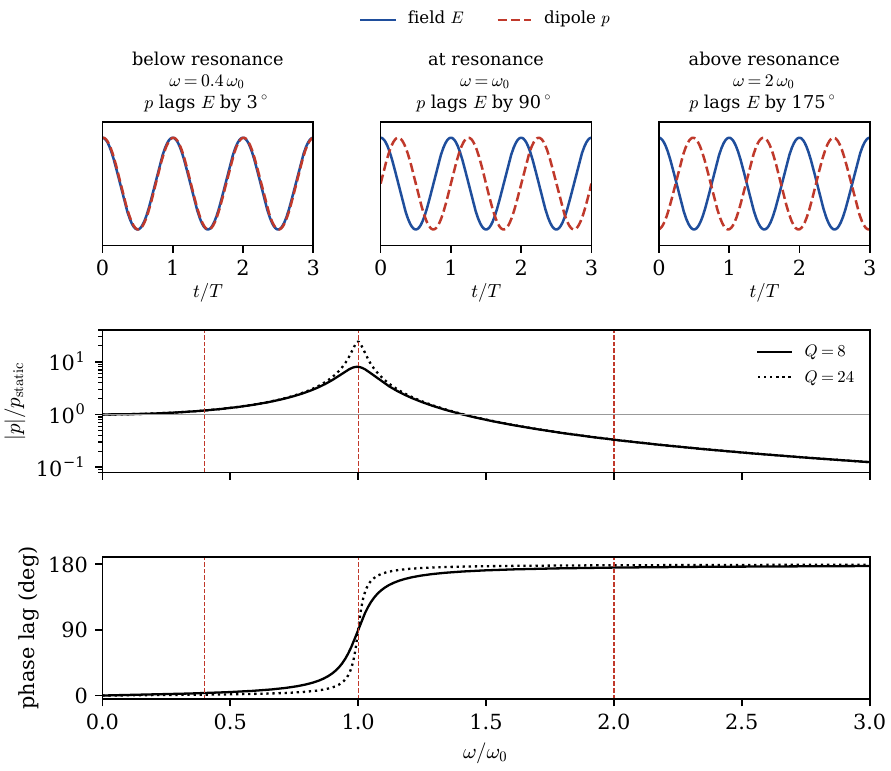}
\caption{The driven electron. Top: the driving field $E$ (solid) and the induced dipole $p$
(dashed) below, at and above resonance, for a quality factor $Q=\omega_0/2\gamma=8$. Middle and
bottom: amplitude, relative to the static response $p_{\rm static}=e^2E_0/m\omega_0^2$, and phase
lag of the dipole as functions of frequency, for $Q=8$ (solid) and $Q=24$ (dotted).}
\label{fig:oscillator}
\end{figure}

Following Lorentz~\cite{Lorentz}, we picture the electron (charge $-e$, mass $m$) as held near
its equilibrium position by an elastic restoring force, with natural angular frequency $\omega_0$,
and slowed down by a small friction, measured by the damping rate $\gamma$. A light wave of
frequency $\omega$ and amplitude $E_0$, polarized along $x$, pushes the electron with the force
$-eE_0\cos\omega t$. The equation of motion is that of a driven, damped harmonic oscillator,
\begin{equation}
\ddot x+2\gamma\dot x+\omega_0^2x=-\frac{e}{m}\,E_0\cos\omega t .
\label{eq:oscillator}
\end{equation}
The wavelength of visible light (400--700~nm) is thousands of times larger than an atom, so the
electron feels a field that is uniform over the atom and changes only in time. Writing
$E_0\cos\omega t=\mathrm{Re}[E_0e^{-i\omega t}]$ and looking for the steady oscillation
$x=\mathrm{Re}[\tilde x\,e^{-i\omega t}]$, one finds $\tilde x=-(e/m)E_0/(\omega_0^2-\omega^2-2i\gamma\omega)$.
The displaced electron and the nucleus form an oscillating electric dipole $p=-e\tilde x$,
\begin{equation}
p=\varepsilon_0\,\alpha(\omega)\,E_0,\qquad
\alpha(\omega)=\frac{e^2}{\varepsilon_0m}\,\frac{1}{\omega_0^2-\omega^2-2i\gamma\omega},
\label{eq:alpha}
\end{equation}
where $\alpha$, the \emph{polarizability} of the atom, has the dimensions of a volume. Its
modulus tells how strongly the atom responds; its phase tells how the response lags behind the
push. Figure~\ref{fig:oscillator} shows the three regimes.
\begin{itemize}
\item \emph{Below resonance} ($\omega<\omega_0$) the spring dominates: the electron follows the
field almost in phase, and the dipole points along the field. This is the regime of ordinary
transparent matter in visible light: the resonances of glass, water and air lie in the
ultraviolet, at wavelengths of order 100~nm.
\item \emph{At resonance} ($\omega=\omega_0$) the amplitude is largest, limited only by the
damping, and the dipole lags the field by a quarter of a period, $90^\circ$.
\item \emph{Above resonance} ($\omega>\omega_0$) the inertia of the electron dominates: it
oscillates in opposition of phase with the field, with an amplitude that decreases as $1/\omega^2$.
Far above all resonances, the electron behaves as if it were free.
\end{itemize}

The phase lag is not a detail: it decides the fate of the energy. The power the field delivers to
the electron, averaged over a period, is
\begin{equation}
\langle P_{\rm abs}\rangle=\tfrac12\,\omega\,\mathrm{Im}\big(p\,E_0^*\big)
=\tfrac12\,\varepsilon_0\,\omega\,\mathrm{Im}\,\alpha(\omega)\,|E_0|^2 .
\label{eq:absorbed}
\end{equation}
Only the part of the dipole that oscillates a quarter of a period behind the field absorbs
energy; the part in phase (or in opposition) with it takes energy during one half of the cycle and
gives it back during the other. Away from resonance $\mathrm{Im}\,\alpha$ is tiny and the atom is
almost a lossless, reactive object. We shall meet the opposite case---a dipole that runs a quarter
of a period \emph{ahead} of the field, and gives energy to it---in Sec.~\ref{sec:beyond}, when we discuss
amplification.

\subsection{How a charge radiates: a kink in the field lines}
\label{sec:kink}

Why should an oscillating charge radiate at all? A beautiful argument, due to
J.J.~Thomson~\cite{Thomson} and presented in this form by Purcell and Morin~\cite{Purcell},
answers with nothing more than Gauss's law and the finite speed of light.

\begin{figure}[!htbp]
\centering
\includegraphics[width=0.95\textwidth]{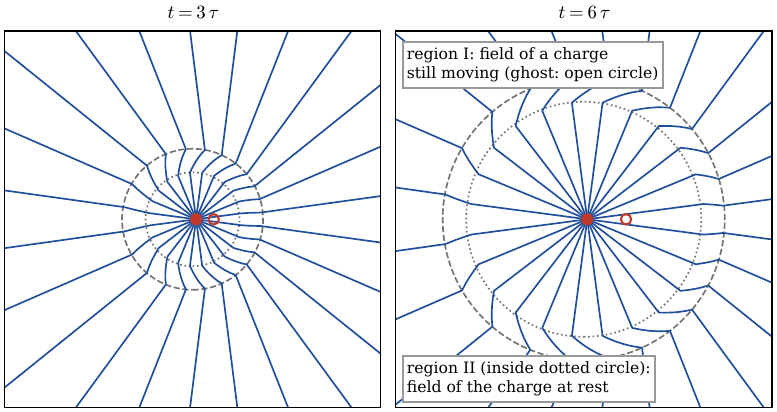}
\caption{Field lines of a charge that moved to the right at $v_0=0.3\,c$ and was stopped during
the time $\tau$, at two later instants. Outside the dashed circle ($R=cT$) the lines still point to
where the charge would be had it not stopped (open circle); inside the dotted circle
($R=c(T-\tau)$) they come from the charge at rest. The kinks in between travel outward at speed
$c$: they are the radiated pulse (\anim{1.2}{anim1_kink.mp4}). Lines drawn with the light-ray
construction of Ref.~\cite{Tsien}.}
\label{fig:kink}
\end{figure}

Take a charge $q$ that has been moving along $x$ at a constant, small speed $v_0$ for a long time.
At $t=0$ it is stopped, with a constant deceleration $a=v_0/\tau$ lasting a short time $\tau$.
Look at its electric field at a later time $T\gg\tau$ (Fig.~\ref{fig:kink}). The news of the
stop travels outward at speed $c$. Observers farther than $R=cT$ have not yet received it: for
them the field is still that of a charge in uniform motion, whose lines point radially away from
the position $x=v_0T$ where the charge would be now had it not been stopped. Observers closer
than $c(T-\tau)$ see the field of a charge at rest near the origin. In between lies a spherical
shell of thickness $c\tau$, and Gauss's law tells us what happens there: the flux inside a cone of
half-angle $\theta$ must be the same in the two regions, so a field line at angle $\theta$
outside must continue as the line at the same angle inside. The two pieces are parallel but
displaced sideways by $v_0T\sin\theta$, and the shell must join them with a \emph{kink}. Within
the kink the field has a radial component $E_r$, which by Gauss's law is the Coulomb field
$q/4\pi\varepsilon_0R^2$, and a transverse component $E_\theta$; from the geometry of the kink,
$E_\theta/E_r=v_0T\sin\theta/c\tau$, so that
\begin{equation}
E_\theta=\frac{q\,a\sin\theta}{4\pi\varepsilon_0c^2R}.
\label{eq:kink}
\end{equation}
This is the heart of the matter. The radial field decreases as $1/R^2$, but the transverse
field generated by the acceleration decreases only as $1/R$: the farther the kink travels, the
more it dominates. Together with a magnetic field $B=E_\theta/c$, it carries energy. Adding the
electric and magnetic energy contained in the shell, whose volume is $4\pi R^2c\tau$, one finds
that the energy is independent of $R$: it travels outward undiminished. Dividing by the
duration $\tau$ of the pulse gives Larmor's formula for the power radiated by an accelerated
charge,
\begin{equation}
P_{\rm rad}=\frac{q^2a^2}{6\pi\varepsilon_0c^3}.
\label{eq:Larmor-q}
\end{equation}

\begin{figure}[!htbp]
\centering
\includegraphics[width=0.95\textwidth]{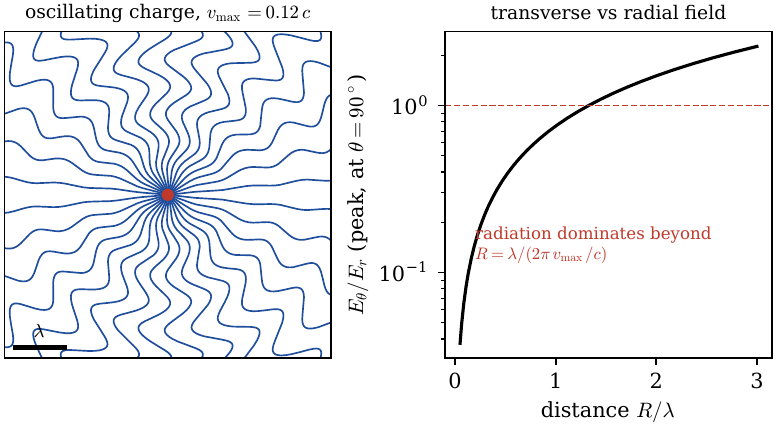}
\caption{Left: field lines of a charge oscillating horizontally with maximum speed $0.12\,c$:
the acceleration imprints transverse ripples on the radial lines, which run outward at speed $c$
(\anim{1.3}{anim1_charge.mp4}). Right: ratio of the transverse (radiation) field to the radial
(Coulomb) field along the direction perpendicular to the motion.}
\label{fig:osccharge}
\end{figure}

Now let the charge oscillate, $x(t)=x_0\sin\omega t$. Its acceleration changes continuously, and
the field lines carry a continuous train of kinks: transverse ripples running outward along the
radial lines (Fig.~\ref{fig:osccharge}, \anim{1.3}{anim1_charge.mp4}). With
$a=-\omega^2x$, Eq.~\eqref{eq:kink} gives a transverse field oscillating at the frequency of the
charge, with the retardation $R/c$, an amplitude proportional to $\sin\theta$ and decreasing as
$1/R$: a spherical light wave. Its ratio to the radial Coulomb field,
\begin{equation}
\frac{E_\theta}{E_r}=\frac{\omega^2x_0R}{c^2}\,\sin\theta=kR\,\frac{v_{\max}}{c}\,\sin\theta,
\end{equation}
grows linearly with distance: near the charge the static field dominates, but beyond
$R\sim\lambda/(2\pi v_{\max}/c)$ only the wave is left.

\paragraph{How does the wave leave the static field?} This point often puzzles students, and
it deserves a careful answer. For a single charge, the field lines never detach. They remain
attached to the charge, like a rope tied to a hand: what travels along them is a transverse
disturbance, as along a rope that is shaken, while the rope itself---the Coulomb field---stays
where it is. The energy and the momentum of the wave are carried by the kinks, and since the
radial field fades as $1/R^2$ and the transverse one as $1/R$, far away the disturbance
\emph{is} the field: a transverse wave that no longer remembers the radial lines on which it was
born.

\begin{figure}[!htbp]
\centering
\includegraphics[width=0.95\textwidth]{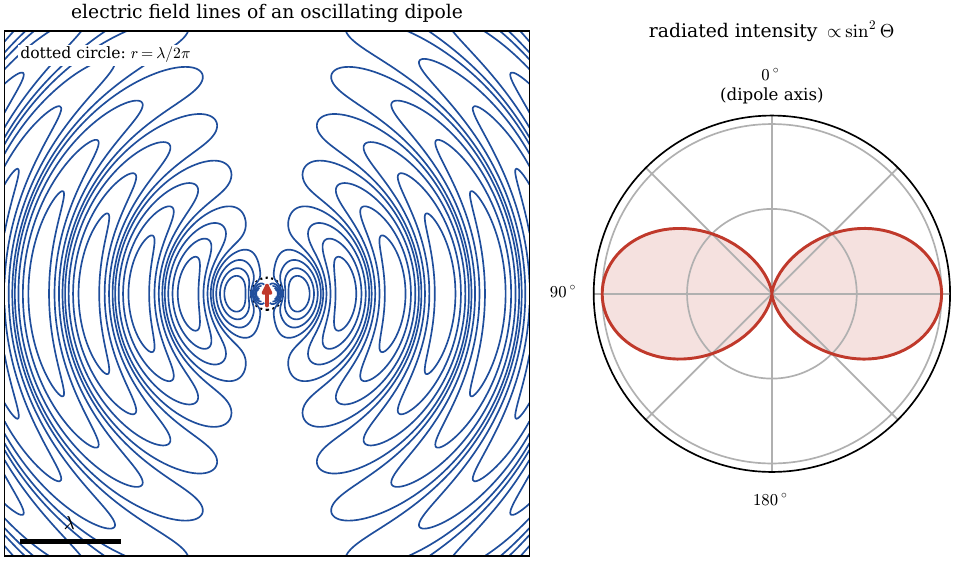}
\caption{Left: electric field lines of an oscillating dipole (vertical arrow, at the centre) at
one instant, computed from the exact near- and far-zone field; the black bar is one wavelength,
the dotted circle has radius $\lambda/2\pi$. Right: angular distribution of the radiated
intensity, $\propto\sin^2\Theta$; no radiation is emitted along the dipole axis.}
\label{fig:dipole}
\end{figure}

\begin{figure}[!htbp]
\centering
\includegraphics[width=0.95\textwidth]{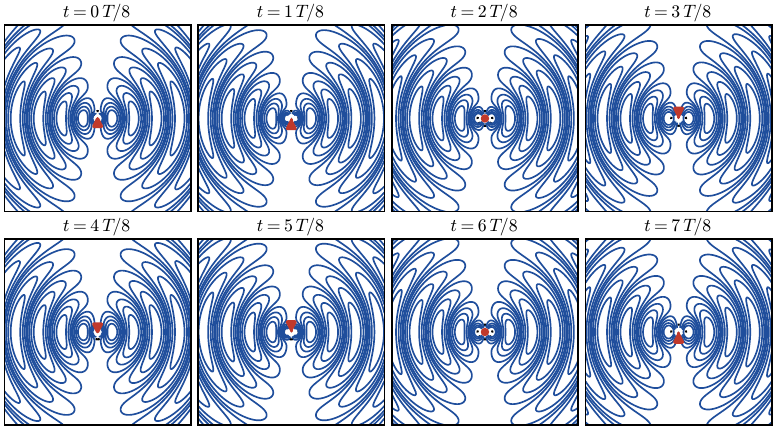}
\caption{Eight instants within one period $T$ of the dipole oscillation: field lines detach from
the dipole and move outward as closed loops (\anim{1.1}{anim1_dipole.mp4}).}
\label{fig:frames}
\end{figure}

An atom, however, is neutral: the oscillating electron is accompanied by a nucleus that hardly
moves. Far away, the radial Coulomb fields of the two charges cancel, while the ripples generated
by the accelerated electron do not. With no net charge, no field line can run to infinity: the
lines must start and end on the two charges, or close on themselves. This is where the
detachment really occurs. Each time the dipole reverses, the transverse ripples that have run
beyond the near zone can no longer be connected to the charges, and they close into loops that
travel away on their own (Figs.~\ref{fig:dipole} and~\ref{fig:frames}, \anim{1.1}{anim1_dipole.mp4}).
For a dipole the static field decreases as $1/r^3$ and the radiation field as $1/r$, both
proportional to $p$, so the crossover no longer depends on the amplitude of the motion: it
occurs at $kr\simeq1$, i.e.\ at $r\simeq\lambda/2\pi$. From there on, the electromagnetic wave
exists independently of its source.

\subsection{The dipole radiates}

The field of the oscillating dipole follows exactly from Maxwell's equations through the retarded
potentials: what is felt at distance $r$ at time $t$ was produced by the dipole at the earlier
time $t-r/c$. Within a distance of order $\lambda/2\pi$ (the \emph{near zone}) the field is
essentially the electrostatic field of the dipole, following its oscillation instant by
instant. Far away (the \emph{radiation zone}, $r\gg\lambda$) only the part that decreases as
$1/r$ survives~\cite{Jackson}:
\begin{equation}
\bE_{\rm rad}(\mathbf r,t)=\frac{k^2}{4\pi\varepsilon_0}\,\big[(\bn\times\bp)\times\bn\big]\,
\frac{e^{i(kr-\omega t)}}{r},\qquad k=\frac{\omega}{c},
\label{eq:Erad}
\end{equation}
where $\bn$ is the direction of observation; it is Eq.~\eqref{eq:kink} for the electron, with
$q\,a=\ddot p=-\omega^2p$. The field is transverse to $\bn$, lies in the plane containing $\bn$
and the dipole, and is proportional to $\sin\Theta$, where $\Theta$ is the angle between $\bn$
and the dipole axis. The dipole does not radiate at all along its own axis: we shall see in
Sec.~6 that this simple fact is the whole secret of Brewster's angle.

Averaging Eq.~\eqref{eq:Larmor-q} over a period, with $q^2\langle a^2\rangle=\omega^4|p|^2/2$,
gives the power radiated by the oscillating dipole,
\begin{equation}
\langle P_{\rm rad}\rangle=\frac{\omega^4|p|^2}{12\pi\varepsilon_0c^3}.
\label{eq:Larmor}
\end{equation}
The steep $\omega^4$ dependence will paint the sky blue in a moment.

\begin{panel}{Radiation damping and the lifetime of an atom}
The radiated energy must come from the oscillator. A free oscillation of amplitude $x_0$ stores
the energy $U=\tfrac12m\omega_0^2x_0^2$ and, by Eq.~\eqref{eq:Larmor} with $|p|=ex_0$, radiates
the power $e^2\omega_0^4x_0^2/12\pi\varepsilon_0c^3$. The energy therefore decays at the rate
\begin{equation*}
\Gamma=\frac{\langle P_{\rm rad}\rangle}{U}=\frac{e^2\omega_0^2}{6\pi\varepsilon_0mc^3}
=\frac23\,\frac{r_e\,\omega_0^2}{c},\qquad r_e=\frac{e^2}{4\pi\varepsilon_0mc^2}=2.82\times10^{-15}~\mathrm{m},
\end{equation*}
which corresponds to a radiative contribution $\gamma_{\rm rad}=\Gamma/2$ to the damping of
Eq.~\eqref{eq:oscillator}. Here $r_e$ is the classical radius of the electron. For the yellow
light of sodium ($\lambda_0=589$~nm) this gives a lifetime $1/\Gamma\simeq16$~ns, remarkably close
to the measured lifetime of the sodium $3p$ level, about 16~ns. The quality factor of the
radiating oscillator, $\omega_0/\Gamma=3\lambda_0/4\pi r_e$, is of order $10^7$--$10^8$: the
atomic oscillator is an extraordinarily sharp resonator. In real matter collisions and other
processes add to $\gamma$, but in a transparent medium, far from resonance, the damping hardly
matters.
\end{panel}

\subsection{Scattering: amplitude and cross section}

\begin{figure}[!htbp]
\centering
\includegraphics[width=0.72\textwidth]{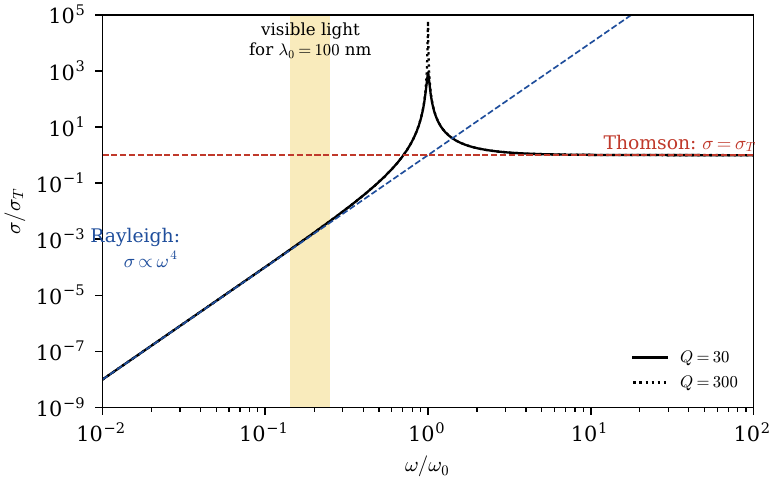}
\caption{Scattering cross section of the Lorentz oscillator, Eq.~\eqref{eq:sigma}, in units of
the Thomson cross section, for $Q=\omega_0/2\gamma=30$ and $300$ (the damping is exaggerated: for
a real atom $Q\sim10^7$). The shaded band shows where visible light falls for an oscillator
resonating in the ultraviolet at $\lambda_0=100$~nm.}
\label{fig:cross}
\end{figure}

Seen from the outside, the atom has taken a little of the incident wave and re-sent it in all
directions: it has \emph{scattered} it. In the radiation zone, in the plane perpendicular to the
dipole, the scattered wave can be written as
\begin{equation}
E_{\rm sc}=f(\omega)\,E_0\,\frac{e^{ikr}}{r},\qquad
f(\omega)=\frac{k^2\alpha(\omega)}{4\pi}=r_e\,\frac{\omega^2}{\omega_0^2-\omega^2-2i\gamma\omega},
\label{eq:f}
\end{equation}
which defines the \emph{scattering amplitude} $f$, a length. (In other directions the amplitude
is multiplied by $\sin\Theta$.) Two of its properties will be essential in what follows.
\begin{itemize}
\item It is tiny: below resonance $f$ is of the order of $r_e$, a few femtometres, while the
wavelength is hundreds of nanometres: the ratio $f/\lambda$ is of order $10^{-8}$.
\item It carries a phase. Below resonance $f$ is real and positive: the scattered wave leaves the
atom in step with the field that drives it. Above resonance $f$ is negative; for a free electron
($\omega\gg\omega_0$) $f\to-r_e$.
\end{itemize}
Squaring the amplitude and integrating over directions gives the total power removed from the
incident beam. Dividing by the incident intensity defines the \emph{scattering cross section},
the area that the atom seems to present to the wave:
\begin{equation}
\sigma(\omega)=\frac{8\pi}{3}\,|f|^2=\sigma_T\,
\frac{\omega^4}{(\omega_0^2-\omega^2)^2+4\gamma^2\omega^2},\qquad
\sigma_T=\frac{8\pi}{3}\,r_e^2=6.65\times10^{-29}~\mathrm{m}^2 .
\label{eq:sigma}
\end{equation}
Figure~\ref{fig:cross} shows the three regimes. Far below resonance
$\sigma\simeq\sigma_T(\omega/\omega_0)^4$ (Rayleigh scattering); far above, $\sigma\to\sigma_T$
(Thomson scattering by a free electron). At resonance, if the damping is purely radiative,
$\sigma(\omega_0)=3\lambda_0^2/2\pi$: an atom, a tenth of a nanometre across, intercepts light
over an area of the order of the square of the wavelength, millions of times its geometric size.

\subsection{Why the sky is blue}

\begin{figure}[!htbp]
\centering
\includegraphics[width=0.95\textwidth]{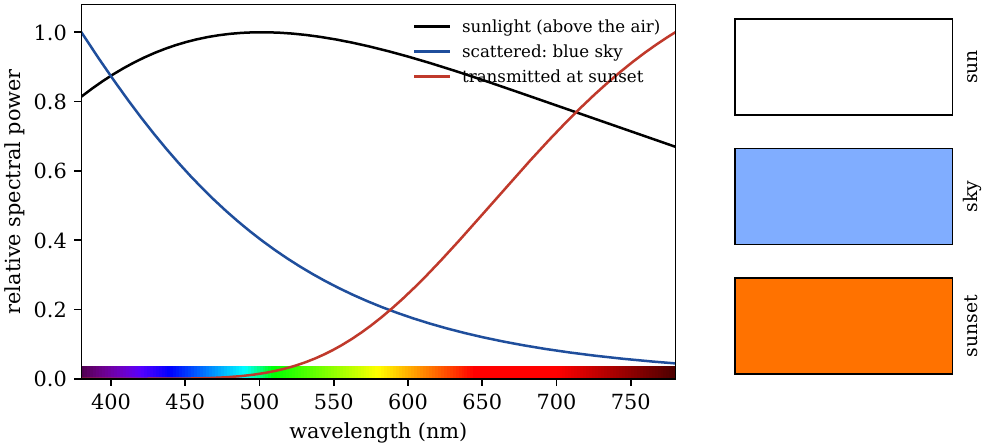}
\caption{Left: spectrum of sunlight outside the atmosphere (black-body at 5778~K), of the light
scattered by air ($\propto\lambda^{-4}$) and of sunlight transmitted at sunset (38 air masses,
Rayleigh optical depth $0.0088\,\lambda_{\mu\rm m}^{-4.05}$ at the zenith); curves normalized to
their maxima. Right: the corresponding colours, computed from the CIE colour-matching
functions~\cite{Wyman} for an eye adapted to sunlight.}
\label{fig:sky}
\end{figure}

Air molecules resonate in the ultraviolet, so visible light is scattered in the Rayleigh regime,
with $\sigma\propto\omega^4\propto\lambda^{-4}$~\cite{Rayleigh,BohrenHuffman}. Blue light at 450~nm is
scattered $(700/450)^4\simeq6$ times more than red light at 700~nm. Looking away from the sun we
see only scattered light, enriched in short wavelengths: the sky is blue. Looking at the setting
sun through a long path in air, we see the light that has \emph{not} been scattered, depleted of
blue: the sun is red (Fig.~\ref{fig:sky}).

Two remarks complete the picture. Why is the sky blue and not violet, if violet is scattered even
more? Because sunlight contains less violet than blue, and our eyes are much less sensitive to it:
the colour on the right of Fig.~\ref{fig:sky} is computed from the scattered spectrum on the
left, and it is sky-blue. And the light of the sky is \emph{polarized}: at $90^\circ$ from the
sun we receive only the radiation of dipoles oscillating perpendicularly to our line of sight,
since those oscillating along it do not radiate towards us. A pair of polarizing sunglasses,
rotated while looking at the sky, shows it at once \rev{(see the panel ``Polarization and polarizers'' at the end of this section)}.

{\revon
\emph{Counting molecules with the sky.} The energy scattered sideways is taken from the direct beam,
which decays as $e^{-N\sigma L}$ (Beer's law): the setting sun is red because the blue has been removed
more. Rayleigh turned this into a way of counting molecules. Eliminating the polarizability with the
index of refraction, Eq.~\eqref{eq:Fermi} of Sec.~3, the attenuation coefficient becomes
\begin{equation*}
N\sigma=\frac{32\pi^3\,(n-1)^2}{3N\lambda^4},
\end{equation*}
so that measuring the transparency of the air and its index gives $N$. With $n-1=2.93\times10^{-4}$ and
$N=2.55\times10^{25}~\mathrm{m^{-3}}$ the attenuation length $1/N\sigma$ is about 80~km at 550~nm and
160~km at 650~nm: from the visibility of distant mountains Rayleigh obtained a number of molecules
close to Loschmidt's. And how large are the oscillations involved? Sunlight, about 1~kW/m$^2$, has a
field of some 870~V/m; an electron bound with a resonance at 100~nm oscillates under it with an
amplitude of $4\times10^{-19}$~m, less than a thousandth of the size of a nucleus, and a free electron in green
light about thirty times more~\cite{Weisskopf}.
\par}

\begin{figure}[!htbp]
\centering
\includegraphics[width=\textwidth]{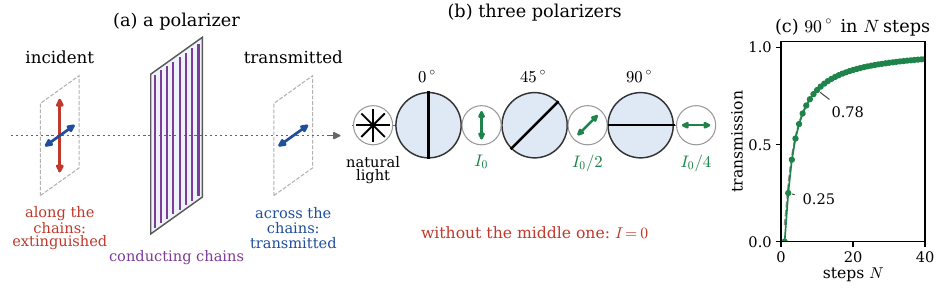}
\caption{\rev{(a) A polarizer as an array of conducting molecular chains: the field along the
chains drives the electrons, whose radiation extinguishes it; only the field across the chains is
transmitted. The transverse plane is drawn in perspective. (b) Natural light through three
polarizers, the field being shown face-on between them: crossed polarizers alone transmit nothing,
a third one at $45^\circ$ between them transmits $I_0/4$. (c) Transmission when the polarization is
turned through $90^\circ$ in $N$ equal steps (dots); the approximation $e^{-\pi^2/4N}$
(dashed) differs visibly only for the first few steps.}}
\label{fig:polarizers}
\end{figure}

\begin{panel}{\rev{Polarization and polarizers}}
{\revon
\emph{What it is.} Far from its source the electric field of a light wave is transverse: it
oscillates in the plane perpendicular to the direction of propagation. The direction along which it
oscillates is the \emph{polarization} of the wave. If it always oscillates along the same line, the
light is \emph{linearly polarized}; this is the light radiated by a single dipole, whose field lies
along the projection of the dipole on that plane. Two perpendicular components oscillating a
quarter of a period apart make the field rotate, and the light is \emph{circularly} (in general
\emph{elliptically}) polarized. The light of the sun or of a lamp is emitted by a huge number of
independent atoms, each with its own direction and phase: its direction of oscillation changes at
random many times in a nanosecond, and on average no direction is preferred. It is
\emph{unpolarized}, or natural, light.

\emph{How a polarizer works.} A polarizing sheet (Land's Polaroid) is a film of long molecules,
aligned by stretching and made conducting by iodine: a dense array of tiny parallel wires
(Fig.~\ref{fig:polarizers}a). The component of the field along the chains drives their electrons,
which radiate a forward wave that cancels it (the same mechanism as the extinction theorem of
Sec.~\ref{sec:extinction}); its energy is absorbed, or, in a wire-grid polarizer, reflected. Across
the chains the electrons cannot move, and that component passes. A polarizer therefore transmits
the field \emph{across} its wires, not along them, contrary to the image of light slipping through
slits. A field at the angle $\phi$ to the transmission axis is projected on it, $E\cos\phi$, and the
intensity is reduced to $I_0\cos^2\phi$ (Malus' law, 1809); natural light, on average, loses half.

\emph{A projection, not a filter.} Two crossed polarizers transmit nothing, but a third one placed
between them at $45^\circ$ lets $I_0/4$ through (Fig.~\ref{fig:polarizers}b): the middle polarizer
adds no light, it projects the field on a new direction: its electrons radiate a new wave,
polarized along its own axis, which has a component along the last axis. With $N$ polarizers turning the direction in equal steps through $90^\circ$ the transmission,
$\cos^{2N}(\pi/2N)\approx e^{-\pi^2/4N}$, tends to one (Fig.~\ref{fig:polarizers}c): a field turned
gently is turned without loss. The twisted-nematic cell of a liquid-crystal display does exactly
this, continuously, between crossed polarizers; a voltage untwists it, and the pixel goes dark.

\emph{One photon at a time.} For a single photon, $\cos^2\phi$ is a probability: the photon either
passes or does not, and Malus' law holds on average. Polarization is the simplest quantum system,
with two states, described by the same Pauli matrices as the spin of an electron, and a polarizer
is a measurement that projects the photon on its axis. The long chain of polarizers that turns the
photon without losing it is a version of the quantum Zeno effect: frequent measurements keep the
state on a prescribed path.
\par}
\end{panel}

But there is a deeper question hidden here, and it is the question with which Weisskopf opens
his article~\cite{Weisskopf}. Every molecule of air scatters sideways, and so does every molecule
of glass or water---more strongly, since they are packed a thousand times more densely. Why,
then, is a window not luminous and blue when the sun shines on it from the side? Why does a
glass of water not glow? Why does light go \emph{through} dense matter, almost undisturbed, in a
straight line? A single atom cannot answer. The answer lies in how the waves scattered by many
atoms add up---and it is the subject of the next chapter.

\begin{panel}{Why classical oscillators?}
Quantum mechanics is, of course, the correct theory of atoms. Yet for the linear response of an
atom in its ground state to a weak field, the classical oscillator of Eq.~\eqref{eq:oscillator}
is not a mere analogy: it is the answer that quantum mechanics gives. The amplitude for elastic
forward scattering of a photon, from the Kramers--Heisenberg formula~\cite{KH,FeynmanQED,Holstein},
is
\begin{equation*}
f(\omega)=r_e\sum_n f_n\,\frac{\omega^2}{\omega_n^2-\omega^2-i\Gamma_n\omega},\qquad
f_n=\frac{2m\,\omega_n}{\hbar}\,|\langle n|\hat x|0\rangle|^2,\qquad \sum_nf_n=1 ,
\end{equation*}
a sum of Lorentz oscillators, one for each transition $0\to n$ of frequency $\omega_n$, each
weighted by its \emph{oscillator strength} $f_n$; the Thomas--Reiche--Kuhn sum rule fixes the total
weight to one per electron. All the phenomena of this text---scattering, refraction, dispersion,
absorption, reflection, polarization, the propagation of pulses---follow from these classical
oscillators, as Feynman stressed in his \emph{Lectures}~\cite{FeynmanI}; see also~\cite{MilonniEberly}.

The classical model fails where the quantum nature of the atom or of the field becomes
essential:
\begin{itemize}
\item \emph{Spontaneous emission.} \rev{A classical oscillator at rest does not radiate; an excited
atom left alone does. It is often said that the vacuum fluctuations of the quantized field
``shake'' the atom out of its excited state, but the picture is incomplete: with a suitable
(symmetric) ordering of atomic and field operators, the decay rate is the sum of two equal
halves, one due to the vacuum fluctuations and one due to radiation reaction, the quantum
counterpart of the classical self-field; in the ground state the two contributions cancel, and
the atom is stable~\cite{DDC1982}. The classical model thus gives the right decay
\emph{rate} (see the panel on radiation damping), but not why an undisturbed excited atom
decays while one in the ground state does not, nor the emission of light in single photons.}
\item \emph{Strong fields.} When the field is strong enough to change the populations of the
levels, the response is no longer linear (saturation).
\item \emph{Inelastic (Raman) scattering}, in which the atom ends in a state different from the
initial one.
\end{itemize}
An atom in an \emph{excited} state, instead, responds to a weak field like an oscillator with a
\emph{negative} oscillator strength: it gives energy to the wave instead of taking it. This is
the classical image of stimulated emission, and of the laser amplifier (Sec.~\ref{sec:beyond}).
\end{panel}

%% file: ch2_sheet.tex
\section{A sheet of atoms}
\label{sec:sheet}

One atom scatters light in all directions. A window pane contains some $10^{23}$ atoms per cubic
centimetre, all illuminated by the same wave, all scattering. If each of them simply added its
little share of light, the pane would glow like a lamp when the sun shines on it from the side.
It does not, and the reason is that waves do not add like heaps of sand: they add with their
phases. This chapter follows the simplest possible collection of atoms---a thin plane sheet---and
shows what comes out of the sum.

\subsection{Order and disorder}
\label{sec:order}

\begin{figure}[!htbp]
\centering
\includegraphics[width=0.95\textwidth]{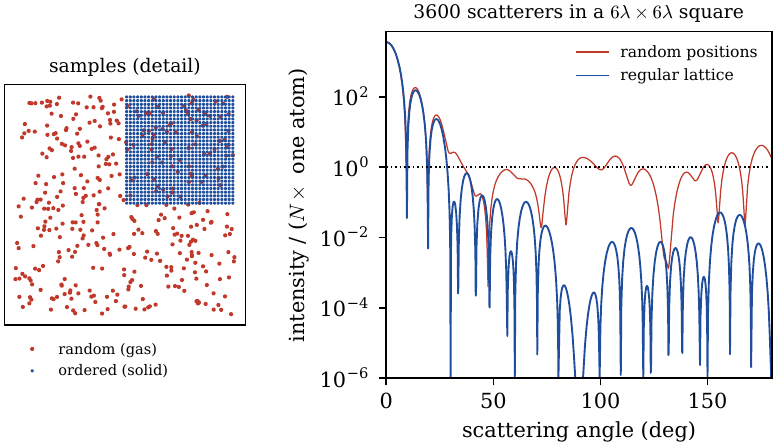}
\caption{Left: the two samples (a detail): atoms at random positions (red) and on a square lattice
of spacing $\lambda/10$ (blue). Right: intensity of the scattered light as a function of the
angle from the incident direction, in units of $N$ times the intensity scattered by one atom.
Forward ($0^\circ$) both samples give $N^2$; sideways the random sample gives $N$ on average,
the ordered one almost nothing.}
\label{fig:order}
\end{figure}

Let a plane wave travel along $x$ and fall on $N$ identical atoms at positions $\mathbf r_j$. The
atom $j$ is driven with the phase $e^{i\mathbf k\cdot\mathbf r_j}$ of the incident wave at its
position, and its wavelet reaches a distant detector in the direction $\mathbf k'$ (with
$|\mathbf k'|=|\mathbf k|$) with the additional phase $-\mathbf k'\cdot\mathbf r_j$. The total
scattered amplitude is that of a single atom multiplied by the \emph{structure factor}
\begin{equation}
F(\mathbf q)=\sum_{j=1}^Ne^{i\mathbf q\cdot\mathbf r_j},\qquad \mathbf q=\mathbf k-\mathbf k'.
\label{eq:structure}
\end{equation}
Three cases cover everything that follows.
\begin{itemize}
\item \emph{Forward direction}, $\mathbf k'=\mathbf k$: all the phases vanish, whatever the
positions of the atoms. The wavelets add in step, $F=N$, and the intensity is $N^2$ times that of
one atom. This is true for a gas as much as for a crystal: \emph{forward scattering is always
coherent}.
\item \emph{Other directions, random positions} (a gas): the phases are random, and the
intensities add, $|F|^2\simeq N$. Each molecule scatters on its own: this is the light of the sky.
\item \emph{Other directions, regular positions} much closer than a wavelength (a solid, and on
average a liquid): for every atom there is another whose wavelet arrives in opposition of phase,
and the sum cancels, $|F|^2\ll N$. Nothing is scattered sideways.
\end{itemize}
Figure~\ref{fig:order} shows the angular distribution of the light scattered by 3600 atoms in a
square six wavelengths wide, arranged either at random or on a regular lattice. The random
sample scatters in all directions, on average with $N$ times the intensity of one atom; the
regular one sends almost everything forward, where both reach $N^2$. What survives sideways from
the regular sample comes only from its edges.

This answers the question left open at the end of Sec.~\ref{sec:one-atom}. Glass and water do not
glow because their atoms are packed closely and evenly: their sideways wavelets cancel. A real
liquid is not perfectly regular, of course. Its density fluctuates a little from place to place,
and the light scattered sideways measures exactly these fluctuations, as Smoluchowski and
Einstein understood~\cite{Smoluchowski,Einstein1910}: for water at room temperature it is more than
ten times weaker, per molecule, than for an ideal gas. \rev{The intensity scattered is proportional to the
mean square fluctuation of the number of molecules in a small volume, $\langle\Delta N^2\rangle=N\,(Nk_BT\kappa_T)$,
where $\kappa_T$ is the isothermal compressibility; for an ideal gas the bracket is one, for water
about $0.06$. Per unit volume, water, a thousand times denser than air, scatters only about a hundred
times more (Weisskopf's estimate in Ref.~\cite{Weisskopf}, five times, is too low, but the point stands).} Sideways scattering is the exception, the mark of
disorder; what the regular arrangement leaves is the light that goes forward. The rest of this
chapter is about that light.

\subsection{Fresnel zones and the quarter-period delay}

\begin{figure}[!htbp]
\centering
\includegraphics[width=0.95\textwidth]{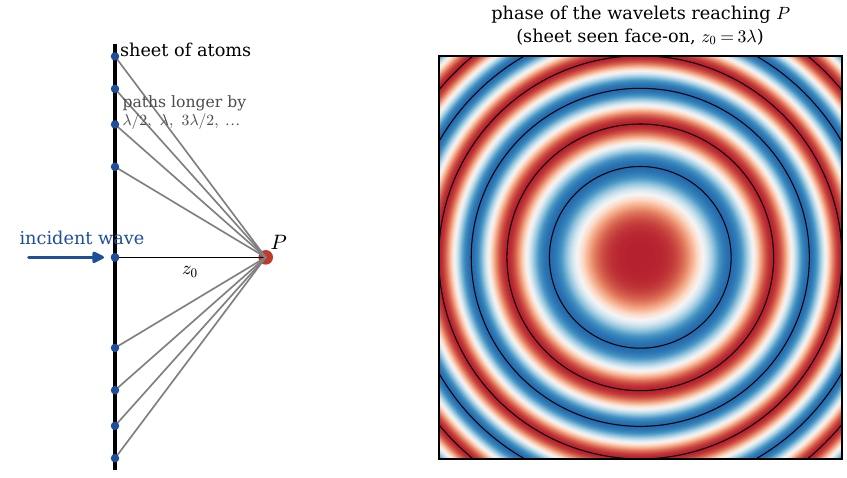}
\caption{Left: the wavelets from a sheet of atoms reach the point $P$ along paths of different
length; the rings whose paths differ by half a wavelength bound the Fresnel zones. Right: the
phase of the wavelets arriving at $P$ from each point of the sheet, seen face-on (red and blue:
opposite phases), with the zone boundaries.}
\label{fig:zones}
\end{figure}

Take now a thin sheet of thickness $\ud z$, perpendicular to the incident wave, containing $N$
atoms per unit volume, and ask what field it produces at a point $P$ at distance $z_0$ behind it
(Fig.~\ref{fig:zones}). All the atoms of the sheet are driven in phase, since the incident wave
reaches them at the same time. But their wavelets reach $P$ along paths of different length: the
atoms on a ring of radius $\rho$ are at the distance $r=\sqrt{z_0^2+\rho^2}$. Fresnel's idea was
to divide the sheet into rings---the \emph{Fresnel zones}---whose distances from $P$ differ by
successive half-wavelengths. For $\rho\ll z_0$ the zone boundaries have radii $\rho_m\simeq\sqrt{m\lambda z_0}$,
and all the zones have the same area $\pi\lambda z_0$. Consecutive zones therefore contribute
equal amounts in opposition of phase.

\begin{figure}[!htbp]
\centering
\includegraphics[width=0.95\textwidth]{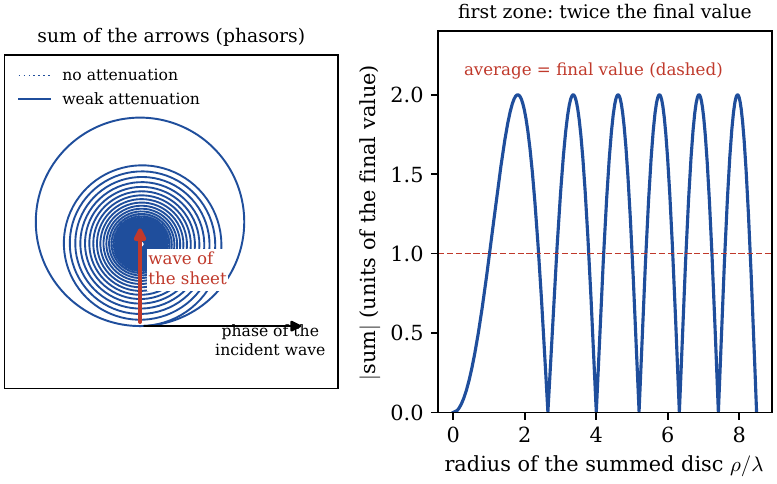}
\caption{Left: the arrows of the wavelets from the sheet, added from the centre outward. Without
attenuation the sum runs round a circle; with any slight attenuation of the far contributions it
spirals into the centre. The resultant (red) is perpendicular to the arrow of the incident wave
(black): a quarter-period phase shift. Right: magnitude of the sum as a function of the radius
of the summed disc; its average is the final value, reached when the far zones are weakened.}
\label{fig:spiral}
\end{figure}

The sum is best done with arrows, as Feynman taught~\cite{FeynmanQED-book}: each small piece of
the sheet contributes a little arrow whose direction is the phase of its wavelet at $P$. As the
summed area grows, the arrow turns steadily, because the path length grows; the arrows of the
first zone curve round half a circle, those of the second complete the circle and bring the sum
back to zero, and so on (Fig.~\ref{fig:spiral}). The sum circles forever and never settles---it
oscillates between zero and twice the diameter of the circle. Its \emph{average} is
well defined: it is the centre of the circle. Any physical reason that makes the far zones count
a little less---the finite width of a real beam, the slight obliquity of the far wavelets, a weak
absorption---turns the circle into a spiral that winds into that centre. The result is the same
in all cases: the sheet produces at $P$ a wave equal to \emph{half the contribution of the first
zone}, and pointing at right angles to the arrow of the incident wave.

\begin{figure}[!htbp]
\centering
\includegraphics[width=0.95\textwidth]{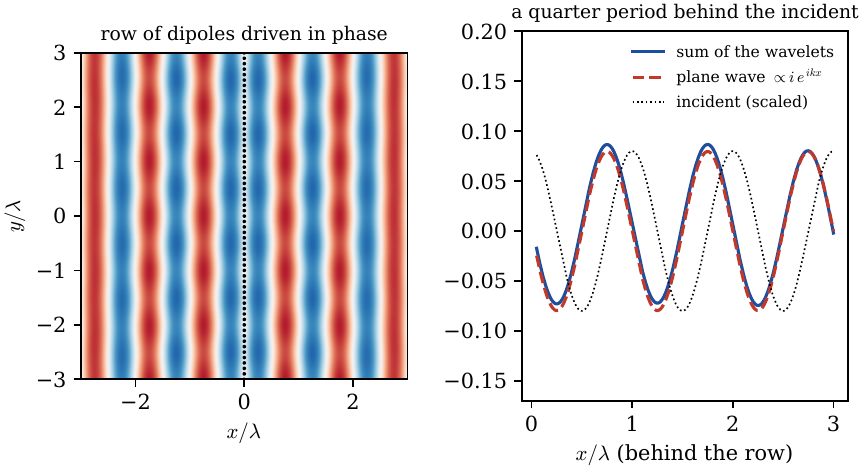}
\caption{Left: field radiated by a long row of dipoles (dots) driven in phase by a wave coming
from the left, computed by adding the exact cylindrical wavelets in two dimensions (spacing
$\lambda/10$; the incident wave is not shown). Close to the row the individual wavelets are
visible; at a fraction of a wavelength they have merged into two plane waves, one forward and one
backward. Right: the field behind the row (blue) coincides with a plane wave lagging a quarter of a
period behind the incident wave (red, dashed; the incident wave, dotted, is shown on an arbitrary
scale) (\anim{2.1}{anim2_row.mp4}).}
\label{fig:row}
\end{figure}

Doing the sum explicitly (panel below) gives the field of the sheet,
\begin{equation}
E_{\rm sheet}(P)=\frac{2\pi i\,Nf\,\ud z}{k}\,E_0\,e^{ikz_0}
=\frac{i\omega}{2\varepsilon_0c}\,P\,\ud z\;e^{ikz_0},
\label{eq:sheet}
\end{equation}
where $f$ is the scattering amplitude of one atom, Eq.~\eqref{eq:f}, and $P=N\varepsilon_0\alpha E_0$
is the polarization, the dipole moment per unit volume. \rev{Four} facts are written in this formula.
\begin{itemize}
\item \rev{\emph{It holds at every distance.} Although we built it from the far-field wavelets of the
atoms, Eq.~\eqref{eq:sheet} is exact at any distance from the sheet, even a small fraction of a
wavelength: computed directly from the retarded potentials of the oscillating charges, without any
far-zone approximation, it gives the same result~\cite{LLW}. The near fields of the individual
dipoles---the quasi-static parts that fall off as $1/r^2$ and $1/r^3$---cancel when summed over a
uniform sheet, just as a plane uniformly polarized along itself produces no static field. This is
what allows us, in Sec.~\ref{sec:extinction}, to add up sheets that are arbitrarily close to one
another.}
\item The spherical wavelets have merged into a \emph{plane wave}: the field of the sheet does
not decrease with the distance $z_0$, exactly like the field of a uniformly charged plane in
electrostatics. Figure~\ref{fig:row} and \anim{2.1}{anim2_row.mp4} show the same merging for a
row of dipoles.
\item The factor $i$: the wave of the sheet lags a quarter of a period behind the wave that
drives it (for $f>0$, below resonance). This quarter period is the key to the refractive index
(Sec.~3). \rev{Its origin is simple: the field of a sheet follows the \emph{velocity} of its
charges, not their displacement (panel ``The sheet in the time domain'' below).}
\item The sheet radiates equally \emph{backward}, with $e^{ik|z_0|}$: a single sheet of atoms
already reflects. This is the seed of reflection (Sec.~5).
\end{itemize}

\begin{panel}{The sum over the sheet}
The atoms in a ring of radius $\rho$ and width $\ud\rho$ number $N\,\ud z\,2\pi\rho\,\ud\rho$ and
are all at the distance $r=\sqrt{z_0^2+\rho^2}$ from $P$; since $\rho\,\ud\rho=r\,\ud r$,
\begin{equation*}
E_{\rm sheet}(P)=E_0\,Nf\,\ud z\int_{z_0}^{\infty}\frac{e^{ikr}}{r}\,2\pi r\,\ud r
=2\pi Nf\,\ud z\,E_0\left[\frac{e^{ikr}}{ik}\right]_{z_0}^{\infty}.
\end{equation*}
The upper limit oscillates without converging: it is the circling arrow of Fig.~\ref{fig:spiral}.
With a convergence factor $e^{-\epsilon r}$, $\epsilon\to0^+$, it vanishes, and the lower limit
gives Eq.~\eqref{eq:sheet}. The convergence factor is a convenient device, not a necessity: as
shown in the next section, at oblique incidence the integral converges by itself, and normal
incidence is recovered as a regular limit. The amplitude $f$ depends on the scattering angle,
but only the forward value $f(0)$ matters: the far zones cancel in pairs, and only the first one,
seen from $P$ almost straight ahead, survives.
\end{panel}

\subsection{Oblique incidence}
\label{sec:oblique}

When the wave falls on the sheet at an angle $\theta_i$, with wave vector
$\mathbf k=(q,0,k_z)$, $q=k\sin\theta_i$, $k_z=k\cos\theta_i$, the atoms are no longer driven in
phase: the atom at the transverse position $\boldsymbol\rho$ oscillates with the phase
$e^{i\mathbf q\cdot\boldsymbol\rho}$. The sum over the sheet becomes \rev{(see the panel below)}
\begin{equation}
\int\ud^2\rho\;e^{i\mathbf q\cdot\boldsymbol\rho}\,\frac{e^{ikr}}{r}
=2\pi\int_{z_0}^{\infty}e^{ikr}\,J_0\!\left(q\sqrt{r^2-z_0^2}\right)\ud r
=\frac{2\pi i}{k_z}\,e^{ik_zz_0},
\label{eq:Weyl}
\end{equation}
where $J_0$ is the Bessel function. This time the integral converges without any help, because
the oscillating phases of the far atoms no longer conspire. The result, used by Lai, Lau and
Wong~\cite{LLW} to generalize the treatment of Feynman and Schwartz to oblique incidence, is
Weyl's representation of the spherical wave read backwards~\cite{Weyl}: a sheet of sources with
the transverse phase $e^{i\mathbf q\cdot\boldsymbol\rho}$ radiates plane waves with the same
transverse wave vector $\mathbf q$. Two directions have it: the forward direction of the incident
wave, and its mirror image, $(q,0,-k_z)$. The sheet sends a transmitted wave straight on and a
\emph{specularly reflected} wave back: the law of reflection is already contained in the
phases. \rev{The same formula answers another question: what if the phase of the sources varies
along the sheet faster than that of any wave in vacuum, $q>k$, as for a pattern of polarization finer
than a wavelength? Then $k_z=i\sqrt{q^2-k^2}$ is imaginary, and the sheet radiates no wave at all: its
field, $e^{iqx}e^{-\sqrt{q^2-k^2}\,|z|}$, clings to the sheet and dies away within a fraction of a
wavelength. Here the near fields of the dipoles no longer cancel. These \emph{evanescent} fields carry
all the details finer than the wavelength, and we shall meet them again in the microscope
(Sec.~\ref{sec:instruments}).} For the polarization perpendicular to the plane of incidence the field of the sheet is
\begin{equation}
E_{\rm sheet}=\frac{i\omega}{2\varepsilon_0c\cos\theta_i}\,P\,\ud z\;e^{i(qx\pm k_zz)},
\end{equation}
the factor $1/\cos\theta_i$ accounting for the longer path through the sheet; for the other
polarization the dipoles are tilted, and the backward wave acquires an angular factor that will
lead us to Brewster's angle (Sec.~6). Normal incidence is the limit $\theta_i\to0$ of
Eq.~\eqref{eq:Weyl}, and it is regular: the convergence factor of the panel above gives the right
answer because it reproduces this limit.

\rev{\paragraph{The order of the limits.} There is a lesson here that goes beyond optics. The two
operations ``add up the wavelets of the whole sheet'' and ``let the angle of incidence go to zero''
do not commute. If we first set the angle to zero and then try to sum, the sum does not exist: the
arrows circle forever. If we first sum, at any angle however small, and then let the angle go to
zero, we obtain a perfectly definite answer. A familiar example of the same behaviour is the
integral $\int_0^\infty\sin(ax)/x\,\ud x$, which equals $\pi/2$ for every $a>0$, however small, but
$0$ for $a=0$: the limit of the integral is not the integral of the limit. Which order is the
physical one? Exactly normal incidence on an infinite sheet is an idealization that no experiment
can realize: every real beam contains a spread of directions, and every real sheet is finite. The
physical answer is the one that is stable against such small imperfections, that is, the limit
taken \emph{after} the sum. The convergence factor $e^{-\epsilon r}$ of the previous panel is
another way of making the same choice: it, too, is removed only after the integral has been done.}
\rev{Causality offers a third way, the most physical of all: a wave switched on at a finite time
has been heard only from a finite disc of the sheet, and the circling contribution of the edge is
a transient that moves ever farther out (panel ``The sheet in the time domain'').}

\begin{panel}{Where the integral comes from}
The middle step of Eq.~\eqref{eq:Weyl} follows by \rev{using polar coordinates on the
sheet, $\mathrm d^2\rho=\rho\,\ud\rho\,\ud\varphi$, and} integrating over the azimuth\rev{:
the atoms of a ring of radius $\rho$ are all at the same distance $r$ from $P$, but the
incident wave gives them the phase $q\rho\cos\varphi$, and the sum of their arrows is}
$\int_0^{2\pi}e^{iq\rho\cos\varphi}\ud\varphi=2\pi J_0(q\rho)$\rev{, which defines the
Bessel function $J_0$ (for $q=0$ the arrows are aligned and $J_0=1$); then
$\rho\,\ud\rho=r\,\ud r$, as in the panel of Sec.~2.2, with $\rho=\sqrt{r^2-z_0^2}$}.
The last step uses the Laplace transform
$\int_{z_0}^{\infty}e^{-pr}J_0\big(q\sqrt{r^2-z_0^2}\big)\ud r=e^{-z_0\sqrt{p^2+q^2}}/\sqrt{p^2+q^2}$,
$\mathrm{Re}\,p>0$; since for $q\neq0$ the integral converges also at $p=-ik$, Abel's continuity
theorem allows one to set $p=-ik$, with $\sqrt{q^2-k^2}=-ik_z$. For $q=0$ the integral
$\int e^{ikr}\ud r$ does not converge: this is the indefinite factor $e^{ik\infty}$ met by
Feynman~\cite{FeynmanI}, and the reason why Schwartz~\cite{Schwartz} had to argue that it averages
to zero. As Lai, Lau and Wong observe, normal incidence is simply a singular case of a regular
problem.
\end{panel}

\begin{panel}{\rev{The sheet in the time domain}}
{\revon
Feynman introduces radiation, in the second volume of his lectures, with an even simpler sheet: a
plane of charge, with surface density $\sigma$, suddenly set into motion along itself at $t=0$ with
velocity $u$~\cite{FeynmanII}. Holstein worked this example out from the retarded potentials and
used it to resolve the paradoxes it seems to raise~\cite{Holstein1995}. Three of his results throw
light on the sum over the sheet of this section.

\emph{Only the ring that is being heard.} At a distance $x$ from the plane and at a time $t>x/c$,
the only charges whose acceleration is seen at $P$ are those on the ring of radius
$\rho=\sqrt{c^2t^2-x^2}$, from which the news of the kick is just arriving; the charges inside the
ring are already moving uniformly and contribute only to the magnetic field. The field of each
charge on the ring falls off as $1/r$, but the number of charges on the ring grows as $r$: the
radiated field does not decrease, and is
\begin{equation*}
E=-\frac{\sigma u}{2\varepsilon_0c}\,\theta\!\left(t-\frac{x}{c}\right),
\end{equation*}
directed against the motion: a uniform step travelling at the speed $c$ on both sides of the
sheet. This is the time-domain
version of the circling arrow of Fig.~\ref{fig:spiral}: in a steady wave the outermost ring keeps
adding a contribution of constant size and turning phase, and the sum never settles. For a wave
switched on at a finite time only a finite disc has been heard from: causality itself provides
the cut-off that the convergence factor imitates.

\emph{Velocity, not acceleration.} The field is proportional to the velocity of the charges, that
is, to the surface current $K=\sigma u$: $E=-K/(2\varepsilon_0c)$. The accelerations of the single
charges, summed over the ring that is arriving, add up to a field that follows the velocity. For
the oscillating sheet of this section the surface current is $K=-i\omega P\,\ud z$, and the same
rule gives exactly Eq.~\eqref{eq:sheet}, $E=i\omega P\,\ud z/(2\varepsilon_0c)$. The factor $i$,
the quarter-period lag that will become the refractive index, is the difference between the
displacement of the charges, in phase with the driving field below resonance, and their velocity.

\emph{The sheet brakes itself.} The sheet feels its own field, which opposes the current: keeping
it in motion takes a power $K^2/(2\varepsilon_0c)$ per unit area, exactly the flux radiated on its
two sides. For a sheet of dipoles this self-field is the radiation reaction of the sheet as a
whole, the collective counterpart of the radiative damping of a single atom
(Sec.~\ref{sec:one-atom}); the energy it requires is supplied by the field that drives the
sheet.
\par}
\end{panel}

The sheet is the elementary brick of everything that follows. Stack many sheets one behind the
other, and the quarter-period delays add up to the refractive index (Sec.~3); the backward waves
add up to reflection (Sec.~5); and the forward waves, together with the incident one, build the
wave that actually travels in the medium (Sec.~4).

%% file: ch3_index.tex
\section{The refractive index}
\label{sec:index}

Light is slower in glass than in vacuum. This is perhaps the most familiar fact of optics, and
the explanation one most often hears is wrong: that the atoms ``absorb the light and re-emit it
with a delay''. In this chapter we derive the refractive index from the sheets of
Sec.~\ref{sec:sheet}, discuss what it depends on, and then return to the delay picture to see
precisely why it fails.

\subsection{Many sheets: the accumulated delay}

\begin{figure}[!htbp]
\centering
\includegraphics[width=0.95\textwidth]{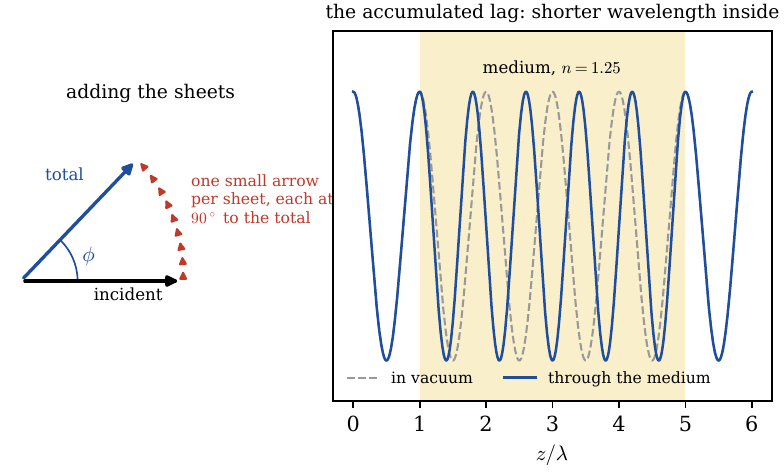}
\caption{Left: each sheet adds to the wave a small arrow at right angles to it; the total arrow
does not grow but turns, by an angle $\phi$ proportional to the thickness crossed. Right: the
result---the wave crossing the medium (solid) falls progressively behind a wave in vacuum
(dashed), and inside the medium its wavelength is shorter.}
\label{fig:arrows}
\end{figure}

Consider a dilute medium---a gas---filling the region beyond $z=0$, and cut it into thin sheets.
Each sheet adds to the wave that reaches it a small wave of its own, Eq.~\eqref{eq:sheet},
proportional to it and a quarter of a period behind it. In the language of arrows
(Fig.~\ref{fig:arrows}): to the arrow of the wave each sheet adds a small arrow at right angles.
Adding a small perpendicular arrow does not change the length of the total, to first order, but
turns it. The next sheet adds another small arrow, perpendicular to the new total, and turns it
further. Sheet after sheet the arrow rotates steadily: the wave arrives later and later with
respect to a wave travelling in vacuum, which is exactly what a wave travelling at a lower speed
does. The quarter-period delay of the wave of each sheet has become a continuous phase delay.

Quantitatively, a sheet of thickness $\ud z$ multiplies the wave by
\begin{equation*}
1+i\,\frac{2\pi Nf}{k}\,\ud z\simeq\exp\Big(i\,\frac{2\pi Nf}{k}\,\ud z\Big).
\end{equation*} A wave in a medium of index $n$ gains, with
respect to vacuum, the extra phase $(n-1)k\,\ud z$. Hence
\begin{equation}
n-1=\frac{2\pi Nf(0)}{k^2}=\frac{N\alpha}{2},
\label{eq:Fermi}
\end{equation}
the formula of Fermi and van de Hulst~\cite{Fermi,vandeHulst}, here in the language of
Feynman's lecture on the origin of the refractive index~\cite{FeynmanI}. With the polarizability
of the Lorentz oscillator, Eq.~\eqref{eq:alpha}, or more generally with the sum over oscillators
of the panel at the end of Sec.~\ref{sec:one-atom},
\begin{equation}
n(\omega)-1=\frac{Ne^2}{2\varepsilon_0m}\sum_n\frac{f_n}{\omega_n^2-\omega^2-i\Gamma_n\omega}.
\label{eq:dispersion}
\end{equation}
This is the dispersion formula of a dilute medium; for a dense one, where the sheets act also on
each other, we shall find in Sec.~4 its generalization, the Lorentz--Lorenz formula.

\begin{panel}{From sheets to a differential equation}
Let $E(z)$ be the wave at depth $z$. The sheet between $z$ and $z+\ud z$ adds the forward wave
$(i\omega/2\varepsilon_0c)\,P\,\ud z$, with $P=N\varepsilon_0\alpha E(z)$ in a dilute medium.
Hence
\begin{equation*}
\frac{\ud E}{\ud z}=ik\,E+i\,\frac{k\,N\alpha}{2}\,E=ik\Big(1+\frac{N\alpha}{2}\Big)E ,
\end{equation*}
whose solution $E\propto e^{iknz}$ is a wave of index $n=1+N\alpha/2$. The term $ikE$ is the
ordinary propagation from one sheet to the next. Neglected here is the backward wave of each
sheet, which builds the reflected wave (Sec.~5), and the action of the sheets on the field that
drives the others, which matters in dense media (Sec.~4).
\end{panel}

\subsection{Dispersion}

\begin{figure}[!htbp]
\centering
\includegraphics[width=0.95\textwidth]{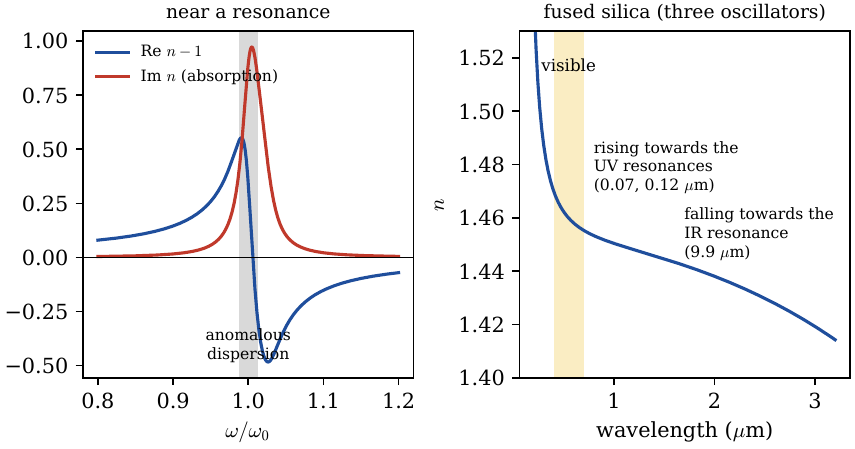}
\caption{Left: real and imaginary parts of the index near a resonance (dilute medium, $Q=40$).
Right: the index of fused silica~\cite{Malitson}, described by three oscillators, two
electronic resonances in the ultraviolet and a vibrational one in the infrared; visible light
falls in the transparent window between them.}
\label{fig:dispersion}
\end{figure}

The refractive index inherits the frequency dependence of the oscillators
(Fig.~\ref{fig:dispersion}). Below a resonance the electrons follow the field, $f>0$, and $n>1$;
approaching the resonance $n$ increases---\emph{normal dispersion}, the reason why blue light is
refracted more than red, and why a prism makes a spectrum. Within the width of the line the
index decreases (\emph{anomalous dispersion}) and the imaginary part of $n$, which describes
absorption, peaks. Above the resonance $f<0$ and $n<1$.

Real transparent materials are described, to remarkable accuracy, by a few such oscillators.
The Sellmeier formula of fused silica~\cite{Malitson}, used every day in optical design, is
nothing but Eq.~\eqref{eq:dispersion} (in its dense-medium form $n^2-1$) with three terms: two
electronic resonances in the ultraviolet, at 68 and 116~nm, and a vibration of the
silicon--oxygen lattice in the infrared, at 9.9~$\mu$m. Visible light lives in the window
between them, where nothing absorbs and the index rises gently towards the blue.

\subsection{\texorpdfstring{\rev{Why the colours separate}}{Why the colours separate}}

\begin{figure}[!htbp]
\centering
\includegraphics[width=0.95\textwidth]{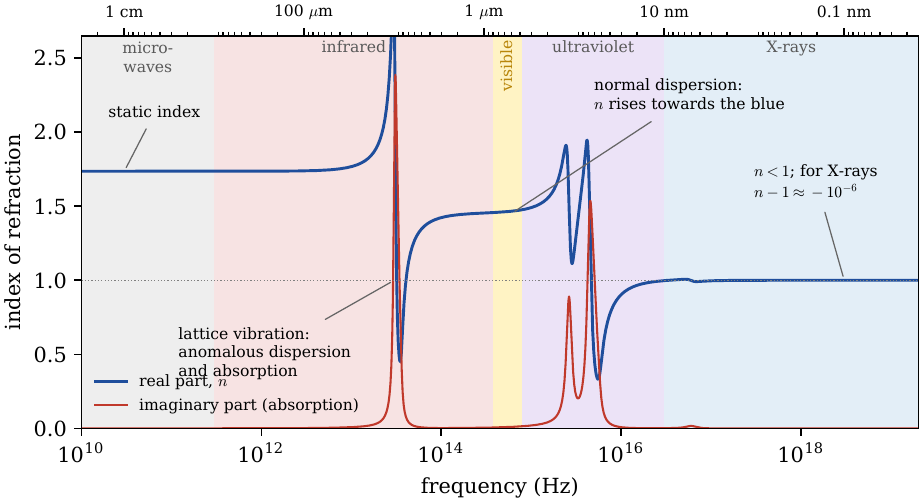}
\caption{\rev{Index of refraction of a glass-like medium from microwaves to X-rays: four oscillators, a
vibration of the lattice at $9.9~\mu$m, valence electrons at 116 and 68~nm (with the strengths of the
Sellmeier formula of fused silica~\cite{Malitson}), and inner electrons near 5~nm (with the strength that
gives the total electron density of silica). Damping exaggerated for visibility. The model gives $n=1.463$
at 550~nm (silica: 1.460) and $n-1=-3\times10^{-6}$ at 0.1~nm.}}
\label{fig:spectrum}
\end{figure}

{\revon
Why should the index depend on the colour at all? In the language of the sheets the answer is
immediate. The forward wavelet that each sheet adds to the light is proportional to the response of its
oscillators, $\alpha(\omega)\propto1/(\omega_0^2-\omega^2)$. Visible light lies below the ultraviolet
resonances of the electrons; as its frequency rises towards them, the oscillators respond more strongly,
every sheet adds a larger wavelet, the arrow of the light turns more per unit length, and the wave is
delayed more. Blue light is slowed more than red because it is closer to the resonance. This is
\emph{normal} dispersion, and it is the tail of the ultraviolet resonances.

Figure~\ref{fig:spectrum} shows the whole picture for a glass-like medium, from microwaves to X-rays.
Below every resonance the index is large: all the oscillators, ionic and electronic, follow the slow
field. Each resonance---a vibration of the lattice in the infrared, the valence electrons in the
ultraviolet, the inner electrons further up---brings a band of absorption, across which the index
\emph{decreases} with frequency (anomalous dispersion), and leaves behind a smaller index, because one kind of
oscillator no longer follows the field. The visible window, between the infrared and the ultraviolet
resonances, is where glass and water are transparent and $n$ rises gently. Above all resonances the
electrons respond as if free, in opposition to the field, and $n<1$, Eq.~\eqref{eq:xray}.
\par}

\begin{figure}[!htbp]
\centering
\includegraphics[width=\textwidth]{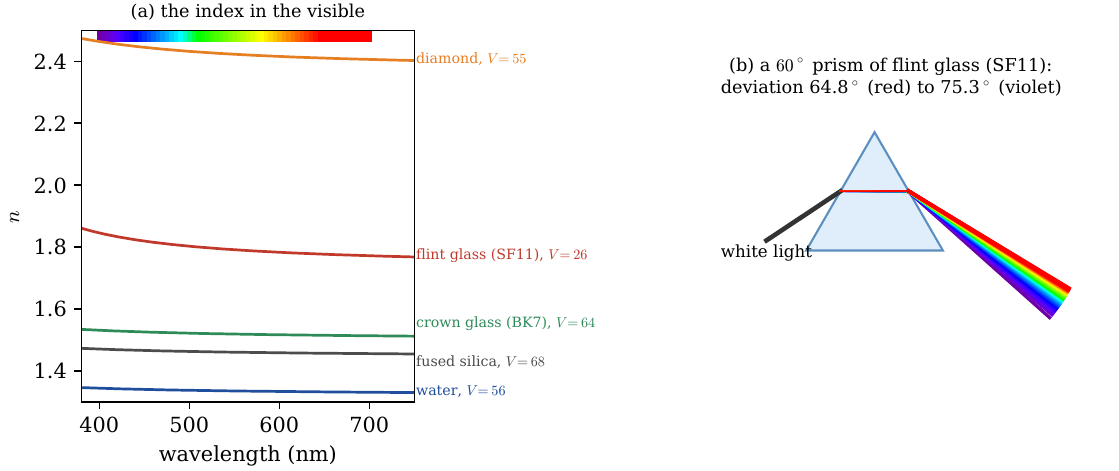}
\caption{\rev{(a) Index of refraction in the visible for water~\cite{DaimonMasumura}, fused
silica~\cite{Malitson}, a crown (BK7) and a flint (SF11) glass (Schott data) and diamond~\cite{Peter},
with their Abbe numbers $V$. (b) White light through a $60^\circ$ prism of SF11 at minimum deviation for
550~nm: exact ray tracing for 31 wavelengths from 400 to 700~nm.}}
\label{fig:prism}
\end{figure}

{\revon
\emph{The prism.} The index of real materials in the visible, computed from their Sellmeier formulas, is
shown in Fig.~\ref{fig:prism}a. The amount of dispersion is measured in optics by the \emph{Abbe number},
$V=(n_d-1)/(n_F-n_C)$, computed from the indices at a yellow (587.6~nm), a blue (486.1~nm) and a red
(656.3~nm) line: the smaller $V$, the more the colours are separated. Crown glasses ($V\approx64$) and fused
silica ($68$) disperse little; flint glasses, loaded with lead or titanium, much more ($V\approx26$ for
SF11); diamond combines a very high index with moderate dispersion, and its ``fire'' comes from both.
A prism turns these small differences into angles. In a $60^\circ$ prism of flint glass
(Fig.~\ref{fig:prism}b) red light is deviated by $64.8^\circ$ and violet by $75.3^\circ$: white light opens
into a fan of colours. This is Newton's experiment of 1666; with a second prism he showed that a colour
selected from the fan is not split again---that the colours are in the light, not made by the
glass~\cite{Newton}. The same dispersion paints the rainbow (Sec.~\ref{sec:nature}), gives lenses their
chromatic aberration (Sec.~\ref{sec:instruments})---corrected in achromatic doublets by pairing a crown
and a flint lens of opposite powers---and, as we shall see in Sec.~\ref{sec:pulse}, separates the speed of a
pulse from the speed of its wave crests.
\par}

Far above all resonances, for X-rays, every electron behaves as if it were free, $f\to-r_e$, and
\begin{equation}
n-1=-\frac{N_er_e\lambda^2}{2\pi},
\label{eq:xray}
\end{equation}
with $N_e$ the number of electrons per unit volume. The index is slightly \emph{less} than one,
by a few parts in a million: X-rays grazing a polished surface from vacuum are totally
reflected, which is how X-ray mirrors and telescopes work. The phase velocity $c/n$ exceeds
$c$---without any conflict with relativity, since it is the velocity of a pattern, not of a
signal.

\subsection{Index and cross section}
\label{sec:KK}

An atom contributes to the index through its forward amplitude $f(0)$, and to the attenuation of
the beam through its cross section $\sigma$. \rev{The connection between the two, the
\emph{optical theorem}, follows from the same thin sheet that gave Eq.~\eqref{eq:Fermi}, and with the
Lorentz oscillator it can be carried through explicitly.}

{\revon
Take a dilute layer of thickness $L$ (so dilute that $Nf L/k$ is small and every atom feels only the
incident wave). Behind it, the wave of unit amplitude has become, by the sheet sum of Sec.~2,
\begin{equation*}
E(z)=e^{ikz}\Big(1+i\,\frac{2\pi NL}{k}\,f(0)\Big)\simeq e^{ikz}\exp\Big(i\,\frac{2\pi NL}{k}\,f(0)\Big).
\end{equation*}
The real part of $f(0)$ gives the phase delay, i.e.\ the index~\eqref{eq:Fermi}. The imaginary part
gives a decay of the intensity,
\begin{equation}
|E(z)|^2=\exp\Big(-\frac{4\pi NL}{k}\,\mathrm{Im}\,f(0)\Big).
\label{eq:slabdecay}
\end{equation}
Now count the same loss atom by atom: each atom removes from the beam the power $\sigma I$, so that the
intensity decays as $e^{-N\sigma L}$ (Beer's law). Comparing the two exponents,
\begin{equation}
\sigma=\frac{4\pi}{k}\,\mathrm{Im}\,f(0)=k\,\mathrm{Im}\,\alpha,\qquad\text{i.e.}\qquad 2k\,\mathrm{Im}\,n=N\sigma .
\label{eq:optthm}
\end{equation}
Two independent calculations meet here: one is coherent, a sum of \emph{amplitudes} over the whole
sheet; the other is a balance of \emph{energy}, atom by atom. The light scattered by the atoms is missing
from the beam because, in the forward direction, the scattered wave is in quadrature with the incident one
and interferes destructively with it.

With the polarizability~\eqref{eq:alpha} of the Lorentz oscillator, $\mathrm{Im}\,f=r_e\,\omega^2\,
2\gamma\omega/[(\omega_0^2-\omega^2)^2+4\gamma^2\omega^2]$, and Eq.~\eqref{eq:optthm} gives
\begin{equation*}
\sigma=\frac{8\pi r_e\,\gamma\,\omega^2/c}{(\omega_0^2-\omega^2)^2+4\gamma^2\omega^2}.
\end{equation*}
If the damping is purely radiative, $\gamma=\gamma_{\rm rad}=r_e\omega^2/3c$ (the panel of Sec.~1,
evaluated at the driving frequency), this is exactly the cross section~\eqref{eq:sigma}, which we
obtained in Sec.~1 from Larmor's formula, i.e.\ from the power radiated by the dipole: Rayleigh
scattering below resonance, Thomson scattering above. The optical theorem is, for the oscillator,
nothing but the conservation of its energy. If the damping contains other, non-radiative, losses,
the same formula gives the total \emph{extinction} cross section, scattering plus absorption---the
starting point of Sec.~\ref{sec:beyond}.

A word on validity. For a \emph{single} scatterer the optical theorem is exact, whatever its size and
index (for a glass sphere it is the familiar Mie formula for $Q_{\rm ext}$, Sec.~\ref{sec:nature}). What requires
$n\simeq1$ is the route through the layer, and with it the relation $2k\,\mathrm{Im}\,n=N\sigma$: it
assumes single scattering and atoms that scatter \emph{independently}. In a dense, ordered medium
each atom keeps its $\sigma$, but the sideways waves cancel (Sec.~2) and the beam is not attenuated by
$N\sigma$; what survives is the scattering by density fluctuations, which in an ideal gas are random and
give back $N\sigma$, and in glass or water are much smaller~\cite{Smoluchowski,Einstein1910}.\par}

The real part is connected to it by causality. A signal entering a medium cannot travel faster
than light: its front moves at speed $c$, whatever the index (Sommerfeld and
Brillouin~\cite{Brillouin}; we shall see it happen in Sec.~\ref{sec:pulse}). This forces $n(\omega)$ to be an
analytic function in the upper half of the complex frequency plane, and Cauchy's theorem then
links its real and imaginary parts---the Kramers--Kronig relation. Written with the cross
section, it reads~\cite{Holstein}
\begin{equation}
n(\omega)-1=\frac{Nc}{\pi}\;P\!\!\int_0^\infty\frac{\sigma(\omega')}{\omega'^2-\omega^2}\,\ud\omega',
\label{eq:KKsigma}
\end{equation}
where $P$ denotes the principal value. The index at frequency $\omega$ is determined by the cross
section at \emph{all} frequencies, and the integral is dominated by the resonances, where
$\sigma$ is huge. The total weight is fixed by the Thomas--Reiche--Kuhn sum rule,
$\int_0^\infty\sigma\,\ud\omega=2\pi^2r_ec$ per electron; for narrow lines,
Eq.~\eqref{eq:KKsigma} gives back Eq.~\eqref{eq:dispersion}. It also shows at once why $n<1$
above all resonances: there the weight $1/(\omega'^2-\omega^2)$ is negative, while $\sigma$ is
positive.

\subsection{The myth of the delayed re-emission}
\label{sec:myth}

\begin{figure}[!htbp]
\centering
\includegraphics[width=0.72\textwidth]{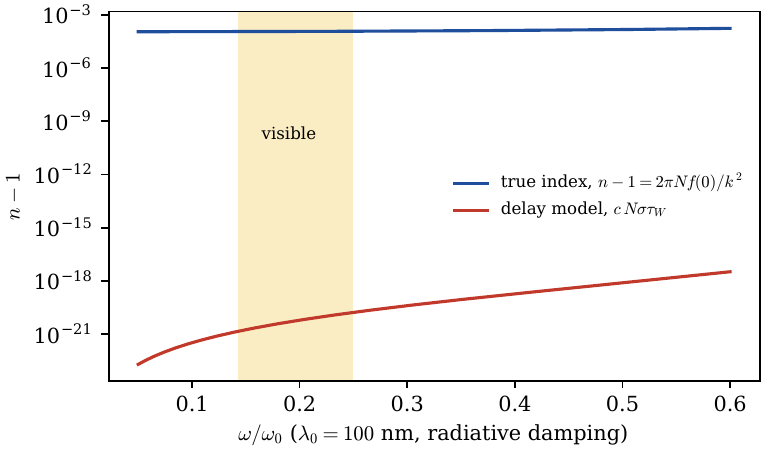}
\caption{Refractivity of an air-like gas ($N=2.5\times10^{25}$~m$^{-3}$, one oscillator at
$\lambda_0=100$~nm with radiative damping): the true index, Eq.~\eqref{eq:Fermi}, and the
prediction of the delay model, Eq.~\eqref{eq:delaymodel} with Wigner's delay.}
\label{fig:delay}
\end{figure}

We can now look at the popular explanation with the eyes of a physicist. In the
delay picture, a photon travels at speed $c$ between atoms, covering on average the mean free
path $\ell=1/N\sigma$, and is held for a time $\tau$ at each encounter. Its average speed is
$\ell/(\ell/c+\tau)$, and the index would be
\begin{equation}
n=1+\frac{c\tau}{\ell}=1+c\,N\sigma\,\tau .
\label{eq:delaymodel}
\end{equation}
The most natural choice for $\tau$ is Wigner's time delay, $\tau_W=\ud\varphi/\ud\omega$, the
derivative of the phase of the scattering amplitude, which measures how late a scattered wave
packet emerges compared with a free one~\cite{Wigner}. Inserting the oscillator's $\sigma$ and
$\tau_W$ gives a definite prediction, plotted in Fig.~\ref{fig:delay} against the true index.
The two differ by some sixteen orders of magnitude, and for reasons of principle:
\begin{itemize}
\item \emph{Order.} The true index is proportional to the amplitude $f$, which interferes with the
incident wave; the delay model goes through the cross section $\sigma\propto|f|^2$, an intensity,
which has lost the phase. It is of second order in the coupling instead of first.
\item \emph{Damping.} $\tau_W$ is proportional to the damping $\gamma$: an undamped medium would not
refract. The true index, away from resonance, does not depend on $\gamma$ at all.
\item \emph{Sign.} Delays are positive, so the model always gives $n>1$. Above resonances---for
X-rays, Eq.~\eqref{eq:xray}---the true index is less than one.
\end{itemize}
The model can also be tested directly against the data, with no assumption on $\tau$. For air
at 550~nm the Rayleigh cross section per molecule is about $5\times10^{-31}$~m$^2$, so that
$\ell\simeq80$~km. To reproduce $n-1=2.8\times10^{-4}$ the photon would have to be held about
80~ns at each encounter, whereas a molecule driven far from resonance responds within
$10^{-16}$~s.

The decisive argument, however, is another. The derivation of the previous section holds for a
layer much thinner than the mean free path: the index is established in a regime where almost
no photon has been scattered at all. The ratio between the phase accumulated over a thickness
$\ud z$ and the probability $N\sigma\,\ud z$ of being scattered on the way is
$(n-1)k/N\sigma\simeq3/4kf$, of order $1/kr_e\sim10^7$--$10^8$. In one millimetre of air light
has already lagged by three radians, while the probability that a photon has been scattered is
about $10^{-8}$. The delay picture builds the index out of collisions; the index is in place long
before any collision has occurred. The slowing of light is a collective, coherent effect of the
forward waves of all the atoms, not the sum of individual detentions.

Something of the delay picture does survive, and Sec.~\ref{sec:pulse} will make it visible. A \emph{pulse} of
light moves through a dispersive medium at the group velocity $c/n_g$, with
$n_g=n+\omega\,\ud n/\ud\omega$, and the reason is that part of its energy is stored, reversibly,
in the oscillation of the electrons: the more energy sits in the matter, the slower the pulse.
But this storage is shared by all the atoms at once, and it concerns the group velocity, not the
phase velocity.

%% file: ch4_extinction.tex
\section{The extinction theorem}
\label{sec:extinction}

In a gas the forward waves of the atoms are a small correction to the incident wave, and it is
natural to think of the light inside as the incident wave, slightly delayed. In glass the
correction is no longer small: the index differs from one by fifty per cent. What happens then to
the incident wave? It has not stopped existing---the field of the distant source still arrives,
at speed $c$, at every point inside the glass. Yet the wave we observe there travels at speed
$c/n$. The answer, found by Ewald~\cite{Ewald} and Oseen~\cite{Oseen} a century ago, is one of the
most beautiful results of optics: the atoms of the medium radiate a wave that \emph{exactly
cancels} the incident wave everywhere inside, and replace it with a wave of their own.

\subsection{The pieces of the field}

Take a medium filling the half-space $z>0$, made of sheets that respond to the field they feel,
$P=\varepsilon_0\chi E$, and let a wave $E_0e^{ikz}$ arrive from the left. The field at any point
is the incident wave plus the waves of all the sheets, Eq.~\eqref{eq:sheet}, each travelling
forward and backward at speed $c$:
\begin{equation}
E(z)=E_0e^{ikz}+\frac{i k\chi}{2}\int_0^\infty E(z')\,e^{ik|z-z'|}\,\ud z'.
\label{eq:selfconsistent}
\end{equation}
This is a \emph{self-consistent} equation: the field $E$ that drives the sheets is the field they
help to create. It is the one-dimensional form of the retarded integral equations of
Refs.~\cite{RealiI,RealiII}, and it can be solved exactly (panel below). Inside the medium the
solution is a single wave $E=A\,e^{iknz}$, with
\begin{equation}
n^2=1+\chi,\qquad A=\frac{2}{n+1}\,E_0,
\label{eq:EO}
\end{equation}
and outside, on the left, a reflected wave $rE_0e^{-ikz}$ with $r=-(n-1)/(n+1)$: the Fresnel
coefficients at normal incidence, obtained without ever writing a boundary condition.

\begin{figure}[!htbp]
\centering
\includegraphics[width=0.92\textwidth]{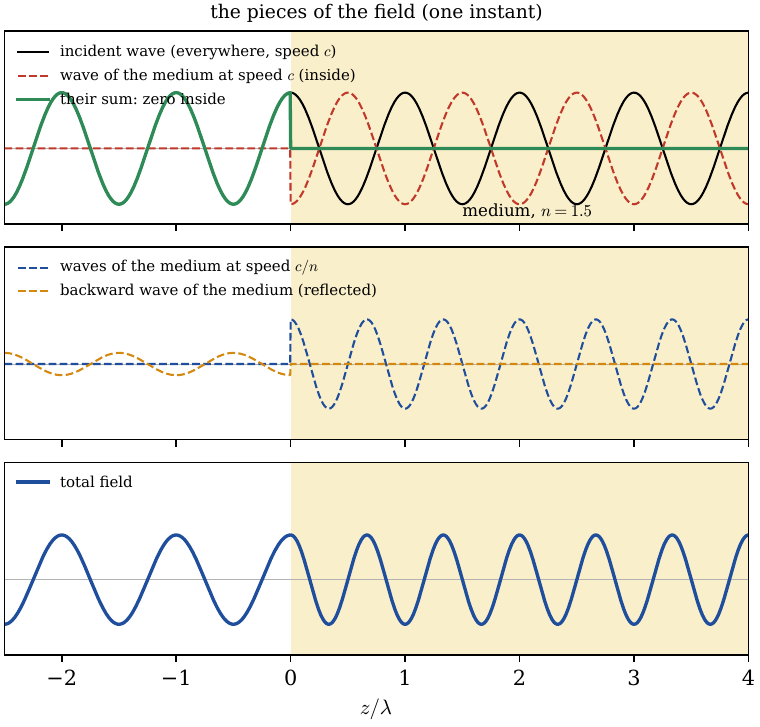}
\caption{The field at one instant for a medium of index $n=1.5$ (shaded) filling $z>0$. Top: the
incident wave continues everywhere at speed $c$ (black); inside, the medium radiates a wave at
speed $c$ that is its exact opposite (red, dashed), so that their sum vanishes (green). Middle: the
waves of the medium travelling at speed $c/n$ inside and backward outside. Bottom: the total field,
i.e.\ the refracted wave inside and the incident plus reflected waves outside
(\anim{4.1}{anim4_extinction.mp4}).}
\label{fig:extinction}
\end{figure}

But the instructive part is how the solution is made. Splitting the integral into the sheets
before and after the point $z$, the field radiated by the medium inside it turns out to be the sum
of three waves (Fig.~\ref{fig:extinction}, \anim{4.1}{anim4_extinction.mp4}):
\begin{equation}
\underbrace{-\,E_0\,e^{ikz}}_{\text{speed }c}\;+\;
\underbrace{\tfrac{n+1}{2}\,A\,e^{iknz}}_{\text{from the sheets behind}}\;-\;
\underbrace{\tfrac{n-1}{2}\,A\,e^{iknz}}_{\text{from the sheets ahead}} .
\label{eq:pieces}
\end{equation}
The first term travels at the speed of light, like the incident wave, and is its exact
opposite: added to it, it gives zero at every point of the medium. This is the \emph{extinction
theorem}. The other two travel at speed $c/n$ and add up to the refracted wave $A\,e^{iknz}$. In
the half-space on the left, the backward waves of all the sheets make up the reflected wave.

\begin{panel}{Solving the self-consistent equation}
Try $E(z')=A\,e^{iknz'}$ in Eq.~\eqref{eq:selfconsistent} for $z>0$. The sheets with $z'<z$ give
\begin{equation*}
\frac{ik\chi A}{2}\,e^{ikz}\int_0^ze^{ik(n-1)z'}\ud z'=\frac{\chi}{2(n-1)}\,A\,\big(e^{iknz}-e^{ikz}\big),
\end{equation*}
and those with $z'>z$ (with a vanishing absorption to make the integral converge)
\begin{equation*}
\frac{ik\chi A}{2}\,e^{-ikz}\int_z^\infty e^{ik(n+1)z'}\ud z'=-\frac{\chi}{2(n+1)}\,A\,e^{iknz}.
\end{equation*}
The equation is satisfied if the coefficients of $e^{iknz}$ and of $e^{ikz}$ balance separately:
\begin{equation*}
A=\Big[\frac{\chi}{2(n-1)}-\frac{\chi}{2(n+1)}\Big]A,\qquad
0=E_0-\frac{\chi}{2(n-1)}\,A .
\end{equation*}
The first condition gives $\chi=n^2-1$, the dispersion relation of the medium; with it,
$\chi/2(n-1)=(n+1)/2$, and the second gives $A=2E_0/(n+1)$---the extinction of the incident wave.
For $z<0$ all the sheets are ahead, and their waves add up to
$-\chi A/2(n+1)\,e^{-ikz}=-\frac{n-1}{n+1}E_0\,e^{-ikz}$, the reflected wave.
\end{panel}

\subsection{Where the extinction comes from}

\begin{figure}[!htbp]
\centering
\includegraphics[width=0.66\textwidth]{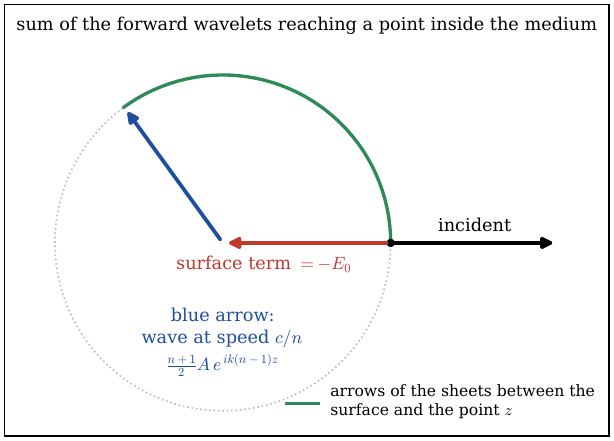}
\caption{The arrows of the forward wavelets reaching a point inside the medium from the sheets in
front of it (green), measured relative to a wave travelling at speed $c$. Their sum splits into a
fixed arrow from the surface, opposite to the incident wave (red), and an arrow that turns with
depth: the wave travelling at $c/n$ (blue).}
\label{fig:phasorsEO}
\end{figure}

Which atoms cancel the incident wave? Figure~\ref{fig:phasorsEO} answers with arrows, as in
Sec.~\ref{sec:sheet}. Stand at a point inside the medium and add the forward wavelets of the
sheets between the surface and that point. Relative to a wave travelling at speed $c$, the wavelet
of a sheet at depth $z'$ has the phase $k(n-1)z'$, so that the arrows, added one after the other,
run round a circle. The sum is an arrow from the starting point to the current point of the circle,
and it can be read as the sum of two arrows: a fixed one, from the start to the centre of the
circle, and a rotating one, from the centre to the current point. The rotating arrow is the wave
travelling at speed $c/n$. The fixed one, equal and opposite to the incident wave, comes from the
lower limit of the integral, the surface: it is the wave that the boundary of the medium sends
forward at speed $c$. \emph{The extinction is the work of the surface}, or more precisely of the
abrupt beginning of the sheets; deep inside the medium, the atoms merely sustain the refracted
wave.

Two misunderstandings are worth avoiding. First, in the steady state the extinction is not
progressive: the incident wave is cancelled exactly at every depth, not little by little over some
``extinction length''. What builds up with depth is only the phase of the refracted wave. Second,
the Ewald--Oseen picture and the gas picture of Sec.~\ref{sec:index} are not in conflict: they are
two ways of reading the same field. Writing $e^{iknz}=e^{ikz}\,e^{ik(n-1)z}$, a weakly refracting
medium can be seen either as an incident wave whose phase is progressively turned, or as an
incident wave extinguished and replaced. The second reading becomes indispensable when $n-1$ is
not small. And in time, when a signal first enters the medium, the extinction does take time to
establish itself: we shall watch it happen in Sec.~\ref{sec:pulse}.

\subsection{Molecules and the local field}

The continuous sheets of this chapter hide one more subtlety. In a dense medium of discrete
molecules, the field that drives a molecule is not the average field $E$ of
Eq.~\eqref{eq:selfconsistent}: the near neighbours, a few nanometres away, contribute their
near fields, which do not average out. For molecules arranged isotropically, the field acting on
a molecule is the \emph{local field} $E+P/3\varepsilon_0$, and since $P=N\varepsilon_0\alpha$ times
that field, the dispersion relation $n^2=1+\chi$ becomes the Lorentz--Lorenz (or Clausius--Mossotti)
formula,
\begin{equation}
\frac{n^2-1}{n^2+2}=\frac{N\alpha}{3}.
\label{eq:LL}
\end{equation}
The extinction theorem in its full molecular form---a lattice of discrete dipoles, each driven by
the incident wave and by the retarded fields of all the others---was solved by Ewald for
crystals, and has been the subject of a long line of pedagogical work~\cite{BornWolf,Reali1982,
RealiJOSA1982,FJM1996,Reali2014}. Numerical simulations of finite assemblies of coupled dipoles
make the whole process visible, molecule by molecule\rev{~\cite{RealiStefanini}}.

\begin{figure}[!htbp]
\centering
\includegraphics[width=\textwidth]{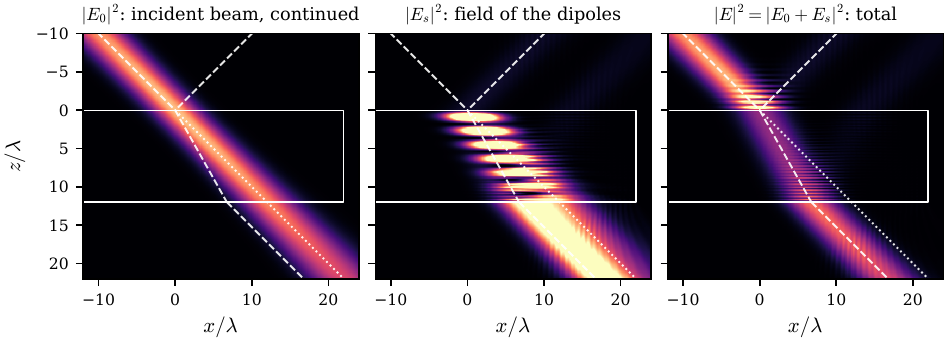}
\caption{\rev{A Gaussian beam (waist $3\lambda$, $45^\circ$) on a slab of 44\,400 coupled line
dipoles (rectangle), computed with the coupled-dipole code of the ancillary files. Intensity of the incident
beam formally continued, $|E_0|^2$, of the field radiated by the dipoles, $|E_s|^2$, and of the
total field. Dashed lines: the centroids of the incident, reflected, refracted and transmitted
beams; dotted: the straight continuation of the incident beam. In the total field there is nothing
along the dotted line: the field of the dipoles cancels the incident beam.}}
\label{fig:cdmmaps}
\end{figure}

\begin{figure}[!htbp]
\centering
\includegraphics[width=\textwidth]{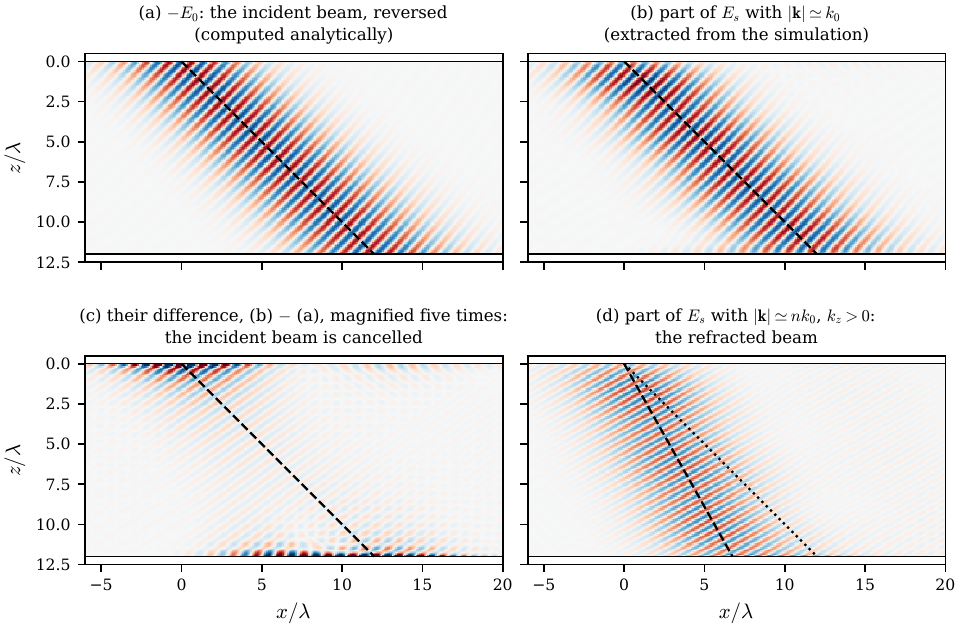}
\caption{\rev{The extinction theorem in a picture: real part of the fields inside the slab of
Fig.~\ref{fig:cdmmaps}. (a) The incident beam with its sign reversed, $-E_0$, computed analytically. (b) The
part of the field radiated by the dipoles whose wave vector has the length of a wave in vacuum,
$|\mathbf k|\simeq k_0$, extracted from the simulation by filtering in Fourier space. Panels (a) and (b) look
the same, and this is the point: the dipoles radiate, inside the slab, a beam equal and opposite to the
incident one. (c) Their difference, magnified five times. The residual, about 9\% (rms), is a numerical artefact of
the analysis: it is concentrated near the faces of the slab, where the Fourier filter applied to a
truncated field is least accurate. In the spectrum of the total field the incident wave vector survives
with an amplitude of only $10^{-3}$: added to the incident beam, the extinguishing wave cancels it. (d) The part with $|\mathbf k|\simeq nk_0$
travelling forward: the refracted beam, at Snell's angle. Dashed: beam centroids (incident in a--c,
refracted in d); dotted: straight continuation of the incident beam.}}
\label{fig:cdmdecomp}
\end{figure}

{\revon
How good is the continuum of sheets? It can be tested numerically, treating the medium as what it is: a large number of point dipoles in vacuum, each
driven by the incident wave and by the retarded fields of all the others---the coupled-dipole
method.

\emph{A beam on a slab of dipoles.} Figures~\ref{fig:cdmmaps} and~\ref{fig:cdmdecomp} show a
two-dimensional version, large enough to contain a whole refracting beam \rev{(the code is included
in the ancillary files)}: 44\,400
line dipoles, parallel to the electric field (s polarization), on a square lattice of spacing
$\lambda/10$ filling a slab $12\lambda$ thick; a Gaussian beam of waist $3\lambda$ falls on it at
$45^\circ$. No index, no continuum, no boundary condition is assumed: only the polarizability of the
dipoles (chosen so that the continuum limit would give $n=1.5$; here the field is along the
dipoles and there is no Lorentz--Lorenz correction) and the propagation of their fields at speed
$c$. The linear system, solved iteratively with the help of the fast Fourier transform, takes a
few minutes on a laptop. Everything we have described in this section emerges from the solution:
\begin{itemize}
\item the incident beam, formally continued inside the slab (left panel of Fig.~\ref{fig:cdmmaps}),
is absent from the total field: the field of the dipoles cancels it, and in the Fourier spectrum of
the total field inside the slab the incident wave vector survives with an amplitude of $10^{-3}$;
\item filtering the field of the dipoles in Fourier space separates it into a part with
$|\mathbf k|\simeq k_0$, which reproduces $-E_0$ (to 9\%, a numerical artefact of filtering a field
truncated at the slab edges), and a part with $|\mathbf k|\simeq nk_0$: the extinguishing and the
refracted waves of Eq.~\eqref{eq:pieces}, separated in a single picture (Fig.~\ref{fig:cdmdecomp});
\item the refracted wave conserves the tangential wave vector, $k_x/k_0=0.7059$ against
$\sin45^\circ=0.7071$, and has $|\mathbf k|=1.443\,k_0$: it travels at $29.29^\circ$, Snell's law with
its own index giving $29.34^\circ$;
\item the index, $1.443$, is a little smaller than the continuum value: the difference is due to the
discreteness of the lattice and decreases, slowly, as the spacing is reduced;
\item outside, the dipoles build a reflected beam and, below the slab, a transmitted beam parallel to
the incident one and shifted sideways, exactly along the rays of geometrical optics.
\end{itemize}
\par}

%% file: ch5_reflection.tex
\section{Reflection}
\label{sec:reflection}

A window pane shows us the street behind it and, faintly, our own face. Both images are made by
the same atoms. The forward waves of the sheets build the transmitted light (Secs.~3 and 4); their
backward waves, which we have so far set aside, build the reflected light. This chapter adds them
up.

\subsection{\texorpdfstring{\rev{Feynman's arrows, made quantitative}}{Feynman's arrows, made quantitative}}

\begin{figure}[!htbp]
\centering
\includegraphics[width=\textwidth]{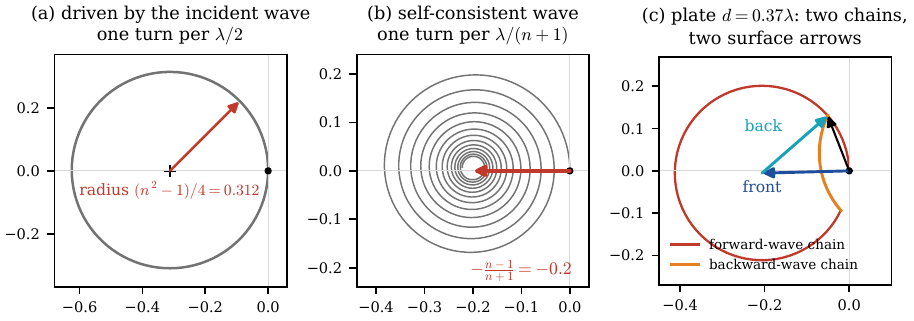}
\caption{\rev{The arrows of the sheets for the reflected light, $n=1.5$. (a) Sheets driven by the
incident wave only: a circle of radius $(n^2-1)/4$, one turn every $\lambda/2$. (b) Self-consistent
wave: one turn every $\lambda/(n+1)$, radius $(n-1)/(n+1)$; with a weak absorption the chain
converges to the centre, and the resulting arrow is Feynman's surface arrow, $-0.2$. (c) A plate
$0.37\lambda$ thick: chain of the forward wave (red) followed by that of the backward wave
(orange). The front (blue) and back (cyan) surface arrows are the initial terms of the two chains;
the black arrow is the reflection of the plate.}}
\label{fig:arrows5}
\end{figure}

\begin{figure}[!htbp]
\centering
\includegraphics[width=0.78\textwidth]{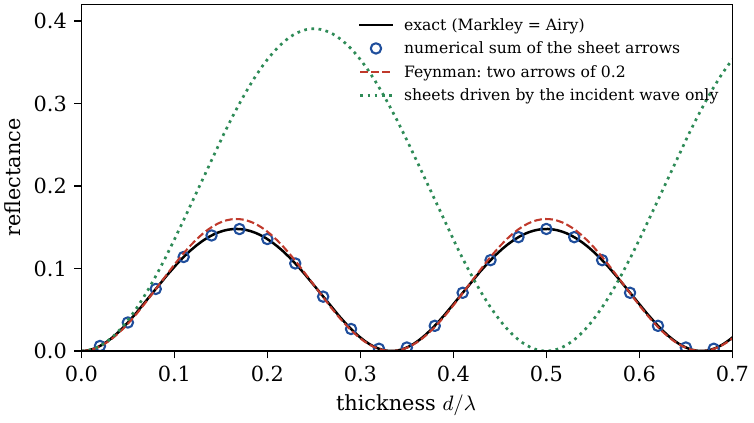}
\caption{\rev{Reflectance of a glass plate ($n=1.5$, normal incidence) as a function of its
thickness. Solid: exact (Markley = Airy); circles: numerical sum of the arrows of the sheets, with
the self-consistent polarization; dashed: Feynman's rule, two arrows of $0.2$ (maximum 16\%);
dotted: sheets driven by the incident wave only.}}
\label{fig:Rd}
\end{figure}

{\revon
In his little book \emph{QED}~\cite{FeynmanQED-book} Feynman explains the partial reflection of
light by a glass plate with \emph{arrows}: amplitudes that turn like the hand of a stopwatch and are
added head to tail. A pane reflects about 4\% from its front surface and as much from its back
surface, but the reflection of the plate is not 8\%: it varies between 0 and 16\% with the
thickness. Feynman reproduces this with two arrows of length $0.2$, one for each surface, the second
one turned by the round trip through the glass. Later in the book he warns that this is a shortcut:
the light is really scattered by all the electrons of the glass, and the arrows of all the layers
combine to leave only two contributions at the surfaces. He does not show the calculation, and the
length $0.2$ remains an empirical datum. The sheets of Sec.~2, made self-consistent as in Sec.~4,
supply the missing calculation. This is the sheet method of Markley~\cite{Markley}, born from a
critical reading of the \emph{Lectures}, and of Hill~\cite{Hill}; Taylor and
co-workers~\cite{TaylorQED} turned Feynman's rules into a course, assigning the surface arrows by
hand.

\emph{The dictionary.} A sheet of thickness $\ud\zeta$ at depth $\zeta$, with polarization
$P(\zeta)$, sends back towards the observer the wave~\eqref{eq:sheet}; measured with respect to the
reference wave $e^{-ikz}$, its \emph{arrow} is
\begin{equation}
\ud A_r(\zeta)=\frac{ik}{2\varepsilon_0}\,P(\zeta)\,e^{ik\zeta}\,\ud\zeta .
\label{eq:sheetarrow}
\end{equation}
The factor $i$ turns every arrow by a quarter of a turn with respect to the polarization that
emits it (the same quarter turn of Sec.~3); the arrow turns from sheet to sheet with the phase of
$P(\zeta)\,e^{ik\zeta}$---the phase of the wave arriving at the sheet plus that of the way back in
vacuum; and its length is proportional to $P=\varepsilon_0(n^2-1)E$, where $E$ is the field that
actually drives the sheet. All the physics is in this last point, and the two possible answers give
two levels of description (Fig.~\ref{fig:arrows5}).

\emph{Sheets driven by the incident wave alone.} If every sheet felt only $e^{ik\zeta}$, as in the
calculation of the \emph{Lectures} that Markley criticized, the arrows would all have the same
length and would turn once every $\lambda/2$: they would run round a circle of radius
$(n^2-1)/4=0.31$ for glass, instead of $0.2$, and with the period of vacuum instead of the
observed one. The reflectance of a plate would reach 39\% (Fig.~\ref{fig:Rd}).

\emph{Self-consistent sheets.} In fact every sheet is driven by the total field, which inside a
semi-infinite medium is the wave $t\,e^{iknz}$, $t=2/(n+1)$, of Sec.~4. The arrows become
\begin{equation*}
\ud A_r=\frac{ik(n^2-1)}{2}\,t\,e^{i(n+1)k\zeta}\,\ud\zeta ,
\end{equation*}
and two things change. They turn once every $\lambda/(n+1)$: the wave reaches the sheet at speed
$c/n$, but the wavelet comes back at speed $c$, because every sheet radiates in vacuum. And the
circle has radius $(n^2-1)t/2(n+1)=(n-1)/(n+1)$. With an infinitesimal absorption the chain becomes
a spiral that converges to the centre (Fig.~\ref{fig:arrows5}b), and the resulting arrow, from the
origin to the centre, is
\begin{equation}
r=-\frac{n-1}{n+1},\qquad R=|r|^2=\Big(\frac{n-1}{n+1}\Big)^2,
\label{eq:Fresnel-normal}
\end{equation}
about $-0.2$ in amplitude and 4\% in intensity for glass. \emph{Feynman's surface arrow is the
radius of the circle of the arrows of the sheets.} No empirical datum is needed, once one knows
that the sheets are driven by the self-consistent wave: the surface arrow already contains the
whole extinction theorem. Reflection is produced by all the atoms of the medium, but the result is
as if it came from the surface alone: deep inside, the backward wavelets of successive layers run
round and round the circle, and only the abrupt beginning of the medium leaves a net contribution.
\par}

{\revon
\emph{The plate: two chains and two surface arrows.} In a plate $0<z<d$ Markley finds two waves,
$E=\alpha\,e^{iknz}+\beta\,e^{-iknz}$, and hence two chains of arrows (Fig.~\ref{fig:arrows5}c): the
chain of the forward wave turns with $(n+1)k$, that of the backward wave with $(1-n)k$, in the
opposite sense and more slowly. Adding the arrows~\eqref{eq:sheetarrow} from $0$ to $d$,
\begin{equation}
r=\underbrace{-\frac{n-1}{2}\,\alpha}_{\text{front surface}}
+\underbrace{\frac{n+1}{2}\,\beta}_{\text{back surface}}
+\underbrace{\Big[\frac{n-1}{2}\,\alpha\,e^{i(n+1)kd}-\frac{n+1}{2}\,\beta\,e^{i(1-n)kd}\Big]}_{=\,0}.
\label{eq:twochains}
\end{equation}
The first two terms come from the beginnings of the two chains, the bracket from their ends, and the
ends cancel: their cancellation is precisely Markley's second extinction condition,
$(n-1)e^{iknd}\alpha=(n+1)e^{-iknd}\beta$, which says that the vacuum wave the back face would send
backward is extinguished. The first condition, $(n+1)\alpha-(n-1)\beta=2$, is the extinction of the
incident wave. \emph{Feynman's two arrows are the initial terms of the two chains.} Eliminating
$\alpha$ with the first condition gives Markley's form, $r=-(n-1)/(n+1)+2n\beta/(n+1)$, and to first
order in the reflections
\begin{equation*}
r\simeq-\frac{n-1}{n+1}+\frac{4n}{(n+1)^2}\,\frac{n-1}{n+1}\,e^{2iknd}:
\end{equation*}
the front arrow $-0.2$ and the back arrow $+0.2$, turned by the round trip, but multiplied by
$t_{01}t_{10}=4n/(n+1)^2=0.96$. Feynman neglects this factor and the multiple reflections, and
therefore finds a maximum of 16\% instead of 14.8\%. With the exact $\alpha$ and $\beta$,
Eq.~\eqref{eq:twochains} is the Airy formula of the panel below. In Fig.~\ref{fig:Rd} the circles
are the numerical sum of the arrows of the sheets of a plate solved self-consistently, without
assuming the form of the solution: the agreement with the exact formula is to one part in $10^6$.
This structure is the one Sivukhin obtained by integrating by parts~\cite{Sivukhin}: the boundary
terms of the integration are the surface arrows. If a transition layer is present near the
surface, the arrows of the first sheets leave the circle, and the surface arrow is corrected by
terms of the order of (thickness of the layer)$/\lambda$.
\par}

\begin{panel}{The plate: summing all the arrows}
Let $r=(1-n)/(1+n)$ and $t=2/(1+n)$ be the amplitudes for reflection and transmission at the
surface from outside, $r'=-r$ and $t'=2n/(1+n)$ from inside, and $\delta=knd$ the phase across
the plate. The reflected light is the sum of the front-surface arrow and of the arrows of all the
paths that enter the plate, bounce $2m+1$ times inside and come out:
\begin{equation*}
r_{\rm plate}=r+tt'r'e^{2i\delta}\sum_{m=0}^\infty\big(r'^2e^{2i\delta}\big)^m
=\frac{r\,(1-e^{2i\delta})}{1-r^2e^{2i\delta}},
\end{equation*}
where $tt'=1-r^2$ has been used. The two-arrow approximation keeps only $m=0$ and sets $tt'\simeq1$.
The plate does not reflect at all when $2\delta$ is a multiple of $2\pi$, i.e.\ when its optical
thickness $nd$ is a multiple of $\lambda/2$.
\end{panel}

\subsection{\texorpdfstring{\rev{Transmission: ``shrink and rotate''}}{Transmission: shrink and rotate}}

\begin{figure}[!htbp]
\centering
\includegraphics[width=0.9\textwidth]{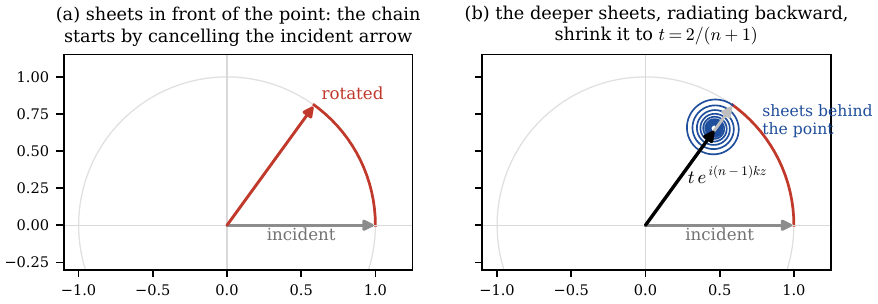}
\caption{\rev{``Shrink and rotate'' in a semi-infinite medium ($n=1.5$), at the depth
$z=0.3\lambda$. (a) The forward arrows of the sheets in front of the point (red) start by cancelling
the incident arrow (grey) and run on a circle of unit radius: the resulting arrow is turned by
$(n-1)kz$. (b) The sheets behind the point add a backward chain (blue) that shrinks it to
$t=2/(n+1)$.}}
\label{fig:shrink}
\end{figure}

{\revon
For the transmitted light Feynman's rule is ``shrink and rotate''. The sheets explain both
words (Fig.~\ref{fig:shrink}). At a depth $z$ inside a semi-infinite medium, the forward arrows come
from the sheets in front of the point, $\zeta<z$. Measured with respect to $e^{ikz}$ each is
$\tfrac{i}{2}k(n^2-1)\,t\,e^{i(n-1)k\zeta}\,\ud\zeta$, and they turn once every $\lambda/(n-1)$.
Their sum,
\begin{equation*}
\frac{n+1}{2}\,t\,\big(e^{i(n-1)kz}-1\big)=e^{i(n-1)kz}-1,
\end{equation*}
runs on a circle as large as the incident arrow. Its first piece, $-1$, cancels the incident arrow:
this is the extinction. Its tip turns with $(n-1)k$: this is the slowing down to $c/n$. The deeper
sheets, $\zeta>z$, radiate backward and add $-\frac{n-1}{n+1}\,e^{i(n-1)kz}$, which shortens the
rotated arrow to $t=2/(n+1)$. The result is $t\,e^{i(n-1)kz}$, i.e.\ $E=t\,e^{iknz}$:
\emph{rotate} is the forward chain, extinction plus slowing down; \emph{shrink} is the contribution
of the sheets behind the point, the same backward scattering that produces the reflection. Energy
conservation, $1-r^2=nt^2$, ties the length of the reflected arrow to how much the transmitted one
shrinks.
\par}

\begin{panel}{\rev{What the arrows hide}}
{\revon
\begin{itemize}
\item \emph{The length $0.2$ is a product of self-consistency.} With electrons driven by the incident
wave only, the arrow would be $(n^2-1)/4=0.31$ and would turn with the wrong period. The surface
arrows already contain the extinction theorem.
\item \emph{The two arrows are boundary terms}: the beginnings of two chains, whose ends cancel by the
extinction condition at the back face.
\item \emph{Every wavelet comes back at speed $c$.} In the arrows of the sheets the phase on the way in
is that of the medium, $nk\zeta$, and the phase on the way back is that of vacuum, $k\zeta$. The
``path of the photon in the glass'' of \emph{QED}, with a stopwatch that slows down, is an
effective description, not the mechanism.
\item \emph{Quantum and classical.} In linear optics the arrow of the photon and the complex amplitude
of the classical field obey the same rules of addition. The sheet calculation is, literally,
Feynman's sum over paths for partial reflection, with the scattering amplitude of each electron
fixed by the condition of self-consistency.
\end{itemize}
\par}
\end{panel}

\subsection{Why the reflection of white light is white}

Weisskopf asked a question that seems to contradict Sec.~\ref{sec:one-atom}~\cite{Weisskopf}. A
single molecule scatters blue light six times more than red. A window pane is made of molecules;
why, then, is the sky reflected in a window, or the sun reflected by a lake, not tinged with blue?

\begin{figure}[!htbp]
\centering
\includegraphics[width=0.95\textwidth]{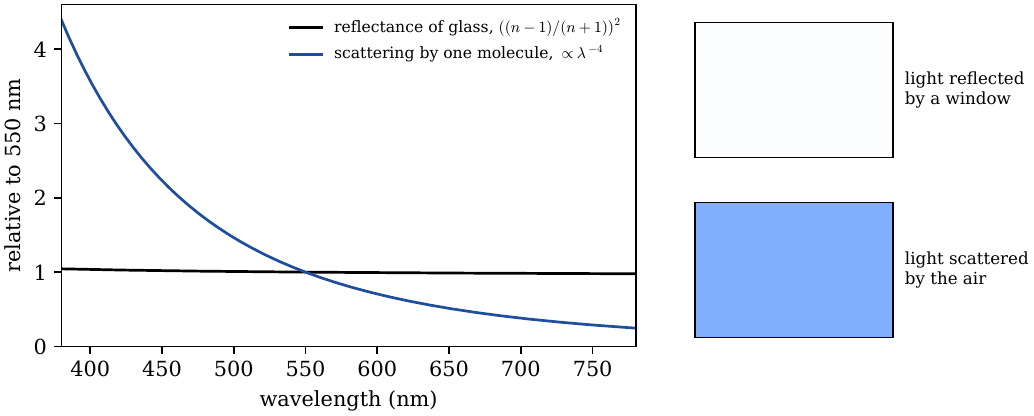}
\caption{Left: reflectance of glass (fused silica) and scattering by a single molecule, as
functions of wavelength, both normalized at 550~nm. Right: the colours of sunlight reflected by a
window and scattered by the air, for an eye adapted to sunlight.}
\label{fig:white}
\end{figure}

The answer lies, once more, in the coherent sum. The amplitude scattered by one molecule is
proportional to $\omega^2$ (Rayleigh, Eq.~\eqref{eq:f}), hence the $\omega^4$ of the intensity. But
in the reflection of a surface two factors of $1/\omega$ come in. The first appears when the
wavelets of a sheet merge into a plane wave: the effective area of the sheet, the first Fresnel
zone, is $\pi\lambda z_0$, proportional to the wavelength (Eq.~\eqref{eq:sheet} carries a $1/k$).
The second appears when the backward waves of the sheets are added in depth: only a layer of
thickness of order \rev{$\lambda/2\pi(n+1)$} contributes a net amount before the round-trip phases start to
cancel---another factor $1/k$. The two factors cancel the $\omega^2$ of the single molecule, and the
reflection, $r=-(n-1)/(n+1)$, depends on frequency only through the index, which in the visible
changes by a percent or so. Figure~\ref{fig:white} compares the two spectra. The coherent
cooperation of the molecules does not only make reflection much stronger than the scattering of
the same molecules in a gas; it also makes it colourless.

\begin{panel}{\rev{Weisskopf's count, in formulas}}
{\revon
Weisskopf's argument is a count of the dipoles that radiate in step towards an observer P at a
distance $d$ from the surface. \emph{Area}: the dipoles whose wavelets reach P with phases differing
by less than $\pi$ fill the first Fresnel zone, of area $S_1=\pi\lambda d$. \emph{Depth}: a wavelet from
the depth $z$ has travelled $z$ in and $z$ out, so that only a layer of thickness $\ell\sim\lambda/4$
contributes in step. The number of coherent dipoles is therefore $N_{\rm coh}\simeq N S_1\ell\propto
\lambda^2d$, and with the field of one dipole, $E_1\propto\omega^2A/d$ (Eq.~\eqref{eq:f}),
\begin{equation*}
E_P\simeq N_{\rm coh}E_1\propto\lambda^2d\cdot\frac{\omega^2A}{d}=(2\pi c)^2A .
\end{equation*}
Two things cancel: the distance $d$, because the reflected beam is a plane wave, and the frequency,
because the $\omega^4$ of a single dipole is compensated by the $\lambda^4$ of $N_{\rm coh}^2$. Only the
amplitude $A$ of the oscillations remains, and in the visible it hardly changes with colour.

The count becomes exact with the sheets. The whole sheet gives half the first Fresnel zone,
Eq.~\eqref{eq:sheet}; the layers, if each is driven by the incident wave alone, give
$\int_0^\infty e^{2ikz}\ud z=i/2k$, an effective thickness $\lambda/4\pi$; together,
$r\simeq-\chi/4\simeq-(n-1)/2$, i.e.\ $|r|=NeA/4\varepsilon_0E_0$: the first-order Fresnel coefficient,
the Born level of Fig.~\ref{fig:arrows5}a. For water it gives $R=2.8\%$ instead of the exact $2.0\%$; the
self-consistent wave of Sec.~4 turns $(n-1)/2$ into $(n-1)/(n+1)$. The reflectance depends on frequency
only through $A(\omega)$, i.e.\ through dispersion: it grows slightly towards the blue, and it becomes
stronger near a resonance.
\par}
\end{panel}

\subsection{Soap bubbles and the colours of thin films}

\begin{figure}[!htbp]
\centering
\includegraphics[width=0.95\textwidth]{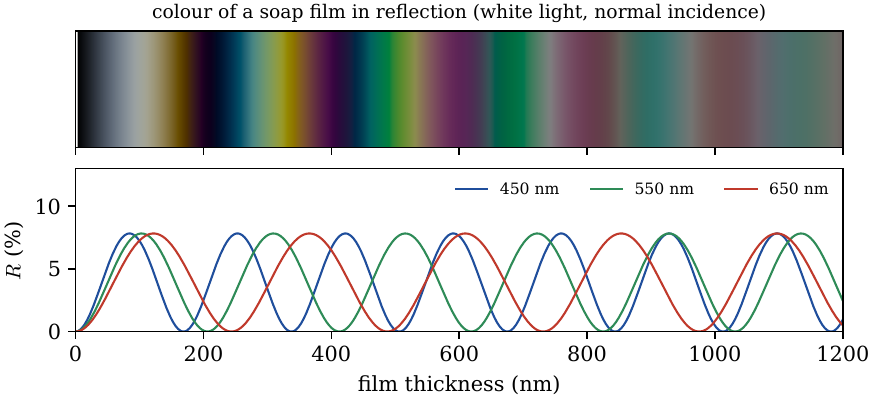}
\caption{Top: the colour of white light reflected by a soap film ($n=1.333$) as a function of its
thickness, computed from the Airy formula and the CIE colour-matching functions. Bottom:
reflectance at three wavelengths.}
\label{fig:soap}
\end{figure}

When a plate is thin, a fraction of a micrometre, its two arrows turn by an amount that depends
strongly on the wavelength, and the reflection becomes coloured. A soap film is the most familiar
example (Fig.~\ref{fig:soap}). As it thins, draining under gravity, it runs through Newton's
sequence of colours: from the iridescent bands of thick films to the yellow, magenta and blue of
the first orders, to a silvery white and finally to black, when the film is much thinner than a
wavelength and the two arrows---equal and opposite---cancel for every colour. Newton measured
these colours with lenses pressed on glass plates, and used them, long before anyone knew what
light was, to measure thicknesses of a fraction of a micrometre.

\subsection{Mirrors made of glass}

\begin{figure}[!htbp]
\centering
\includegraphics[width=0.66\textwidth]{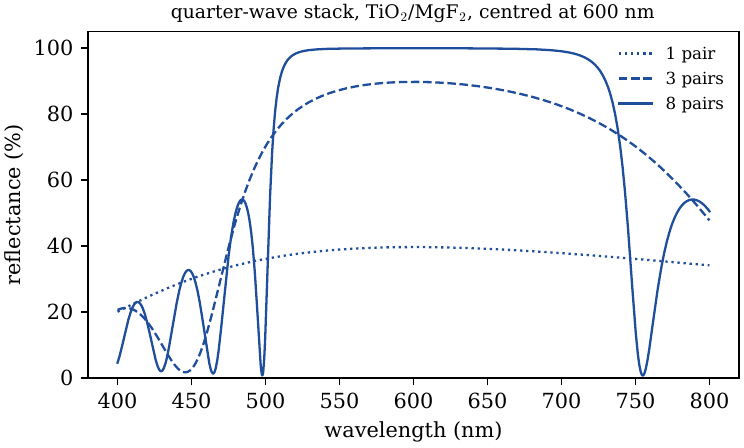}
\caption{Reflectance of a quarter-wave stack of TiO$_2$ ($n=2.35$) and MgF$_2$ ($n=1.38$) on glass,
designed for 600~nm, with one, three and eight pairs of layers.}
\label{fig:stack}
\end{figure}

If one thin layer gives a weak, coloured reflection, many layers can give a strong one. Stack
alternate layers of high and low index, each a quarter of a wavelength thick in optical
path: the arrows reflected at all the interfaces then point the same way, and they add.
With a handful of pairs the reflectance approaches 100\% over a band of wavelengths
(Fig.~\ref{fig:stack})---a mirror made entirely of transparent materials, without a single free
electron. Such stacks are the mirrors of lasers, and nature invented them long before us: the
blue of the \emph{Morpho} butterfly and the green of many beetles are reflections from
multilayers of chitin and air (Sec.~\ref{sec:nature}).

\rev{The calculation of such structures sheet by sheet---each layer seen as the sum of the arrows of
its dipoles, as in Figs.~\ref{fig:arrows5} and~\ref{fig:Rd}---is exactly the sheet method of
Markley and Hill: Feynman's arrows, turned into a quantitative tool.}

\subsection{\texorpdfstring{\rev{Metals and semiconductors}}{Metals and semiconductors}}
\label{sec:metals}

\begin{figure}[!htbp]
\centering
\includegraphics[width=\textwidth]{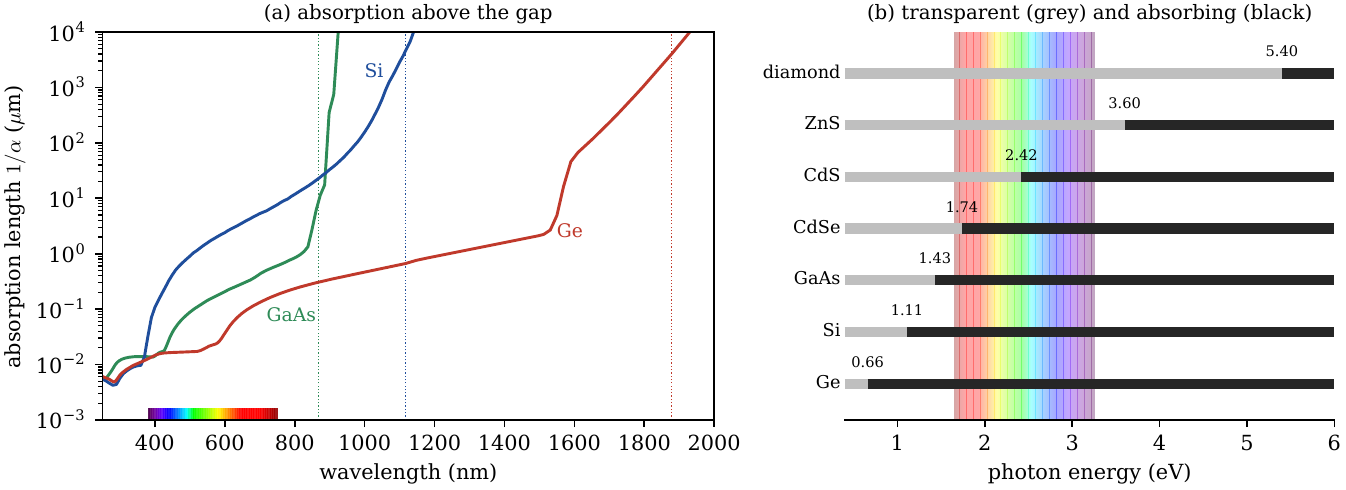}
\caption{(a) Absorption length $1/\alpha$ of gallium arsenide, silicon and germanium against
wavelength~\cite{GreenSi,Papatryfonos,Nunley}; the dotted lines mark the band gaps. (b) Band gaps of
some crystals at room temperature~\cite{Kittel}, against the visible spectrum (coloured band): each
crystal is transparent to photons below its gap (grey) and absorbs those above it (black).}
\label{fig:semi}
\end{figure}

{\revon
A mirror at home is not a quarter-wave stack: it is a sheet of glass with a film of silver or
aluminium behind it, a few tens of nanometres thick. What makes a metal so good at reflecting?

\emph{Free electrons.} In a metal one or more electrons per atom are not bound to any atom: they
move freely through the crystal. For them the restoring force of Eq.~\eqref{eq:oscillator}
vanishes, $\omega_0=0$, and every frequency is ``above resonance''. They oscillate in
antiphase with the field, $f<0$ (Sec.~\ref{sec:one-atom}), and since they are spread over the
whole crystal the field that drives them is the average field, with no local-field correction.
The index, in its dense-medium form, is that of Drude~\cite{AshcroftMermin},
\begin{equation}
n^2=1-\frac{\omega_p^2}{\omega^2+2i\gamma\omega},\qquad
\omega_p^2=\frac{Ne^2}{\varepsilon_0m},
\label{eq:drude}
\end{equation}
where $N$ is the number of free electrons per unit volume and $\omega_p$ is the \emph{plasma
frequency}. For silver, $N=5.9\times10^{28}$~m$^{-3}$ and $\hbar\omega_p=9$~eV, which corresponds to a
wavelength of 140~nm, in the ultraviolet; for aluminium, with three free electrons per atom,
$\hbar\omega_p=16$~eV.

Below $\omega_p$---for visible light in every good metal---$n^2$ is negative and $n$ is almost
purely imaginary, $n\simeq i\kappa$. There is no refracted wave: the field decays inside the metal
over a distance of order $c/\omega_p$, about 20~nm in silver, which is why a film of a few tens of
nanometres is enough for a mirror. And the reflection coefficient $r=-(n-1)/(n+1)$, with $n$
imaginary, has modulus one: all the light comes back (Fig.~\ref{fig:metals}a). In the language of
this text, the electrons are so many and respond so strongly that the forward waves of the first few
layers cancel the incident wave completely within a fraction of a wavelength. The extinction theorem
is at work as in glass, but the field that should replace the incident wave cannot propagate. The
light that has been removed from the forward direction has nowhere to go but back. In the
picture of Fig.~\ref{fig:arrows5}, the chain of arrows no longer winds round a circle: the arrows
of successive layers shrink so fast that they hardly turn, and they add up to a single arrow of
length one. Absorption is not needed for this, and in a good metal it is small: silver reflects
about 98\% of visible light, and what it absorbs is due to the damping $\gamma$, the collisions of
the electrons with the lattice.

Above $\omega_p$ the index is real and smaller than one, and the metal becomes \emph{transparent}:
the alkali metals, whose plasma frequencies are lower, let ultraviolet light through, as Wood
discovered in 1933~\cite{Wood1933}. At still higher frequencies Eq.~\eqref{eq:drude} becomes
$n\simeq1-\omega_p^2/2\omega^2$, which is just the X-ray index~\eqref{eq:xray}: far above all
resonances every electron is free. The same formula describes the ionosphere, a dilute plasma with
about $10^{12}$ free electrons per cubic metre: its plasma frequency is about 9~MHz, so short radio
waves are reflected back towards the ground, and carry broadcasts beyond the horizon, while the
higher frequencies of FM radio, television and satellite links pass through.

\emph{The colours of metals.} If free electrons were the whole story, all metals would look like
silver: their plasma frequencies all lie in the ultraviolet. The colours come from the
\emph{bound} electrons, the oscillators of the inner shells, which absorb wherever their resonances
fall (Fig.~\ref{fig:metals}b). In silver the transitions of the $d$ electrons begin only at
about 4~eV, in the ultraviolet. They also screen the free electrons and pull the edge of the
reflectance down to 3.8~eV, where silver becomes almost transparent for a narrow band near
320~nm~\cite{EhrenreichPhilipp}. In copper and gold the same transitions begin within the visible,
around 2.1 and 2.4~eV~\cite{EhrenreichPhilipp,JohnsonChristy}: blue and green light are partly
absorbed, and the reflected light is reddish or yellow. Aluminium has a weak absorption band near
800~nm and otherwise reflects well from the infrared to the deep ultraviolet~\cite{Rakic}; this is
why the mirrors of astronomical telescopes are aluminized.

\begin{figure}[!htbp]
\centering
\includegraphics[width=\textwidth]{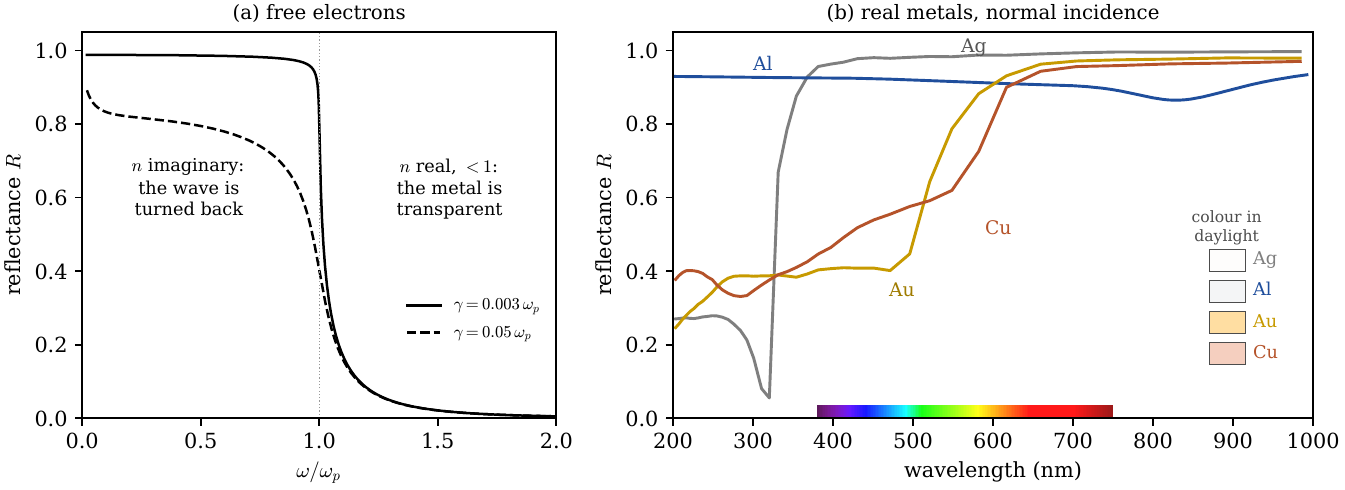}
\caption{(a) Reflectance at normal incidence of a gas of free electrons, Eq.~\eqref{eq:drude}, for
two values of the damping: total below the plasma frequency, small above it. (b) Measured reflectance
of silver, aluminium, gold and copper~\cite{JohnsonChristy,Rakic}, and the colour of each surface in
daylight, computed from the same data. The coloured strip marks the visible spectrum.}
\label{fig:metals}
\end{figure}

\emph{Semiconductors.} Between the metal and the glass lies the semiconductor. Its electrons are
bound, as in glass, but its lowest resonance is low. In a crystal the resonances of the atoms
merge into a continuous band of transitions that starts at an energy $E_g$, the \emph{band gap}:
1.1~eV for silicon, 1.4~eV for gallium arsenide, 0.66~eV for germanium~\cite{Kittel}. Below the gap
the crystal is transparent, and since a low resonance makes a large index---the oscillators are
driven closer to their natural frequency---the index is high: 3.5 for silicon and 4.0 for germanium
in the infrared, where lenses for thermal cameras are made of germanium. Above the gap light is
strongly absorbed (Fig.~\ref{fig:semi}a): silicon absorbs green light within about 1.5~$\mu$m, but
needs some 150~$\mu$m for the near infrared at 1000~nm, close to its gap~\cite{GreenSi}. This is
why a crystalline silicon solar cell is a fraction of a millimetre thick. In the visible a silicon
wafer reflects about a third of the light, because of its index of about 4, and absorbs the rest:
it looks grey and metallic, although it is not a metal.

The colour of a semiconductor is decided by the position of its gap with respect to the visible
spectrum (Fig.~\ref{fig:semi}b)~\cite{Nassau}. If the gap lies above the violet end, at 3.1~eV, the
crystal is colourless, like diamond (5.4~eV) and zinc sulfide. If it falls inside the visible, all
the light above the gap is absorbed and the crystal takes the colour of what remains: cadmium sulfide
(2.4~eV) absorbs blue and violet and is the yellow pigment of painters; cadmium selenide (1.7~eV)
leaves only the deep red and looks almost black. If the gap lies below the red end, as in
silicon and gallium arsenide, the whole visible spectrum is absorbed. The sharp threshold in photon
energy is the trace of the quantum: each photon gives its energy to a single electron, and it can
lift it across the gap only if it carries enough.

Finally, a semiconductor can be given free electrons of its own by doping. With a density some
tens of times smaller than in a metal, and in a medium of higher index, the plasma frequency of Eq.~\eqref{eq:drude} moves into the
infrared. Tin-doped indium oxide (ITO) reflects the infrared like a metal and conducts electricity,
but it is transparent in the visible~\cite{HambergGranqvist}: it is the transparent electrode of
touch screens and solar cells, and the coating of the ``low-emissivity'' windows that keep the heat
of a room inside.
\par}

%% file: ch6_brewster.tex
\section{Polarization and Brewster's angle}
\label{sec:brewster}

Look at the reflection of the sky on a lake, or the glare of the sun on a wet road, through
polarizing sunglasses \rev{(panel in Sec.~\ref{sec:one-atom})}, and turn your head: the glare comes and goes. At a particular angle the
light reflected by water or glass, polarized in the plane of incidence, disappears altogether.
Malus discovered the polarization of reflected light in 1808 looking at the windows of the Palais
du Luxembourg through a calcite crystal, and Brewster found in 1815 the law of the vanishing
angle, $\tan\theta_B=n$. In the picture of this text it has a very simple cause: an oscillating
dipole does not radiate along its own axis (Sec.~\ref{sec:one-atom}).

\subsection{The dipoles that cannot reflect}

\begin{figure}[!htbp]
\centering
\includegraphics[width=0.72\textwidth]{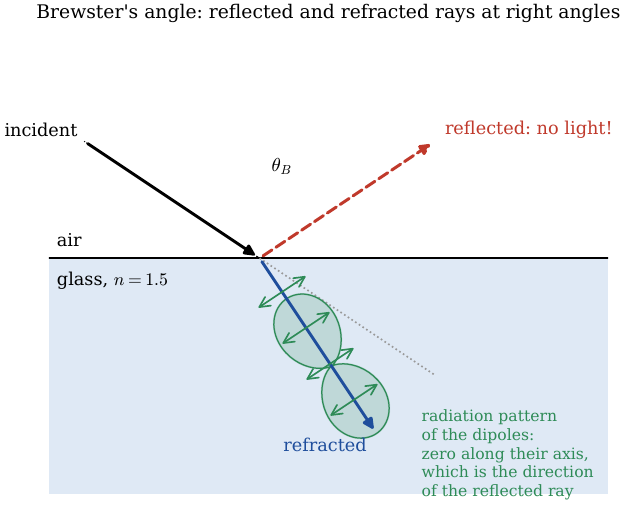}
\caption{At Brewster's angle the refracted and reflected rays are perpendicular. The dipoles of
the glass, driven by the refracted wave polarized in the plane of incidence, oscillate along the
direction of the reflected ray, where their radiation pattern (shaded) vanishes.}
\label{fig:brewster-geom}
\end{figure}

The reflected wave is radiated by the dipoles of the medium, and these oscillate along the field
that actually exists inside, the refracted wave. For the polarization perpendicular to the plane of
incidence (\emph{s}), the dipoles oscillate perpendicularly to that plane, and radiate equally in
all directions within it: they always send some light along the reflected direction. For the
polarization in the plane of incidence (\emph{p}), the dipoles oscillate in that plane,
perpendicularly to the refracted ray. If the reflected ray happens to point exactly along their
axis, they cannot radiate in that direction, and there is no reflected wave
(Fig.~\ref{fig:brewster-geom}). The reflected ray is then perpendicular to the refracted one,
$\theta_i+\theta_t=90^\circ$, and with Snell's law, $\sin\theta_i=n\sin\theta_t=n\cos\theta_i$:
\begin{equation}
\tan\theta_B=n .
\label{eq:brewster}
\end{equation}
For glass, $\theta_B=56.3^\circ$; for water, $53.1^\circ$.
\rev{The same dipoles also emit, into the glass, the wave that extinguishes the incident one
(Sec.~\ref{sec:extinction}), and how strongly they must oscillate to do so is set by their
projection $\cos(\theta_i-\theta_t)$ on that wave. Hence, apart from a sign that depends on the
convention for the $p$ field, $r_p/r_s=\cos(\theta_i+\theta_t)/\cos(\theta_i-\theta_t)$
exactly~\cite{FeynmanI,Reali2014}: the numerator says where the dipoles cannot radiate, the
denominator how strongly they must oscillate.}

\begin{figure}[!htbp]
\centering
\includegraphics[width=0.95\textwidth]{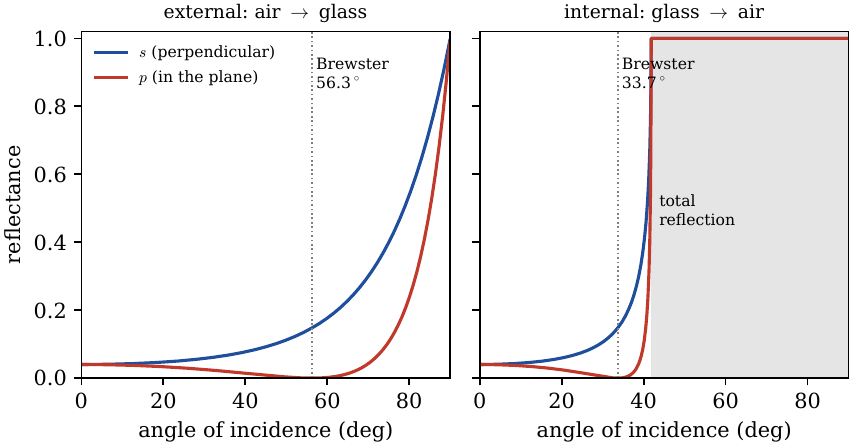}
\caption{Reflectance of a glass surface ($n=1.5$) for the two polarizations, as a function of the
angle of incidence. Left: light coming from air. Right: light coming from inside the glass; beyond
the critical angle ($41.8^\circ$, shaded) the reflection is total.}
\label{fig:fresnel}
\end{figure}

\begin{figure}[!htbp]
\centering
\includegraphics[width=0.95\textwidth]{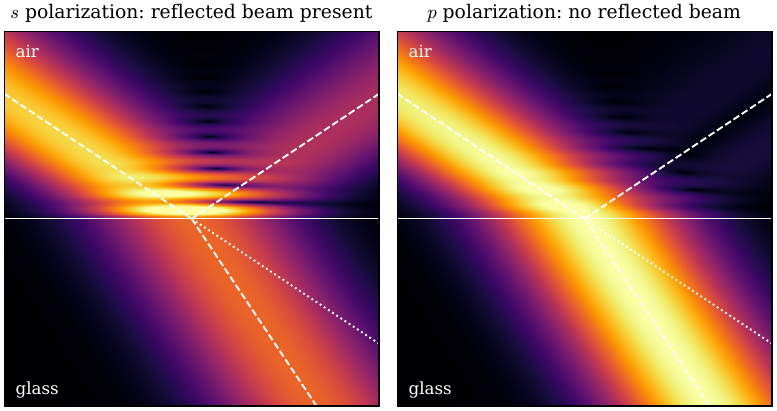}
\caption{Amplitude of the field of a light beam (waist $2.5\lambda$) falling from air on glass at
Brewster's angle, computed exactly as a superposition of plane waves. Dashed lines: centroids of
the incident, reflected and refracted beams; dotted: straight continuation of the incident beam.
Left: \emph{s} polarization; the reflected beam interferes with the incident one near the
surface. Right: \emph{p} polarization; no reflected beam (\anim{6.1}{anim6_brewster.mp4}).}
\label{fig:beams}
\end{figure}

Figure~\ref{fig:fresnel} shows the reflectances for the two polarizations, given by Fresnel's
formulas (panel below), for external reflection (from air into glass) and internal reflection
(from glass towards air). Figure~\ref{fig:beams} shows a light beam meeting glass at Brewster's
angle: in \emph{s} polarization a reflected beam leaves the surface, in \emph{p} polarization it
does not (\anim{6.1}{anim6_brewster.mp4} sweeps the angle of incidence across Brewster's).

\begin{panel}{Fresnel's formulas from the sheets}
For an \emph{s} wave the sheets of Sec.~\ref{sec:oblique} radiate forward and backward with the
same amplitude, and repeating the self-consistent calculation of Sec.~\ref{sec:extinction} with
$k_z=k\cos\theta_i$ and $k_z^{(m)}=nk\cos\theta_t$ gives
\begin{equation*}
r_s=\frac{\cos\theta_i-n\cos\theta_t}{\cos\theta_i+n\cos\theta_t}.
\end{equation*}
For a \emph{p} wave the dipoles have a component normal to the surface, so that a surface charge
appears on the boundary, and the backward wave of each sheet carries the angular factor of the
dipole radiation; as shown by Lai, Lau and Wong~\cite{LLW}, the result is
\begin{equation*}
r_p=\frac{n\cos\theta_i-\cos\theta_t}{n\cos\theta_i+\cos\theta_t}
\end{equation*}
(for the magnetic field), which vanishes when $\cos\theta_t/\cos\theta_i=n$, i.e.\ for
$\tan\theta_i=n$.
\end{panel}

\subsection{Internal reflection}

\begin{figure}[!htbp]
\centering
\includegraphics[width=0.72\textwidth]{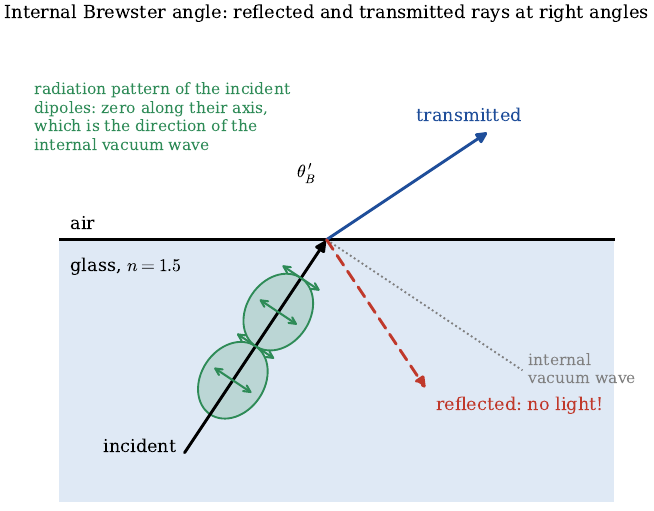}
\caption{\rev{At the internal Brewster angle the reflected and transmitted rays are perpendicular.
The dipoles of the incident wave oscillate along the direction of the internal vacuum wave, the
mirror image of the transmitted ray, where their radiation pattern (shaded) vanishes: the wave that
the reflected dipoles would have to cancel is not emitted, and there is no reflected wave.}}
\label{fig:brewster-int}
\end{figure}

Light travelling inside glass and meeting its surface is also partly reflected, and for the
\emph{p} polarization the reflection vanishes again, at the internal Brewster angle
$\tan\theta_B'=1/n$ ($33.7^\circ$ for glass). The two Brewster angles are complementary, and the
condition is the same: reflected and transmitted rays at right angles. Here, however, the dipoles
are on the side of the incident light, and the transmitted wave travels in air, where there are
none. \rev{Each polarization wave in the glass, the incident and the reflected one, emits from the
surface two waves travelling at the speed of light in vacuum: the outward ones add up to the
transmitted light; the inward ones must cancel, because no such wave survives in the glass
(Sec.~\ref{sec:extinction}). With no outside wave to extinguish, the reflected wave exists precisely
to cancel the inward wave of the incident dipoles. When the reflected and transmitted rays are at
right angles, that inward wave, the mirror image of the transmitted ray, points along the axis of
the incident dipoles (Fig.~\ref{fig:brewster-int}): they do not emit it, there is nothing to
cancel, and no reflected wave is needed~\cite{Sotskii,Reali2014}. At the external Brewster angle
the reflected light is missing because nobody produces it; at the internal one, because nobody
needs it.} Beyond the critical angle,
$\sin\theta_c=1/n$, no transmitted wave can propagate, and the reflection is total; only an
evanescent field, decaying within a wavelength, leaks into the air.

\begin{figure}[!htbp]
\centering
\includegraphics[width=0.72\textwidth]{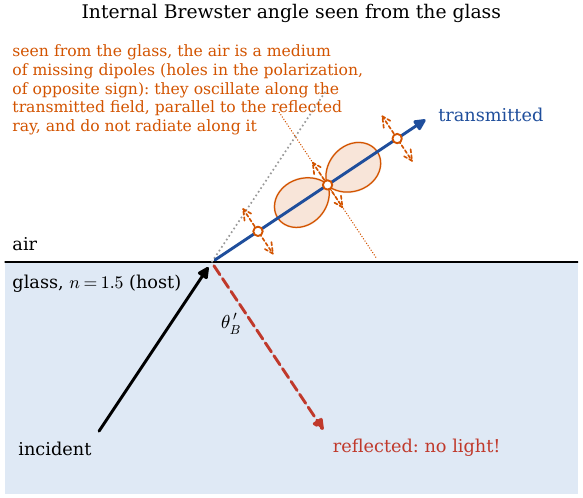}
\caption{\rev{The internal Brewster angle seen from the glass, taken as the host medium. The air
is described as glass with missing dipoles (holes in the polarization, of opposite sign), which
oscillate along the transmitted field. At $\theta_B'$ this field is parallel to the reflected ray,
along which the missing dipoles do not radiate (shaded pattern), and there is no reflected wave.
Compare Fig.~\ref{fig:brewster-geom}, the same picture for the external angle, and
Fig.~\ref{fig:brewster-int}, the same angle in the vacuum reading.}}
\label{fig:missing}
\end{figure}

{\revon
\emph{A second way of seeing it.} This explanation rests on a choice, made silently: the vacuum
as the background in which all the dipoles radiate. The extinction theorem can equally be written
taking the glass itself as the background, the \emph{host} medium. A second medium is then
described by its polarization in excess of that which the host would have in the same field,
$\varepsilon_0(n_2^2-n_1^2)\mathbf E_t$, and the reflected wave is the wave that these excess
dipoles radiate into the host. Sotskii's own final formula~\cite{Sotskii} contains this reading,
although he does not draw it. Seen from the glass, the air is a medium of \emph{missing} dipoles,
$\varepsilon_0(1-n^2)\mathbf E_t$: holes in the polarization, of opposite sign, oscillating along
the transmitted field. At the internal Brewster angle this field is parallel to the reflected ray,
and the missing dipoles cannot radiate along their own axis (Fig.~\ref{fig:missing}): the picture
of the external Brewster angle returns, with the sign of the source reversed. The same reading
covers any pair of media: the excess dipoles oscillate along the transmitted field, and the
condition is again that reflected and transmitted rays be at right angles. In the vacuum reading,
instead, between two dielectrics neither contribution to the inward vacuum wave vanishes, and the
reflection disappears only because the two cancel each other. The two readings are equivalent,
both being identities on the same solution of Maxwell's equations, and neither is \emph{the}
explanation. The vacuum one treats all media alike, and the molecular theory follows from it with
no further assumption; the host one involves only waves that actually exist, and makes the two
Brewster angles one and the same phenomenon.
\par}

{\revon
\emph{External and internal Brewster angles, seen with the arrows.} The two Brewster angles are
not independent: at the external one, $\tan\theta_B=n$, the refraction angle $\psi$ satisfies
$\tan\psi=1/n$, which is the Brewster angle of the glass--air face. The arrows of the sheets of
Sec.~5 show why this matters for a plate~\cite{Reali1992,Reali2014}. At oblique incidence the field
radiated by a sheet is slightly more complicated than at normal incidence; the panel below derives it.
The arrows turn with $nk\cos\psi\pm k\cos\theta$, and for p polarization the length of each arrow is
multiplied by the cosine of the angle between $\mathbf P$ and the direction of the electric field of
the reflected ray.
\par}

\begin{panel}{\rev{The field of a sheet at oblique incidence}}
{\revon
Let the incident wave arrive at the angle $\theta$, in the plane $xz$. A sheet at depth $\zeta$ is then
polarized as $\mathbf P(\zeta)\,e^{ik_xx}$, with $k_x=k\sin\theta$: it carries the phase that the
incident wave imprints along the surface. Three remarks fix the field it radiates.

\emph{1. Which waves.} The sheet is invariant under translations along $x$, apart from the phase
$e^{ik_xx}$; outside it, therefore, it can only radiate plane waves with the same $k_x$. In vacuum the
wave vector has length $k$, and only two wave vectors are allowed,
\begin{equation*}
\mathbf k_\pm=k\,(\sin\theta,\,0,\,\pm\cos\theta),\qquad \hat{\mathbf u}_\pm=\mathbf k_\pm/k,
\end{equation*}
one forward ($+$, the direction of the transmitted light) and one backward ($-$, the direction of the
reflected light): the oblique version of the two plane waves of Sec.~2.

\emph{2. How large.} At normal incidence the factor was $ik/2\varepsilon_0$, Eq.~\eqref{eq:sheet}; at
oblique incidence it becomes $ik/(2\varepsilon_0\cos\theta)$. The $1/\cos\theta$ is the $2\pi i/k_z$ of
the Weyl integral of Sec.~2; put simply, a beam of given cross section meets an area of the sheet
larger by $1/\cos\theta$.

\emph{3. In which direction.} A dipole does not radiate along its own axis: in the direction
$\hat{\mathbf u}_\pm$ only the part of $\mathbf P$ perpendicular to $\hat{\mathbf u}_\pm$ radiates,
\begin{equation*}
\mathbf P_\perp=\mathbf P-\hat{\mathbf u}_\pm\,(\hat{\mathbf u}_\pm\cdot\mathbf P).
\end{equation*}
Putting the three together,
\begin{equation}
\ud\bE_\pm=\frac{ik}{2\varepsilon_0\cos\theta}\;\mathbf P_\perp(\zeta)\;e^{ik\cos\theta\,|z-\zeta|}\,\ud\zeta .
\label{eq:obliquesheet}
\end{equation}
(The compact form found in the literature~\cite{LLW}, with the dyad $\mathbf k_\pm\mathbf k_\pm$, says
the same thing: $(k^2\mathsf I-\mathbf k_\pm\mathbf k_\pm)\cdot\mathbf P=k^2\mathbf P-\mathbf
k_\pm(\mathbf k_\pm\cdot\mathbf P)=k^2\mathbf P_\perp$.)

\emph{s polarization}: $\mathbf P$ is along $y$, already perpendicular to both $\mathbf k_\pm$, and nothing
is removed. \emph{p polarization}: $\mathbf P$ lies in the plane of incidence, and only its component along
the direction of the radiated electric field survives. For the forward wave in the glass, travelling
along $nk(\sin\psi,0,\cos\psi)$, $\mathbf P$ points along $(\cos\psi,0,-\sin\psi)$; the electric field
of the reflected wave points along $(\cos\theta,0,\sin\theta)$, and the factor is
$\cos\theta\cos\psi-\sin\theta\sin\psi=\cos(\theta+\psi)$, which vanishes at Brewster's angle,
$\theta+\psi=\pi/2$. For the backward wave in the glass the factor is $\cos(\theta-\psi)$, which never
vanishes.
\par}
\rev{\emph{4. Inside the sheet.} Within its own thickness the sheet also produces the field
$-P_z\hat{\mathbf z}/\varepsilon_0$ of its two charged faces (Gauss's law), to be added to
Eq.~\eqref{eq:obliquesheet}. It radiates nothing, but it fixes the direction of the field that
drives the dipoles. Summing only the far-zone dipole field loses it, and the $p$ polarization comes
out wrong: this is the ``hazard'' of volume integration pointed out by Turtle~\cite{Turtle}.}
\end{panel}

\begin{figure}[!htbp]
\centering
\includegraphics[width=\textwidth]{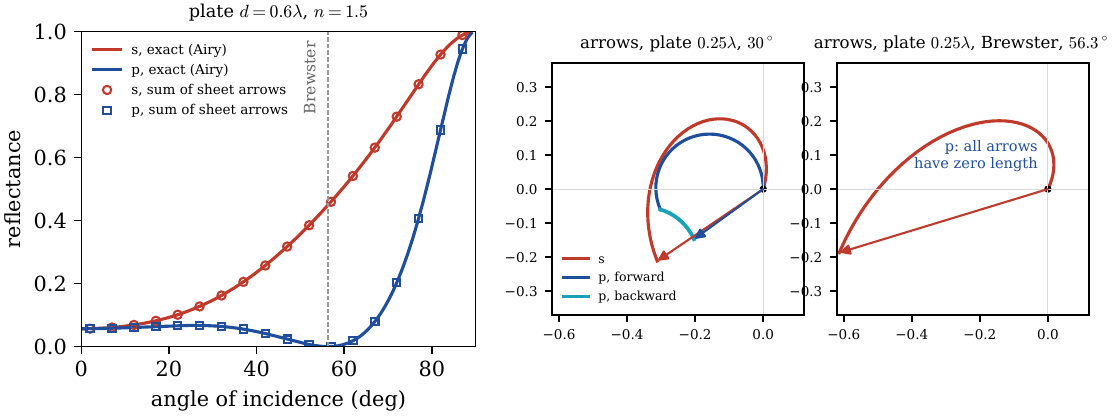}
\caption{\rev{Oblique incidence on a glass plate. Left: reflectance of a plate $0.6\lambda$ thick
for the two polarizations (lines: Airy; symbols: sum of the arrows of the sheets). Centre and right:
the chains of arrows of a plate $0.25\lambda$ thick at $30^\circ$ and at Brewster's angle. There
the p arrows all have zero length: those of the forward chain because of the dipole factor
$\cos(\theta+\psi)$, those of the backward chain because at the refraction angle $\psi$ (the
internal Brewster angle) there is no internally reflected wave.}}
\label{fig:arrowsobl}
\end{figure}

{\revon
For the forward
wave in the plate this factor is $\cos(\theta+\psi)$, which vanishes when $\theta+\psi=\pi/2$: at
the external Brewster angle \emph{all the arrows of the forward chain have zero length}. The chain
of the backward wave has a non-vanishing direction factor, but at that angle $\psi$ is the internal
Brewster angle, the internal reflection vanishes, and there is no backward wave, $\beta=0$: the
second chain disappears as well. The reflection of the plate vanishes for every thickness because
every single arrow vanishes (Fig.~\ref{fig:arrowsobl}). The Brewster angle is not a cancellation
between the two faces, but a property of each dipole of the glass.
\par}

{\revon
\emph{Volume and surface.} There is a difference between the two polarizations that the arrows
hide in the projector. For s polarization $\mathbf P$ is parallel to the surface, and the reflected
and transmitted waves are a pure volume effect of the polarization currents. For p polarization
$\mathbf P$ has a component normal to the surface, and the discontinuity of $\mathbf P\cdot\hat{\mathbf
n}$ at the faces is a bound surface charge. \rev{There are two equivalent bookkeepings. Lai, Lau and
Wong~\cite{LLW} take the current of each sheet, not projected, plus this surface
charge, which then contributes a term of its own, of the same order as the volume one.
Eq.~\eqref{eq:obliquesheet} instead gives each sheet its own charges: the in-plane ones and the
double layer on its two faces turn the current into the transverse projection $\mathbf P_\perp$, and
the double layer adds the contact field of remark~4. Summed over the sheets, the double layers of
neighbouring sheets cancel, and only the charge on the faces of the medium remains.} This surface term is not the local-field correction of Sec.~4: that
one exists in an unbounded medium and comes from the discreteness of the atoms; this one exists
only because the medium ends, and would be there even in a continuous dielectric.
\par}

\subsection{\texorpdfstring{\rev{Birefringence}}{Birefringence}}
\label{sec:birefringence}

\begin{figure}[!htbp]
\centering
\includegraphics[width=\textwidth]{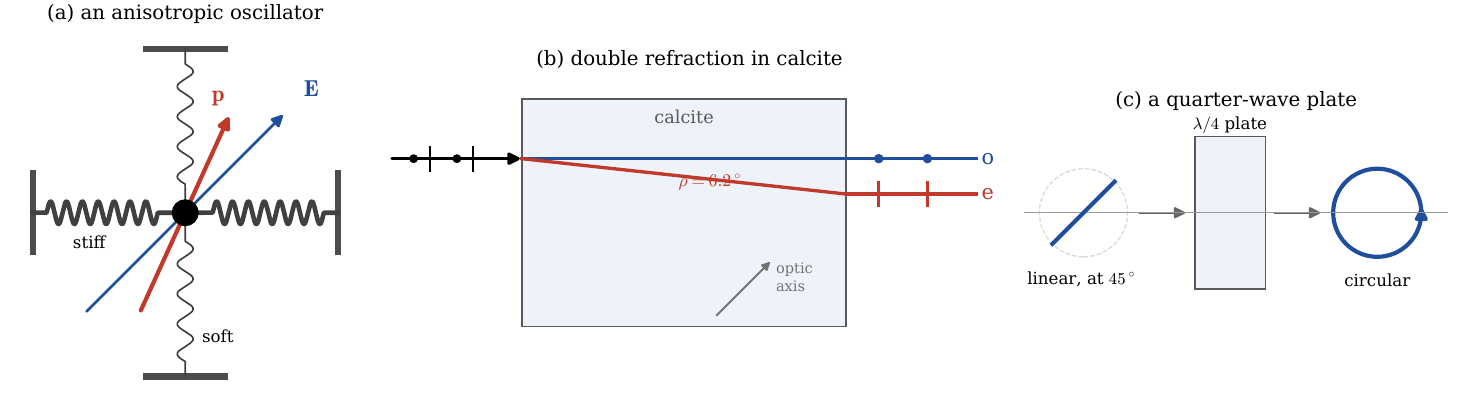}
\caption{\rev{Birefringence. (a) An electron held by springs of different stiffness: the dipole
$\mathbf p$ induced by a field $\mathbf E$ is pulled towards the soft direction. (b) A beam of
unpolarized light at normal incidence on a calcite plate, with the optic axis at $45^\circ$ to the
faces: the ordinary beam (polarization perpendicular to the page, dots) goes straight; the
extraordinary beam (polarization in the page, bars) walks off by $\rho=6.2^\circ$. (c) A quarter-wave
plate turns light linearly polarized at $45^\circ$ to its axis into circularly polarized light.}}
\label{fig:biref}
\end{figure}

{\revon
In glass every dipole oscillates along the field that drives it, whatever the direction of the
field: the medium is \emph{isotropic}. In a crystal this need not be so. The electrons are held by
restoring forces that depend on the direction, as if the spring of Eq.~\eqref{eq:oscillator} were
stiffer along one axis than along another. A field along a stiff direction displaces the electron
less than the same field along a soft one, so the polarizability, and with it the index, depends on
the direction of the field: the crystal is \emph{birefringent}. A field oblique to the axes, moreover,
induces a dipole that is pulled towards the soft direction and is no longer parallel to the field
(Fig.~\ref{fig:biref}a).

The classic example is calcite, CaCO$_3$, whose double images Bartholin described in 1669 and
Huygens explained in 1690 with wavelets of two shapes. In calcite the carbonate groups are flat
triangles of oxygen atoms, all parallel to one plane, perpendicular to the \emph{optic axis} of the
crystal. A field in that plane induces dipoles in the oxygen atoms that reinforce one another; a
field along the axis induces dipoles that partly oppose one another. W.~L.~Bragg computed the two
indices of calcite in 1924 from the mutual action of the induced dipoles of the oxygen
atoms~\cite{Bragg1924}---the self-consistent method of Sec.~\ref{sec:extinction}, applied within
the unit cell of the crystal. Light polarized perpendicular to the axis (the \emph{ordinary} wave)
sees $n_o=1.658$; light polarized along it (\emph{extraordinary}) sees $n_e=1.486$, at 589~nm. For
intermediate directions the extraordinary index lies between the two, and since the dipoles are not
parallel to the field the flow of energy is not parallel to the wave vector either: at normal
incidence on a plate cut obliquely to the axis, the ordinary beam goes straight on, while the
extraordinary one walks off sideways, by up to $6.2^\circ$ in calcite (Fig.~\ref{fig:biref}b). The
two beams leave the plate parallel, polarized at right angles: an object seen through the crystal
appears twice. It was by looking through such a crystal at the windows of the Palais du Luxembourg
that Malus discovered the polarization of reflected light (Sec.~\ref{sec:brewster}).

A plate cut with the optic axis in its face does not separate the beams, but it delays one
polarization with respect to the other by the phase $2\pi(n_o-n_e)d/\lambda$. A \emph{quarter-wave
plate} delays it by a quarter of a period and turns light linearly polarized at $45^\circ$ to the
axis into circularly polarized light (Fig.~\ref{fig:biref}c); a half-wave plate rotates the plane of
polarization. Such plates are usually made of quartz or mica, whose weaker birefringence makes
them easier to make thick enough: a quarter-wave plate of quartz is about 16~$\mu$m thick in the
yellow. Birefringence can also be induced. A transparent plastic under stress becomes
birefringent, and between crossed polarizers (panel in Sec.~\ref{sec:one-atom}) a ruler or a
transparent box shows coloured fringes that map its internal stresses: engineers use them to study
the stresses in models of structures. In Sec.~\ref{sec:beyond} the birefringence of a crystal will
be used to make two waves of different frequency travel at the same speed.

\emph{Liquid crystals.} The molecules of a nematic liquid crystal are rods, more polarizable
along their length than across it, and in the liquid they tend to align with one another: the
liquid is birefringent, with $n_e-n_o$ of the order of 0.1--0.2. A few volts across a layer a few
micrometres thick are enough to turn the rods. In the twisted-nematic cell~\cite{SchadtHelfrich}
the rods are anchored at the two walls in perpendicular directions and twist through $90^\circ$
across the layer; light that has passed a first polarizer follows the twist and passes a second,
crossed polarizer. With the voltage on, the rods stand along the field, the twist is gone, and the
light is stopped. Every pixel of a liquid-crystal display is such a cell: a birefringent layer
between two polarizers, controlled by a voltage.
\par}

\subsection{Polarized light in nature}

The same dipole pattern polarizes the light of the sky: at $90^\circ$ from the sun we see only the
light of dipoles oscillating across our line of sight (Sec.~\ref{sec:one-atom}). Bees, ants and
many birds see this polarization and use the pattern of the sky as a compass, even when the sun
is hidden. The glare of a lake or a road, mostly \emph{s}-polarized because the \emph{p} component
is weakly reflected near Brewster's angle, is removed by sunglasses that transmit only the vertical
polarization; photographers use polarizing filters to darken the sky and to see through the
reflections on water. And Brewster windows---plates set at Brewster's angle---let laser light
polarized in the plane of incidence through without any reflection loss, which is why so many
lasers emit linearly polarized light.

%% file: ch6b_mirrors.tex
\section{\texorpdfstring{\rev{Mirrors, lenses and eyes}}{Mirrors, lenses and eyes}}
\label{sec:instruments}

{\revon
So far we have asked what the atoms of a piece of glass do to light. Optical instruments turn the
question around: given what the atoms do, how should a piece of glass or of silvered metal be
\emph{shaped} so that the wavelets of its dipoles add up where we want them? Mirrors, lenses,
telescopes, microscopes and our own eyes are, from this point of view, devices for managing the
phases of the arrows of Secs.~2--6. Geometrical optics---rays, focal points, the laws of reflection
and refraction---comes out as a limiting case, and the same arrows tell us where it stops working.
\par}

\begin{figure}[!htbp]
\centering
\includegraphics[width=\textwidth]{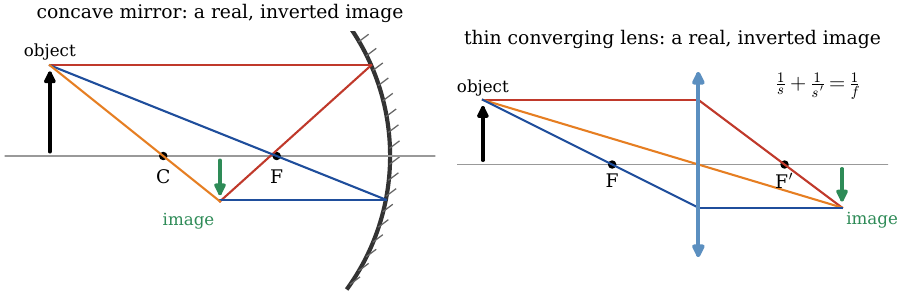}
\caption{\rev{Image formation by a concave mirror (C: centre of curvature, F: focus, $f=R/2$) and by a
thin converging lens (F, F$'$: foci), with the three principal rays from the tip of the object.}}
\label{fig:rays}
\end{figure}

{\revon
Figure~\ref{fig:rays} recalls the familiar pictures of geometrical optics: a concave mirror and a thin
converging lens forming a real, inverted image of an object. Three rays from the tip of the object
are enough: one parallel to the axis, which leaves through the focus; one through the focus, which
leaves parallel to the axis; one through the centre of curvature of the mirror, or through the centre
of the lens, which is not deviated. Where they meet is the image, at the distance $s'$ given by
$1/s+1/s'=1/f$ (with $f=R/2$ for a mirror of radius $R$). The rest of this section explains what a
ray is: the direction along which the arrows of the wavelets add up, and the image the place where
\emph{all} the arrows from one point of the object arrive in step.
\par}

\subsection{\texorpdfstring{\rev{The mirror: where the time is stationary}}{The mirror: where the time is stationary}}

\begin{figure}[!htbp]
\centering
\includegraphics[width=\textwidth]{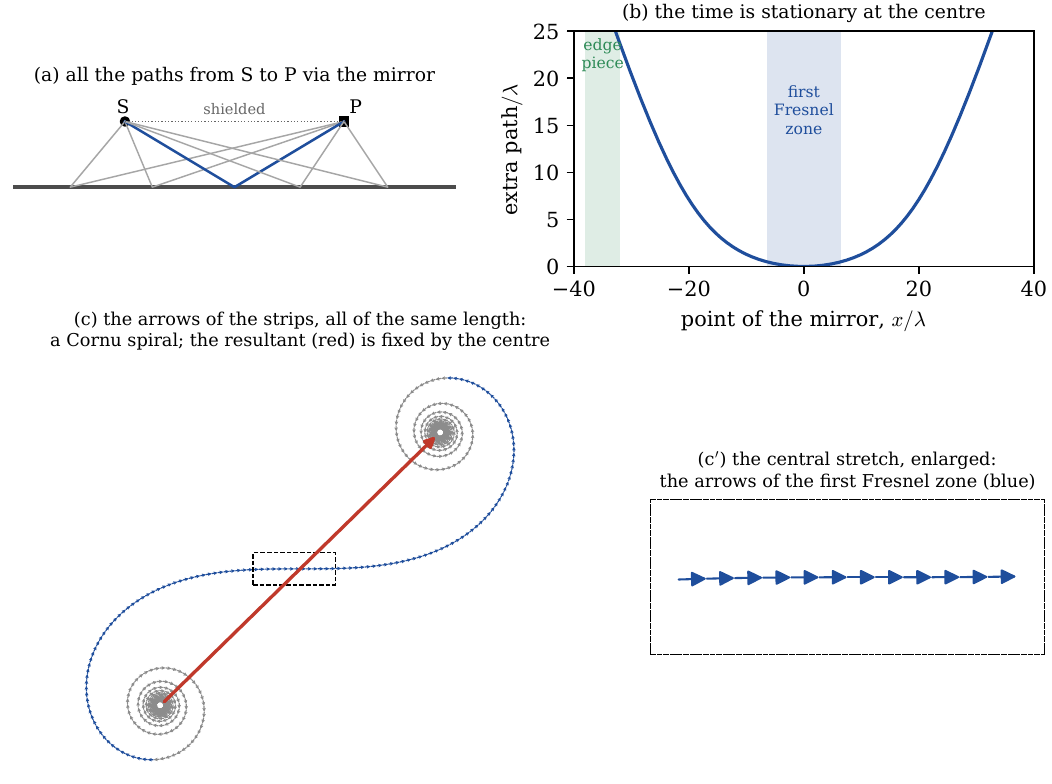}
\caption{\rev{Feynman's mirror~\cite{FeynmanQED-book}, computed (lengths in wavelengths; S and P
$40\lambda$ apart, $12\lambda$ above the mirror). (a) Some of the paths. (b) The extra length of the
path via each point of the mirror: it is stationary at the centre; shaded, the first Fresnel zone
(extra path below $\lambda/2$) and the edge piece of Fig.~\ref{fig:grating}. (c) The arrows of strips
$\lambda/10$ wide, added head to tail: a Cornu spiral. The ends curl up into the two eyes, and the
resultant (red) comes from the centre. (c$'$) The central stretch, enlarged: the arrows of the first
Fresnel zone (blue) are almost aligned. All the arrows are drawn with the same length, as in Feynman's
figure; the true amplitudes decrease slowly away from the centre, but the arrows that differ most are
those that cancel anyway.}}
\label{fig:mirror}
\end{figure}

\begin{figure}[!htbp]
\centering
\includegraphics[width=\textwidth]{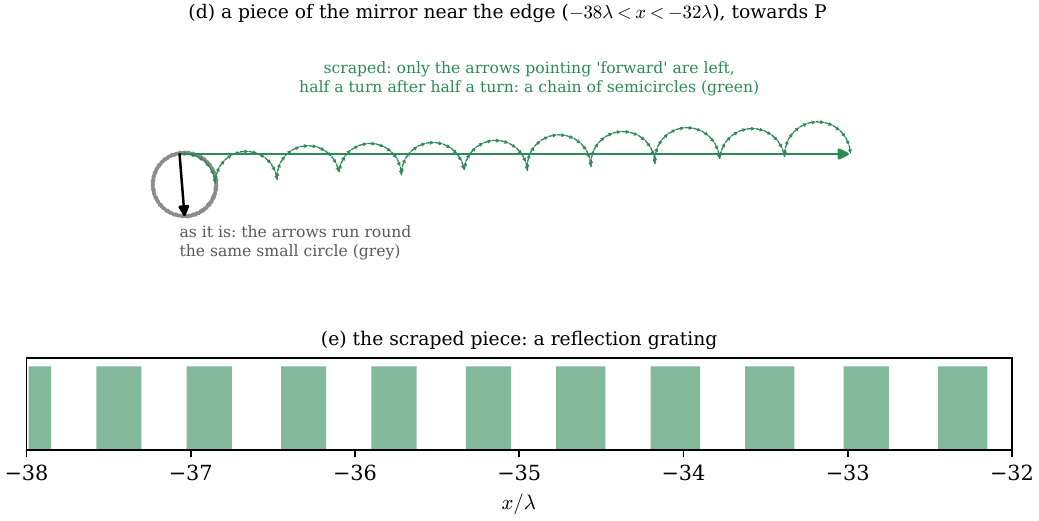}
\caption{\rev{A piece of the mirror near its edge (shaded in Fig.~\ref{fig:mirror}b). (d) As it is, its
arrows run round the same small circle and give almost nothing towards P (grey; resultant in black).
Scraping away the strips whose arrows point backwards leaves half a turn after half a turn: a chain of
semicircles, whose resultant (green) is ten times larger. (e) The scraped piece: a reflection
grating.}}
\label{fig:grating}
\end{figure}

{\revon
In the second chapter of \emph{QED} Feynman turns to an ordinary mirror~\cite{FeynmanQED-book}. Light goes from a source S
to a detector P by way of the mirror (Fig.~\ref{fig:mirror}a). Which part of the mirror does the
reflecting? The rule of Sec.~5 says: every strip of the mirror sends a wavelet to P, each with an
arrow that turns with the length of its path. The path is shortest via the centre, where the angle of
incidence equals the angle of reflection, and there it is \emph{stationary}: near the centre the time
hardly changes from strip to strip (Fig.~\ref{fig:mirror}b), the arrows point the same way and add
up. Far from the centre the time changes rapidly, the arrows turn quickly, and they curl up in tight
spirals that contribute almost nothing (Fig.~\ref{fig:mirror}c). The law of reflection---and, in
general, Fermat's principle of least time---is the statement that only the paths near the
stationary one survive the sum.

The edges of the mirror do not contribute because their arrows cancel, not because they are idle:
every strip radiates. Feynman's proof is to scrape away, at the edge of the mirror, the strips whose
arrows point ``backwards'', leaving only those pointing ``forwards''. The arrows that cancelled no
longer have partners, and the scraped piece, which as it was sent nothing to P, now reflects
strongly in that direction (Fig.~\ref{fig:grating}). It has become a \emph{reflection grating}:
the colours of a compact disc are made this way. It is the same sum as for the row of dipoles of
Sec.~2, where only the directions with all phases in step survive.
\par}

{\revon
A curved mirror applies the same principle to bring light to a point. For a parabolic mirror all the
paths from a distant star to the focus have exactly the same length, so all the arrows arrive in step:
this is why the large telescopes are parabolic mirrors. Total internal reflection (Sec.~6) is a
mirror with no metal at all, and an optical fibre is a thread of glass that guides light by
reflecting it, without loss, thousands of times.
\par}

\begin{panel}{\rev{The arrows as quantum amplitudes: Feynman's sum over paths}}
{\revon
Figure~\ref{fig:mirror} is, computed, the figure with which Feynman explains the mirror in the
second chapter of \emph{QED}~\cite{FeynmanQED-book}. There the arrows are not the wavelets radiated by
the strips of the mirror: each of them is the \emph{probability amplitude} for a photon to go from S to
P along one path. Its direction is given by a stopwatch that turns once per period of the light, i.e.\
by the time taken along that path; the amplitude of the event ``a click at P'' is the sum of the arrows
of all the paths, and the probability of the click is the square of the length of the sum.

The curve traced by the arrows is a \emph{Cornu spiral}. Near the centre of the mirror the extra
length of the path grows as the square of the distance from the centre, the phases change slowly,
and the arrows line up in the almost straight middle stretch of the curve. Farther out the phase
changes faster and faster, and the arrows curl up into the two eyes of the spiral, where they cancel
one another. The resulting arrow goes from one eye to the other, and what fixes it are the paths
close to the one for which the time is \emph{stationary}: Fermat's principle, obtained as a limit
instead of being postulated. Stationary, not necessarily least: for a mirror shaped as an ellipse with
S and P at its foci all the paths take the same time, and all the arrows line up.

The same construction is at the heart of quantum mechanics. Dirac~\cite{Dirac1933} and
Feynman~\cite{Feynman1948} showed that for a particle of mass $m$ the amplitude to go from one point
to another is the sum, over all paths, of arrows $e^{iS/\hbar}$, where $S$ is the classical action
along the path. When $S$ is much larger than $\hbar$, only the paths near the one of stationary action
survive, and the classical trajectory emerges, just as the ray emerges from Fermat's principle:
geometrical optics is to wave optics what classical mechanics is to quantum mechanics.

The arrows of Fig.~\ref{fig:mirror} have thus been read twice: in the text, as the classical
wavelets of the dipoles of the mirror; in this panel, as quantum amplitudes. The arithmetic is the same
(see the panel \emph{What the arrows hide} in Sec.~\ref{sec:reflection}). What quantum theory adds is not a different sum,
but its meaning: the square of the final arrow is the probability of a click, and the clicks come one
at a time. This is the second of Marburger's photons~\cite{Marburger}: an imagined particle that
explores all the paths, not a small ball of light travelling along one of them; what is observed is
only the click at the end.
\par}
\end{panel}

\subsection{\texorpdfstring{\rev{The lens: a phase delay of the right shape}}{The lens: a phase delay of the right shape}}

\begin{figure}[!htbp]
\centering
\includegraphics[width=\textwidth]{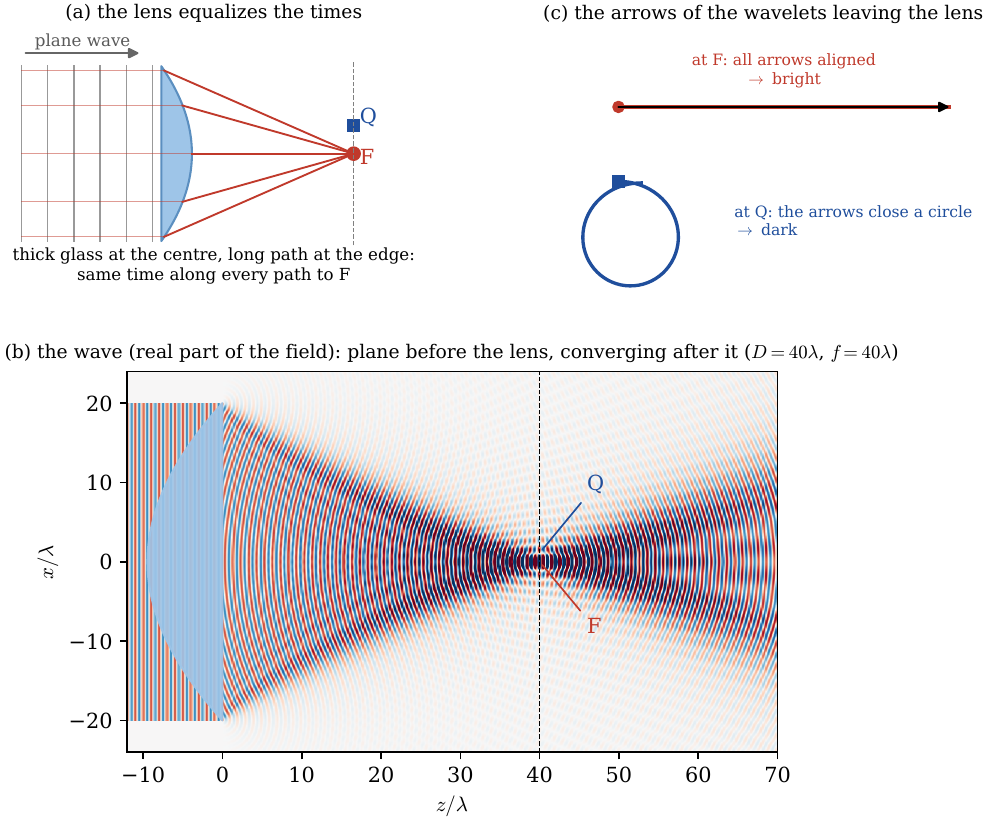}
\caption{\rev{A lens as a phase delay. (a) The paths from a plane wave to the focus F: the glass,
thicker at the centre, compensates the longer geometrical paths through the edge. (b) Real part of
the field of a plane wave crossing a lens with $D=40\lambda$ and $f=40\lambda$ (two-dimensional
calculation, sum of the wavelets leaving the lens; the lens is drawn shaded): plane wavefronts before
the lens, converging ones after it. (c) The arrows of the wavelets from the different points of the
lens: at the focus F they are all aligned; at the point Q of the focal plane, at a distance $\lambda f/D$
from the axis, they close a circle and the field vanishes.}}
\label{fig:lens}
\end{figure}

{\revon
A lens uses the other property of glass: it delays the wave. A thickness $t$ of glass turns the arrow
of the wave by $(n-1)kt$ with respect to vacuum (Sec.~3): the forward wavelets of the sheets turn it,
and the extinction theorem makes the delayed wave the only one. The lens is a plate whose thickness
varies across its face so that the delay compensates the different geometrical paths: the rays
through the edge travel further to the focus, those through the centre less, but those through the
centre cross more glass. If
\begin{equation*}
(n-1)\,t(x)+\sqrt{f^2+x^2}=\text{const},
\end{equation*}
all the paths from a plane wave to the point at distance $f$ take the same time, all the arrows
arrive aligned, and the light is focused. Near the axis $\sqrt{f^2+x^2}\simeq f+x^2/2f$, so the
thickness must decrease as $x^2/2(n-1)f$: a parabola, i.e.\ near the axis a spherical cap. For a
lens bounded by two spheres of radii $R_1$ and $R_2$ this gives the lensmaker's formula,
\begin{equation*}
\frac1f=(n-1)\Big(\frac1{R_1}-\frac1{R_2}\Big).
\end{equation*}
Figure~\ref{fig:lens} shows a plane wave focused by such a lens, computed by adding the
wavelets that leave it: the plane wavefronts entering the lens leave it as converging circles, which
shrink to the focus and open up again beyond it. At the focus all the arrows are aligned; a little to
the side, at a point Q of the focal plane, they turn, and at a distance $\lambda f/D$ from the axis
they close a full circle and the light vanishes. The focus is not a
point: it is a spot about a wavelength across, divided by the angular aperture $D/f$. Since $n$
depends on the colour (Sec.~3), so does $f$: a simple lens has a blue focus closer than the red one
(chromatic aberration), which is why Newton built a reflecting telescope.
\par}

\begin{figure}[!htbp]
\centering
\includegraphics[width=0.82\textwidth]{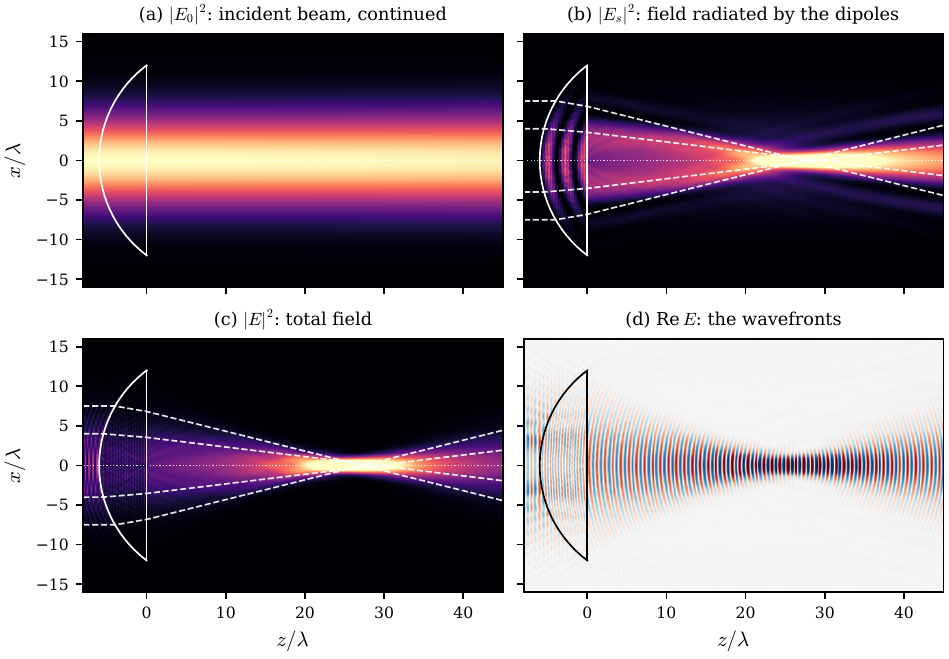}
\caption{\rev{A plano-convex lens made of 14\,492 coupled line dipoles (aperture $24\lambda$, radius of
curvature $15\lambda$, thickness $6\lambda$, lattice spacing $\lambda/12$, index of the lattice $1.44$),
illuminated by a Gaussian beam (waist $8\lambda$); s polarization, two-dimensional calculation with the code
of the ancillary files. (a) The incident beam, formally continued; (b) the field radiated by the dipoles;
(c) the total field; (d) its real part. Dashed: rays of geometrical optics at $\pm4\lambda$ and
$\pm7.5\lambda$ from the axis; dotted: the axis.}}
\label{fig:lenscdm}
\end{figure}

{\revon
A lens can also be built in the computer exactly as nature builds it: as a set of dipoles, each driven
by the incident wave and by the waves of all the others, with no index, no Snell's law and no focal
length put in by hand. Figure~\ref{fig:lenscdm} shows a plano-convex lens made of 14\,492 line dipoles
(the method of Sec.~\ref{sec:extinction}, Figs.~\ref{fig:cdmmaps} and~\ref{fig:cdmdecomp}, with the
dipoles filling the shape of the lens). The field radiated by the dipoles does two things at once: inside
the lens it cancels the incident beam, and behind it it builds a converging wave. The focus appears at
$24.9\lambda$ from the flat face, close to the rays of geometrical optics traced with the index of the
lattice, $1.44$: the paraxial rays cross at $29.8\lambda$, the outer ones, because of spherical aberration,
at $27.2\lambda$, and the focus of a beam a few Fresnel zones wide is pulled further towards the lens.
\par}

\subsection{\texorpdfstring{\rev{Telescopes: the size of the focus}}{Telescopes: the size of the focus}}

\begin{figure}[!htbp]
\centering
\includegraphics[width=\textwidth]{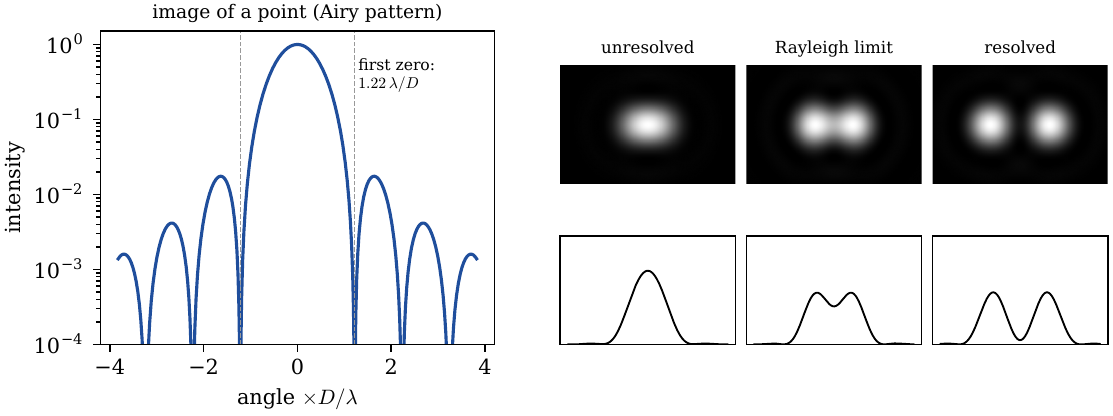}
\caption{\rev{The image of a point formed by a circular aperture of diameter $D$ (Airy pattern,
logarithmic scale) and the images of two points at 0.6, 1 and 1.6 times the Rayleigh separation
$1.22\lambda/D$, with their profiles.}}
\label{fig:airy}
\end{figure}

{\revon
For a circular lens or mirror of diameter $D$ the image of a star is the Airy pattern: a bright disc
surrounded by faint rings, whose first dark ring is at the angle
\begin{equation*}
\theta_{\min}=1.22\,\frac{\lambda}{D}
\end{equation*}
(Fig.~\ref{fig:airy}). Two stars closer than about this angle merge into a single blob; Rayleigh's
criterion~\cite{RayleighCrit} takes it as the limit of resolution. A telescope does two things,
both proportional to the size of its objective: it collects more light, in proportion to $D^2$, and
it separates finer details, in proportion to $D$. For the Hubble Space Telescope, $D=2.4$~m, the limit
is $0.05''$ in visible light; on the ground the turbulence of the air blurs the image to about
$1''$ unless it is corrected by adaptive optics. The magnification---the ratio of the focal lengths of
objective and eyepiece---only enlarges what the objective has resolved.
\par}

\begin{figure}[!htbp]
\centering
\includegraphics[width=\textwidth]{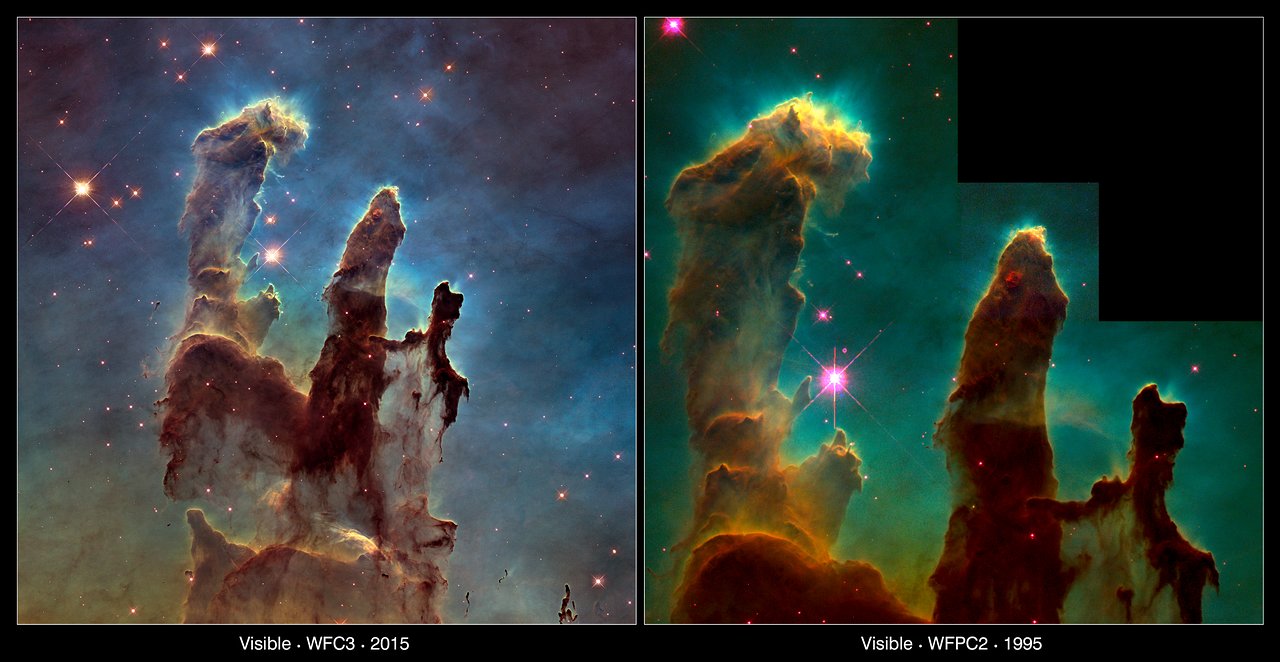}
\caption{\rev{The ``Pillars of Creation'' in the Eagle Nebula (M16), photographed by the Hubble Space
Telescope with the Wide Field Camera~3 in 2015 (left) and with the Wide Field and Planetary Camera~2 in
1995 (right; the stepped outline is the shape of that camera's mosaic of detectors). Credit: WFC3: NASA,
ESA/Hubble and the Hubble Heritage Team; WFPC2: NASA, ESA/Hubble, STScI, J.~Hester and P.~Scowen
(Arizona State University). Licence: CC~BY~4.0.}}
\label{fig:pillars}
\end{figure}

{\revon
Figure~\ref{fig:pillars} shows what a large aperture outside the atmosphere can do: the ``Pillars of
Creation'' in the Eagle Nebula, columns of gas and dust a few light years tall, photographed by the
Hubble Space Telescope twenty years apart. At a distance of some 6500 light years, Hubble's resolution
of $0.05''$ corresponds to details about a hundred times the distance from the Earth to the Sun. The
two pictures were taken with the same mirror, 2.4~m across, yet the newer one is visibly sharper: in
2009 the old camera was replaced by one with finer pixels and better detectors. The optics forms the
image; the detector must sample it finely enough not to waste it---the same matching between optics and
retina that we shall meet in the eye.
\par}

\subsection{\texorpdfstring{\rev{Microscopes: Abbe's theory of the image}}{Microscopes: Abbe's theory of the image}}

\begin{figure}[!htbp]
\centering
\includegraphics[width=0.95\textwidth]{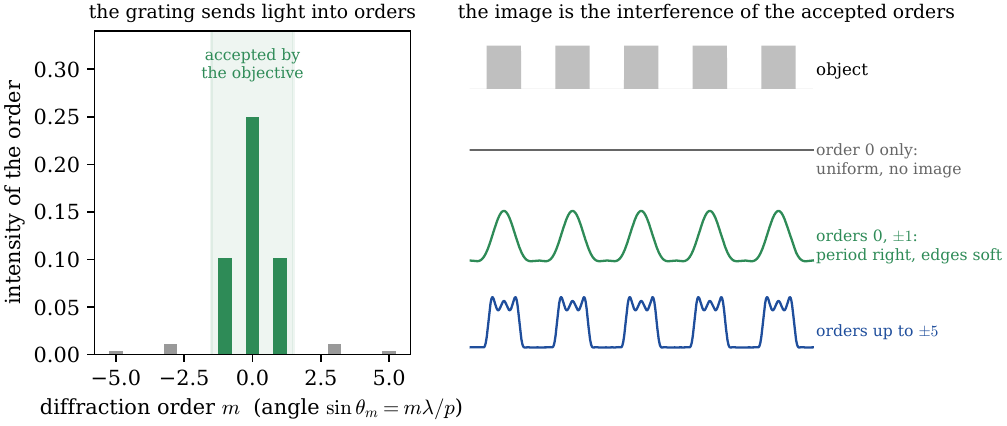}
\caption{\rev{Abbe's theory of the microscope. Left: the intensity sent by a grating (opaque and
transparent bars of equal width) into its diffraction orders; the objective accepts only the
shaded ones. Right: the image formed with the order 0 alone, with the orders $0,\pm1$, and with the
orders up to $\pm5$.}}
\label{fig:abbe}
\end{figure}

{\revon
In the microscope the problem is reversed: the object is small and close, and the question is how
fine a detail can be seen. Abbe's answer~\cite{Abbe} is a beautiful application of Sec.~2. A fine
periodic object, for instance a grating of period $p$ illuminated by a plane wave, sends light only
into the directions where the wavelets of its rows are in step: the diffraction orders,
$\sin\theta_m=m\lambda/p$. The objective collects some of these orders, and the image is their
interference in the image plane (Fig.~\ref{fig:abbe}). With the undeviated order alone the image is
uniform: the grating is invisible. With the orders $0$ and $\pm1$ the image already has the right
period; the higher orders sharpen its edges. The finest period that can be seen is the one whose
first order still enters the objective. With oblique illumination, so that the order 0 and one first
order fall at opposite edges of the objective,
\begin{equation*}
d_{\min}=\frac{\lambda}{2\,\mathrm{NA}},\qquad \mathrm{NA}=n\sin\alpha ,
\end{equation*}
where $\alpha$ is the half-angle of the cone accepted by the objective and $n$ the index of the
medium between the object and the objective. Filling that space with oil ($n\simeq1.5$) raises the
numerical aperture to about $1.4$ and the resolution to about $200$~nm in green light. Detail finer
than this sends its light into orders that leave the object at angles that no lens can collect, or
into evanescent waves that never leave it at all.
\par}

\begin{figure}[!htbp]
\centering
\includegraphics[width=0.62\textwidth]{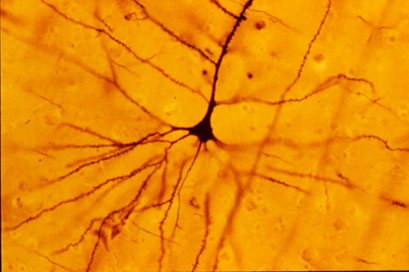}
\caption{\rev{A pyramidal neuron of the human neocortex stained with Golgi's method. Photo: Bob Jacobs,
Laboratory of Quantitative Neuromorphology, Department of Psychology, Colorado College, via Wikimedia
Commons; licence CC~BY-SA~3.0 (\texttt{creativecommons.org/licenses/by-sa/3.0}).}}
\label{fig:golgi}
\end{figure}

{\revon
The limit has a famous illustration close to home. In 1873 Camillo Golgi, later professor in Pavia,
discovered that a silver stain---his ``black reaction''---blackens a few nerve cells out of a hundred,
whole, with all their branches~\cite{Golgi1873}. For the first time single neurons could be seen under
the microscope (Fig.~\ref{fig:golgi}). Golgi saw in them a continuous network; Santiago Ram\'on y Cajal,
using the same stain, saw separate cells in contact. They shared the Nobel prize of 1906 without
agreeing. The light microscope could not settle the question: the gap between two neurons at a synapse
is some 20~nm wide, ten times below Abbe's limit. Only the electron microscope, in the 1950s, showed the
gap, and Cajal was right. Today the finest details of neurons are studied with super-resolution
microscopes that, by switching fluorescent molecules on and off, get around Abbe's limit.
\par}

\begin{figure}[!htbp]
\centering
\includegraphics[width=\textwidth]{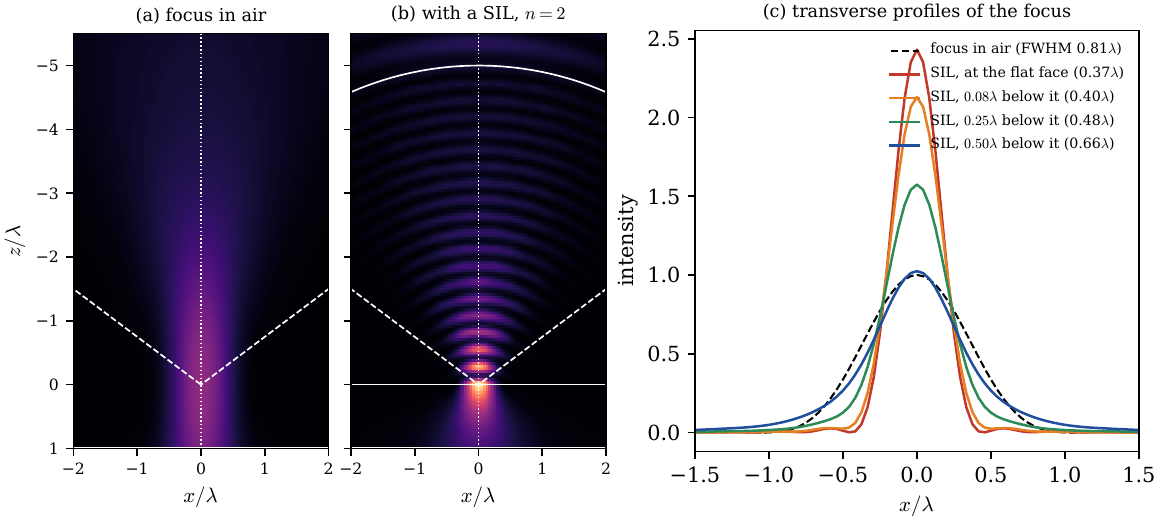}
\caption{\rev{A solid immersion lens made of 22\,622 coupled line dipoles (half disc of radius $5\lambda$,
continuum index $n=2$, lattice spacing $\lambda/24$), illuminated by a beam converging on the centre of its
flat face (numerical aperture in air $0.8$). (a) The same beam focused in air; (b) with the lens (white
outline); dashed: the edges of the converging beam; dotted: the axis. (c) Transverse profiles of the
intensity: in air (dashed) and with the lens, at the flat face and at increasing distances below it, with
their full widths at half maximum.}}
\label{fig:sil}
\end{figure}

{\revon
\emph{Beyond the limit: the near field.} Abbe's limit holds for light that travels; it does not hold
for light that stays. A detail finer than the wavelength sends its light into evanescent waves
(Sec.~\ref{sec:sheet}), which do not leave the object but cling to it, decaying within a distance
comparable to the size of the detail. To catch them one must go there. In near-field scanning optical
microscopy a tip, or an aperture a few tens of nanometres wide, is scanned a few nanometres above
the sample; the resolution is then set by the size of the tip and by its distance, not by the
wavelength. The idea was proposed by Synge in 1928~\cite{Synge} and realized in visible light by Pohl
and co-workers in 1984~\cite{Pohl}. A gentler version is the \emph{solid immersion lens}: a hemisphere of
glass whose flat face touches the sample. A beam converging on the centre of the flat face meets the
dome at normal incidence and is not deviated, but inside the glass its wavelength is $\lambda/n$, and the
focus shrinks by the factor $n$. Figure~\ref{fig:sil} shows it, again computed with dipoles only. The
narrow focus exists at the flat face; just below it, in air, its finest part survives only as an
evanescent wave: its width grows from $0.37\lambda$ at the face to $0.48\lambda$ a quarter of a wavelength
below and $0.66\lambda$ half a wavelength below, approaching the $0.81\lambda$ of the focus in air.
The parts of the beam that strike the flat face beyond the critical angle are totally reflected (Sec.~6):
they make the standing waves inside the dome and feed, below the face, the evanescent field.
\par}

\subsection{\texorpdfstring{\rev{The eye}}{The eye}}

\begin{figure}[!htbp]
\centering
\includegraphics[width=\textwidth]{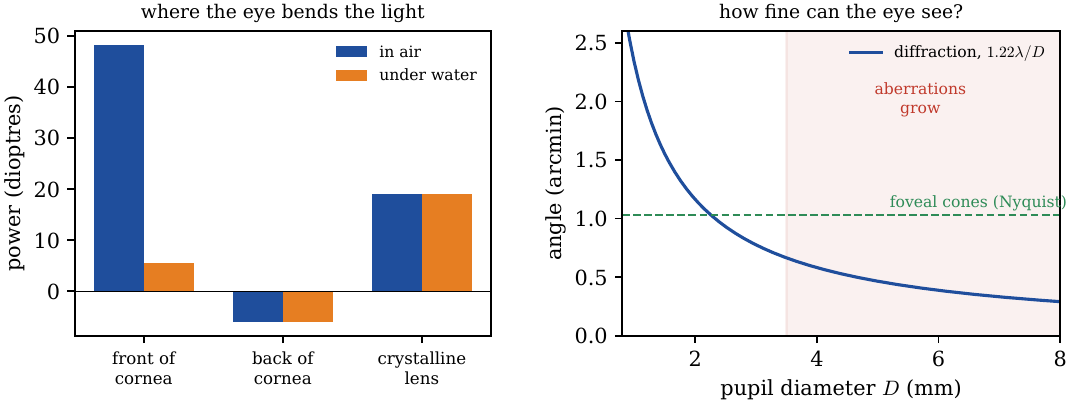}
\caption{\rev{The eye as an optical instrument. Left: power of the refracting surfaces of the
eye, in air and under water (front of the cornea computed from its radius, 7.8~mm, and index,
1.376; back of the cornea and crystalline lens, typical values). Right: the diffraction limit of
the pupil compared with the spacing of the cones in the fovea (two cones per period); for pupils
larger than about 3~mm the aberrations of the eye dominate.}}
\label{fig:eye}
\end{figure}

{\revon
The eye is a camera with a variable lens~\cite{AtchisonSmith}. Its total power is about $60$~dioptres
(a focal length of $17$~mm in air), but most of it is not in the lens: it is in the front surface of
the cornea, where the light passes from air ($n=1$) to a tissue of index $1.376$. With a radius of
curvature of $7.8$~mm that surface alone has a power $(n-1)/R\simeq48$~dioptres. The crystalline
lens, immersed in liquids of index about $1.34$, adds only some 20 dioptres, but it is the part that
can change: by changing its shape it focuses near or far objects (accommodation). Under water the
cornea loses almost all its power, because the step of index from water to cornea is tiny, and the
world becomes blurred (Fig.~\ref{fig:eye}, left); a mask, by putting air back in front of the cornea,
gives back a sharp vision. Fishes, which have no use for the cornea, have a spherical lens of high
index.

How fine can the eye see? The pupil, from 2 to 8~mm across, is the aperture, and diffraction alone
would allow an angle $1.22\lambda/D$, about $0.8'$ (minutes of arc) for a 3~mm pupil. With a larger
pupil the diffraction limit improves, but the aberrations of cornea and lens grow, and the best
optical image is formed with a pupil of 2--3~mm. The retina samples this image with its cones,
which in the centre of the fovea are about $2.5~\mu$m apart. Two cones per period of the finest
pattern means an angle of about $1'$---remarkably close to the optical limit, and indeed to the
measured acuity of a normal eye, about $1'$ (Fig.~\ref{fig:eye}, right). Optics and retina are matched:
finer cones would see nothing more, coarser ones would waste the optics. The same three numbers---
aperture, focal length and pixel size---describe the camera of a phone.
\par}

\subsection{\texorpdfstring{\rev{The most sensitive optical instrument}}{The most sensitive optical instrument}}

\begin{figure}[!htbp]
\centering
\includegraphics[width=\textwidth]{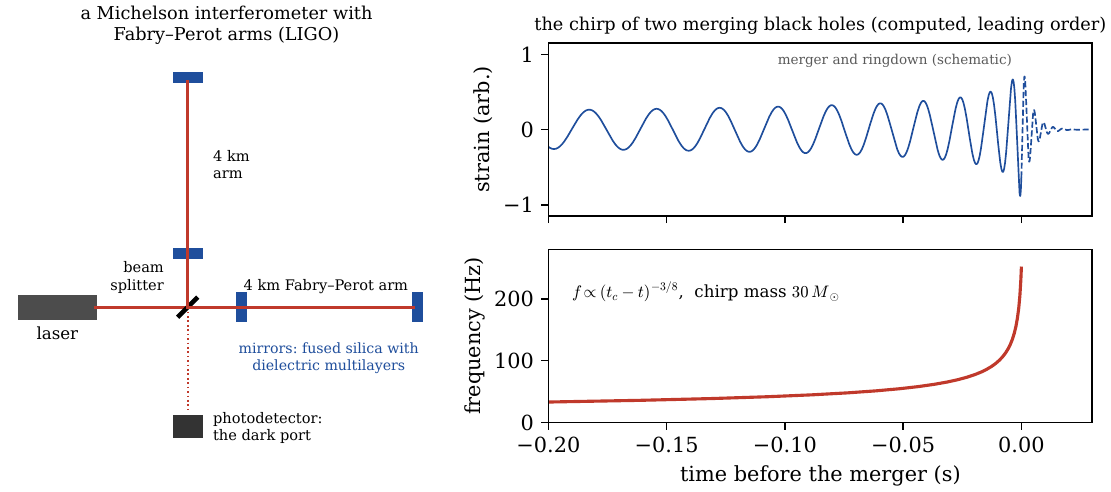}
\caption{\rev{Left: the scheme of LIGO, a Michelson interferometer whose arms are Fabry--Perot
cavities 4~km long; the photodetector watches the dark port. Right: the chirp of two merging black
holes, computed at the leading (Newtonian) order for a chirp mass of $30\,M_\odot$, as for
GW150914~\cite{LIGO2016}: wave form (merger and ringdown only sketched) and frequency, which grows as
$(t_c-t)^{-3/8}$ from about 35~Hz to some 250~Hz in the last fifth of a second.}}
\label{fig:ligo}
\end{figure}

{\revon
On 14 September 2015 the two detectors of LIGO, in Hanford and Livingston, recorded the gravitational
wave emitted by two black holes, of about 36 and 29 solar masses, spiralling into each other 1.3
billion light years away~\cite{LIGO2016}. The signal is a \emph{chirp} (Fig.~\ref{fig:ligo}): an
oscillation whose frequency and amplitude grow, from 35 to about 250~Hz in a fifth of a second, until the
two holes merge.

The instrument that heard it is a Michelson interferometer, and it uses nearly everything in this text.
A laser beam is split in two; the two halves travel along perpendicular arms 4~km long, are reflected
back by mirrors, and are recombined. The output port is kept \emph{dark}: there the two beams arrive
in opposition of phase, and their arrows cancel, as at the point Q of Fig.~\ref{fig:lens}. A
gravitational wave stretches one arm and squeezes the other, the arrows no longer cancel, and light
appears at the dark port. Each arm is a Fabry--Perot cavity, in which the light bounces some three
hundred times, so that the sum of the arrows of the multiple reflections (Sec.~\ref{sec:reflection})
multiplies the effect of the change of length. The mirrors are 40~kg cylinders of fused silica coated
with dielectric multilayers---the mirrors made of glass of Sec.~\ref{sec:reflection}---reflecting all
but a few parts per million of the light. The change of length to be detected is about $10^{-18}$~m, a
thousandth of the radius of a proton; compare it with the amplitude, $4\times10^{-19}$~m, of an electron
shaken by sunlight (Sec.~\ref{sec:one-atom}). At this level even the quantum nature of light matters:
the arrival of photons one by one sets a noise floor, which LIGO now lowers by injecting ``squeezed''
light into the dark port.
\par}

%% file: ch7_pulse.tex
\section{Time: a pulse through a slab}
\label{sec:pulse}

Everything so far has been a steady state: a wave that has always been there, oscillating
forever. Real light comes in pulses, and the most interesting questions are questions about time.
When light enters glass, how does the medium ``learn'' that it must slow it down? What happens to
the incident wave, which was cancelled by the extinction theorem, during the first instants?
Where is the energy while the pulse is inside? With a computer we can follow the field step by
step, as it builds up. This chapter uses the calculations of Refs.~\cite{RealiI,RealiII}: a pulse
only 2.2 optical cycles long (5.9~fs at 800~nm) crosses a slab of Lorentz oscillators with
resonance at 120~nm, whose index at the carrier is $n=1.45$ and group index $n_g=1.47$.

\subsection{\texorpdfstring{\rev{Local and retarded: two ways to compute the field}}{Local and retarded: two ways to compute the field}}

The field can be computed in two ways that look entirely different. The \emph{local} way solves
Maxwell's equations in the form of a wave equation, coupled at each point to the equation of motion
of the oscillators there; on a grid, step by step in time (the finite-difference time-domain
method). Nothing in it speaks of retardation: the finite speed of light is hidden in the
structure of the differential equations. The \emph{non-local} way is the one of this text: the field
at each point and instant is the incident wave plus the retarded waves of all the sheets of the
slab, each emitted at the earlier time $t-|z-z'|/c$---a self-consistent integral equation, the
time-dependent version of Eq.~\eqref{eq:selfconsistent}. Paper~I~\cite{RealiI} proves that the two
formulations are mathematically equivalent; Paper~II~\cite{RealiII} solves both numerically and
finds the same field, within the numerical accuracy, and the same reflected and transmitted
energies, which agree in turn with the exact spectral (Airy) result. The figures of this chapter
come from the local solution; the retarded one would be indistinguishable.

\subsection{The pulse in the slab}

\begin{figure}[!htbp]
\centering
\includegraphics[width=0.62\textwidth]{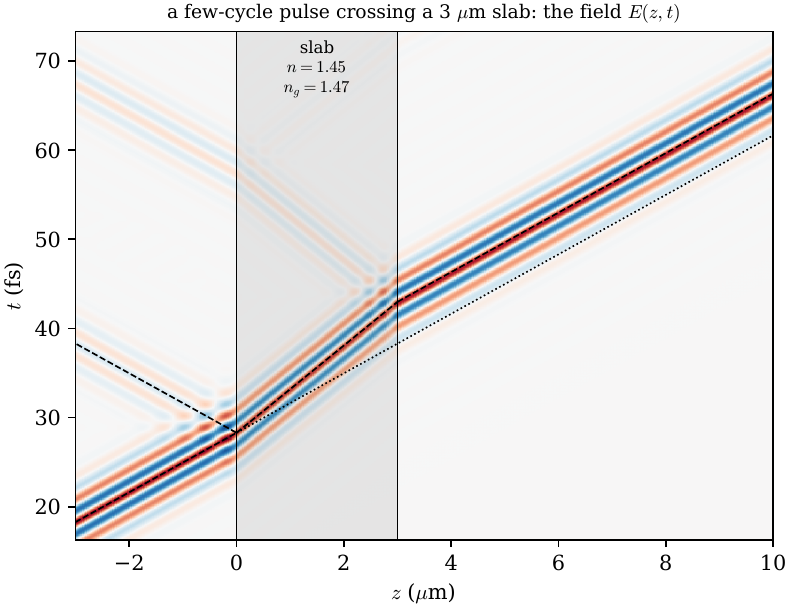}
\caption{Space--time map of the field of a few-cycle pulse crossing a slab 3~$\mu$m thick
(between the vertical lines). Dashed lines: centroids of the incident, reflected, internal and
transmitted pulses; dotted: straight continuation of the incident pulse at speed $c$. The delay of
the transmitted pulse is $(n_g-1)L/c\simeq4.7$~fs.}
\label{fig:spacetime}
\end{figure}

\begin{figure}[!htbp]
\centering
\includegraphics[width=0.66\textwidth]{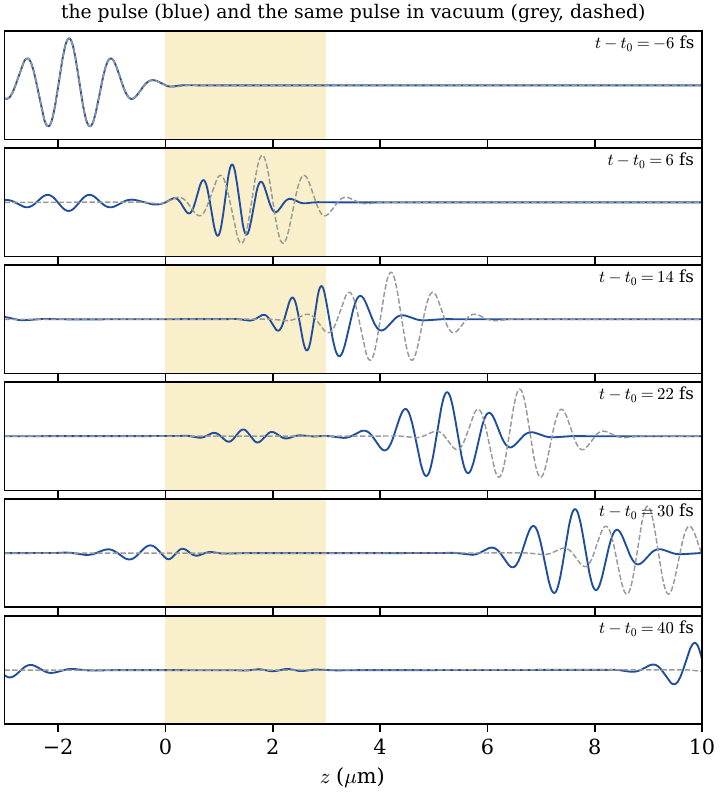}
\caption{Six snapshots of the pulse of Fig.~\ref{fig:spacetime} (blue), with the same pulse
travelling in vacuum (grey, dashed). Inside the slab (shaded) the pulse falls behind and its
wavelength shortens; a weak reflected pulse travels back to the left
(\anim{\rev{8.1}}{anim7_pulse.mp4}).}
\label{fig:snapshots}
\end{figure}

Figure~\ref{fig:spacetime} shows the field $E(z,t)$ as a map in space and time: each horizontal
line is a snapshot, each vertical line the history of one point. The incident pulse arrives from
the left along a line \rev{travelled at speed $c$}. At the front surface a weak reflected pulse goes back, and the
main pulse enters the slab, where it travels more slowly, along a steeper line\rev{, at the speed $c/n_g$}. It
leaves the slab from the back surface (with a second, weaker reflection) and resumes the speed of
light, parallel to---but later than---the straight continuation of the incident pulse (dotted).
Figure~\ref{fig:snapshots} shows six snapshots of the same process, with a pulse travelling in
vacuum for comparison, and \anim{\rev{8.1}}{anim7_pulse.mp4} shows the whole of it.

\subsection{\texorpdfstring{\rev{The ghost and the real pulse}}{The ghost and the real pulse}}
\label{sec:ghost}

{\revon
The extinction theorem of Sec.~\ref{sec:extinction} becomes strikingly visible when a short pulse
of finite width meets the slab at an angle (Fig.~\ref{fig:oblique}, \anim{\rev{8.2}}{anim7_oblique.mp4})\rev{; here
the slab has $n=1.5$ and a larger dispersion, $n_g=1.72$, so that the real pulse falls clearly behind
the ghost}.
The figure shows, at six instants, three maps of the field. On the left is the incident pulse
$E_0$ as it would travel if the slab were not there: straight on, at speed $c$, through the glass as
through vacuum. In the middle is the field $E_s$ radiated by the dipoles of the slab, and on the
right the total field $E=E_0+E_s$, the only one that can actually be observed.

The middle column tells the whole story. As soon as the pulse enters the slab, the dipoles radiate,
inside it, a pulse exactly opposite to the incident one, travelling with it at speed $c$ along its
straight path: it is the extinguishing wave of Sec.~\ref{sec:extinction}, now a pulse. Added to
$E_0$, it cancels it. The incident pulse travels on, undisturbed, as a \emph{ghost}: it is present
in each of the two pieces of the field, but absent from their sum. At the same time the dipoles
radiate a second pulse, the real one, which travels more slowly, at the group velocity $c/n_g$, in
the direction given by Snell's law. When the ghost reaches the back surface the dipoles stop
cancelling it: beyond the slab $E_s$ no longer contains $-E_0$, and the ghost would reappear,
were it not that the field of the dipoles beyond the slab is exactly what is needed to replace it
by the real pulse, emerging later and displaced sideways, parallel to the incident direction.
Weak reflected pulses leave the front and back surfaces. In the total field (right) we see only
what an observer would see: a pulse that bends, slows down, and comes out late.

\begin{figure}[!htbp]
\centering
\includegraphics[width=0.6\textwidth]{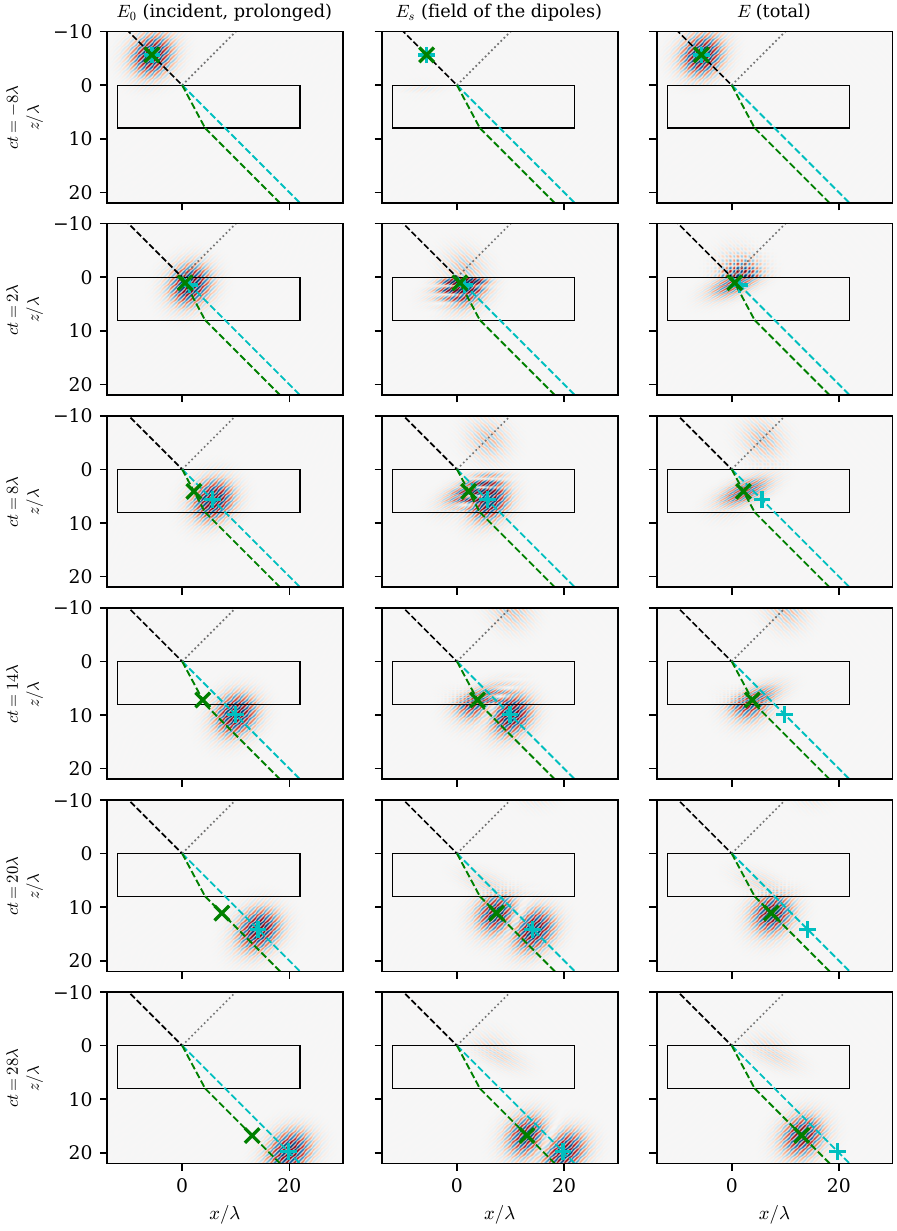}
\caption{\rev{A short pulsed beam (field FWHM 6 cycles, waist $4\lambda$, s polarization) meets at
$45^\circ$ a slab $8\lambda$ thick of Lorentz oscillators ($n=1.5$, $n_g=1.72$ at the carrier),
computed exactly as a superposition of plane waves. Rows: six instants. Columns: the incident pulse
prolonged as if the slab were absent ($E_0$); the field radiated by the dipoles ($E_s$); the total
field ($E$). Dashed lines: centroids of the incident pulse (black), of the ``ghost'' continuing at
speed $c$ (cyan, marker $+$) and of the real pulse, refracted at the group velocity and then
transmitted (green, marker $\times$); dotted: reflected pulse. Inside the slab the dipoles radiate
$-E_0$ along the ghost path, cancelling the incident pulse, and the real pulse along the Snell path
(\anim{\rev{8.2}}{anim7_oblique.mp4}).}}
\label{fig:oblique}
\end{figure}
}

\subsection{The front travels at the speed of light}

\begin{figure}[!htbp]
\centering
\includegraphics[width=0.72\textwidth]{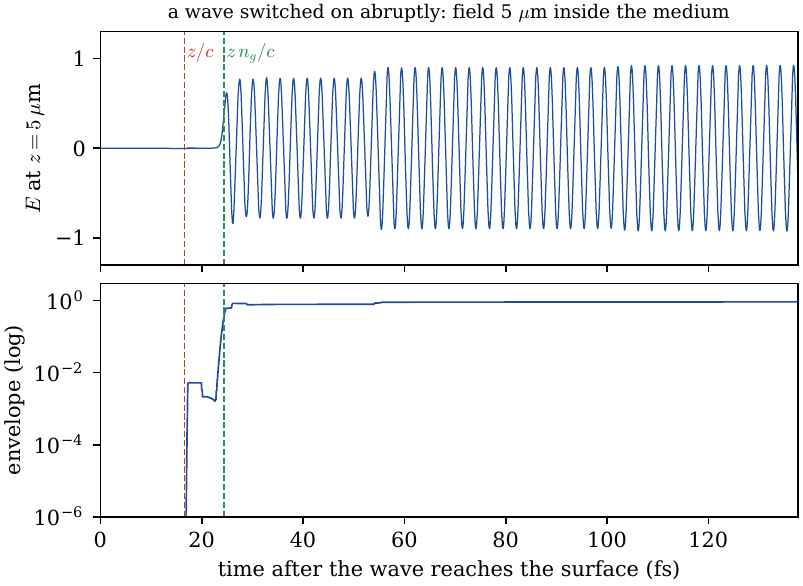}
\caption{A carrier wave at 800~nm switched on abruptly enters the medium; the field 5~$\mu$m
inside, as a function of time. Top: the field. Bottom: its envelope on a logarithmic scale.
Nothing arrives before $z/c$ (red line); then a weak forerunner, and the main signal at $zn_g/c$
(green line). The steady amplitude is the transmission coefficient $2/(n+1)$.}
\label{fig:front}
\end{figure}

What happens at the very beginning? Suppose that a wave is switched on abruptly and enters the
medium. Inside, at a depth $z$, nothing at all can happen before the time $z/c$: the oscillators
there can be set in motion only by fields coming from the source, or from other oscillators, and
all these fields travel at speed $c$. At $t=z/c$ the \emph{front} arrives, and the field starts---but
very weakly (Fig.~\ref{fig:front}). The first oscillators to be reached see a field that has just
begun, too fast for them to follow: like electrons above all resonances, they hardly respond, and
the medium is at first almost transparent, with index close to one. This weak forerunner is
Sommerfeld's precursor~\cite{Brillouin}. Only gradually do the oscillators build up their
steady oscillation, and with them the waves that extinguish the incident one and replace it with
the refracted wave. The bulk of the signal arrives at the time $zn_g/c$, carried at the group
velocity.

This is the time-dependent form of the extinction theorem of Sec.~\ref{sec:extinction}: the
cancellation of the incident wave is not instantaneous, but takes the time needed for the oscillators
to respond---a few periods of the light, here a few femtoseconds. And it is the physical content of
the causality argument of Sec.~\ref{sec:KK}: the front of any signal travels at $c$, whatever the
index, and from this alone follow the Kramers--Kronig relations between the real and imaginary
parts of $n$.

\subsection{Where the energy is}

\begin{figure}[!htbp]
\centering
\includegraphics[width=0.9\textwidth]{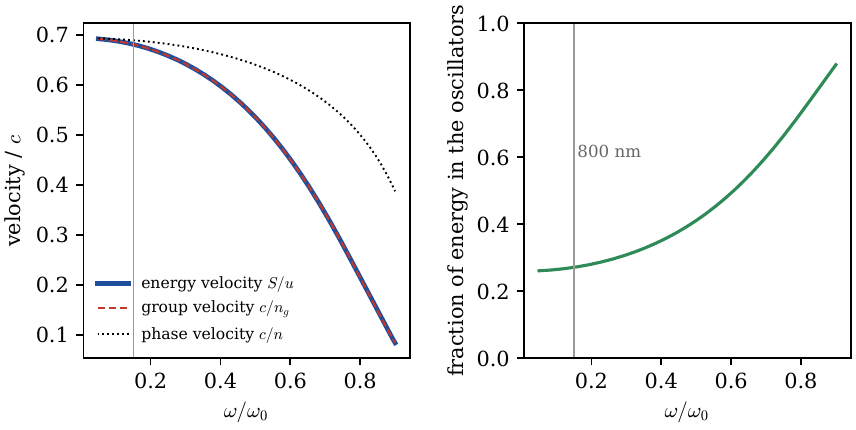}
\caption{Left: in a transparent Lorentz medium the energy velocity $\langle S\rangle/\langle
u\rangle$ (thick) coincides with the group velocity $c/n_g$ (dashed), and is lower than the phase
velocity $c/n$ (dotted). Right: fraction of the energy of the wave stored in the oscillators. The
vertical line marks the carrier of Figs.~\ref{fig:spacetime}--\ref{fig:front} (800~nm).}
\label{fig:energy}
\end{figure}

While the pulse is inside the slab, part of its energy is not in the electromagnetic field: it is in
the motion of the electrons, as kinetic energy of their oscillation and potential energy of their
springs. The energy balance of the coupled field--oscillator system (Paper~I, Appendix~A) reads
\begin{equation}
\frac{\partial}{\partial t}\big(u_{EM}+u_{\rm osc}\big)+\frac{\partial S}{\partial z}=
-\,\text{(power dissipated by the damping)},
\end{equation}
where $u_{EM}$ is the energy density of the field, $u_{\rm osc}$ that of the oscillators and $S$
the Poynting flux. In a transparent medium the energy travels at the speed
\begin{equation}
v_E=\frac{\langle S\rangle}{\langle u_{EM}+u_{\rm osc}\rangle},
\end{equation}
and a short calculation shows that this is exactly the group velocity $c/n_g$
(Fig.~\ref{fig:energy}). At 800~nm, in our medium, more than a quarter of the energy of the pulse
travels as motion of the electrons; closer to the resonance the fraction grows, and the pulse
slows down further. This is the correct core of the ``delayed re-emission'' picture
(Sec.~\ref{sec:myth}): the pulse is slowed because energy is stored, reversibly, in the matter.
But the storage is collective---shared by all the oscillators the pulse covers at a given
moment---and it concerns the group velocity of a pulse, not the phase velocity of a wave. Pushed to
the extreme, near a sharp resonance prepared by a second laser beam, the same mechanism slows light
to a few metres per second: the ``slow light'' of electromagnetically induced transparency\rev{~\cite{Hau1999}}.

\needspace{8\baselineskip}
\subsection{\texorpdfstring{\rev{Faster than light in matter: \v{C}erenkov radiation}}{Faster than light in matter: Cerenkov radiation}}
\label{sec:cerenkov}

\begin{figure}[!htbp]
\centering
\includegraphics[width=0.92\textwidth]{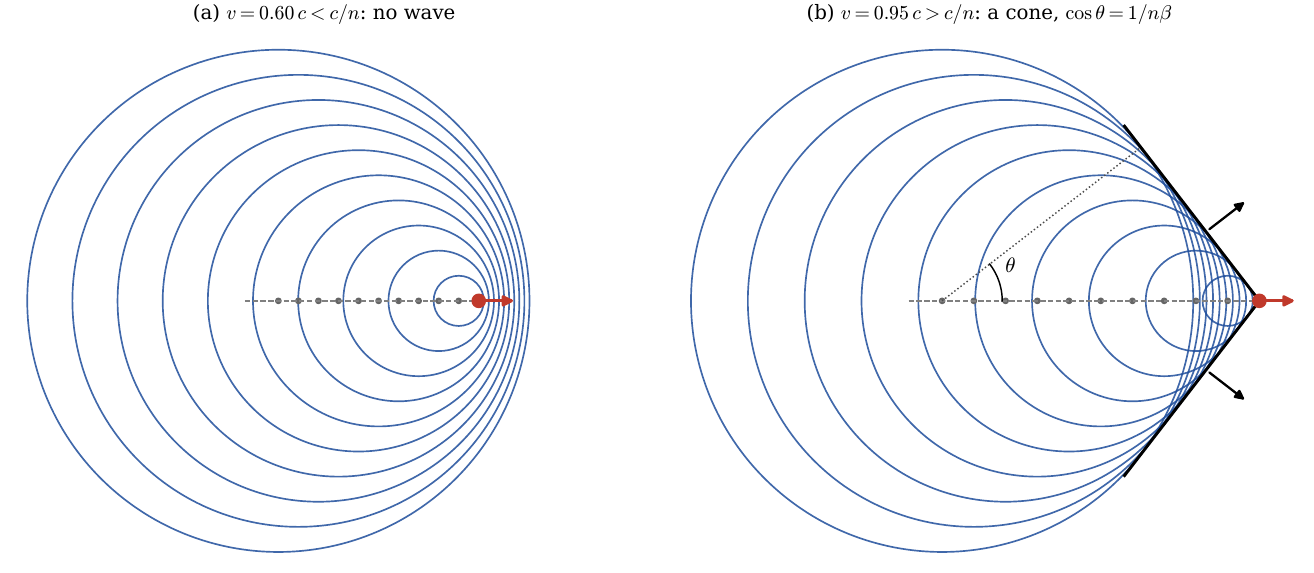}
\caption{\rev{Huygens construction for a charge (red) crossing water, $n=1.33$. The wavelets are
emitted by the atoms polarized as the charge passes (grey dots, at equal time intervals) and travel
at $c/n$. (a) A charge slower than $c/n$: the wavelets are nested, and there is no radiation.
(b) A charge faster than $c/n$: the wavelets have a common envelope, a cone; the light travels at the
angle $\theta$ of Eq.~\eqref{eq:cerenkov} to the path ($37.7^\circ$ here).}}
\label{fig:cerenkov}
\end{figure}

{\revon
Nothing can travel faster than $c$, but a particle can travel faster than light \emph{in a
medium}, faster than $c/n$. What happens then is a last, striking demonstration that the light in a
medium is made by the atoms.

A charge moving uniformly in vacuum does not radiate: its field travels along with it. In a
medium, the charge polarizes the atoms it passes: each of them becomes, for an instant, a dipole
driven by the passing field, and sends out a wavelet. These wavelets, combined with those of all
the other atoms as in Secs.~\ref{sec:index} and~\ref{sec:extinction}, travel at the speed $c/n$. If
the charge is slower than $c/n$, the wavelets left behind it are nested inside one another
(Fig.~\ref{fig:cerenkov}a); their contributions cancel on average, and the polarization simply
accompanies the charge without radiating. If the charge is faster than $c/n$, it outruns its own
wavelets, and they all have a common envelope: a cone, along which they add up in phase, as the
waves left by a boat add up in its bow wave and the sound of a supersonic aircraft in its shock
wave (Fig.~\ref{fig:cerenkov}b). The light leaves at the angle
\begin{equation}
\cos\theta=\frac{c/n}{v}=\frac{1}{n\beta},\qquad \beta=\frac vc ,
\label{eq:cerenkov}
\end{equation}
with respect to the path. Feynman uses exactly this construction for bow waves and shock waves in
Chap.~51 of the first volume of the \emph{Lectures}~\cite{FeynmanI}, where he also mentions \v{C}erenkov radiation.

The glow was observed by \v{C}erenkov in 1934, as a faint blue light from liquids exposed to the
gamma rays of radium, which set fast electrons in motion; Frank and Tamm explained it in 1937, and
the three shared the Nobel prize of 1958~\cite{Jelley}. Their theory gives the number of photons
emitted per unit path length and per unit wavelength,
\begin{equation*}
\frac{\ud^2N}{\ud x\,\ud\lambda}=\frac{2\pi\alpha_f}{\lambda^2}\Big(1-\frac1{n^2\beta^2}\Big),
\end{equation*}
where $\alpha_f\simeq1/137$ is the fine-structure constant. The $1/\lambda^2$ explains the colour:
\v{C}erenkov light is blue, and the pools of water around the cores of nuclear reactors glow with
it. In water ($n=1.33$) electrons must have a kinetic energy of more than 0.26~MeV to emit; a fast
particle emits about two hundred visible photons per centimetre, at angles up to $41^\circ$. Equation
\eqref{eq:cerenkov} also says where the emission ends: only at frequencies where $n>1/\beta$, so never
in the X-ray region, where $n<1$ (Sec.~\ref{sec:index}). And it turns the angle into a measure of the
speed. The Super-Kamiokande detector in Japan, a tank of 50\,000 tonnes of pure water lined with
photomultipliers, detects neutrinos through the rings of \v{C}erenkov light of the charged particles
they produce~\cite{SuperK}.
\par}

%% file: ch8_nonlinear.tex
\section{Beyond the linear oscillator}
\label{sec:beyond}

Two simple modifications of the electron on a spring open two new worlds: give the oscillator a
negative strength, and matter amplifies light instead of absorbing it---the principle of the laser;
make the spring slightly imperfect, and matter changes the colour of light---nonlinear optics.

\subsection{Absorption and amplification}

\begin{figure}[!htbp]
\centering
\includegraphics[width=0.95\textwidth]{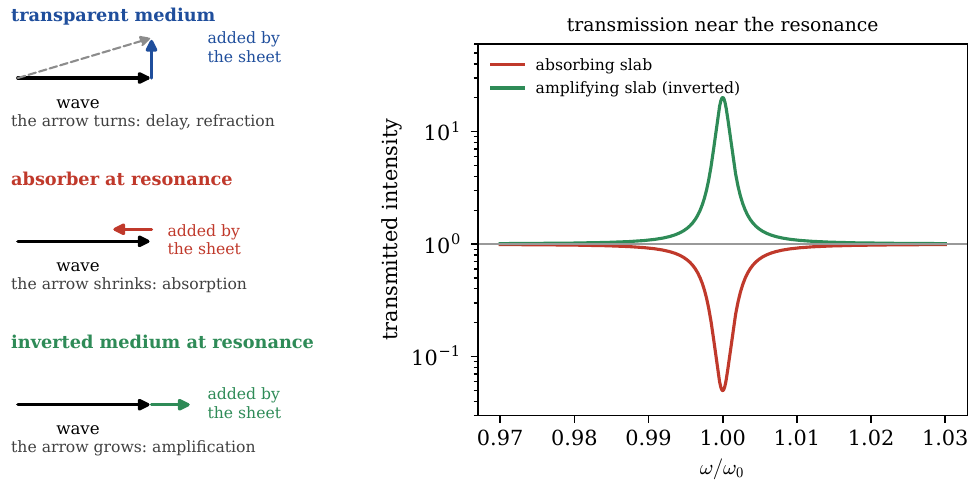}
\caption{\rev{Left: the small wave added by a sheet (coloured) to the wave crossing it (black): at right
angles in a transparent medium (the arrow turns: delay, refraction), opposite in an absorber at
resonance (the arrow shrinks), in phase in an inverted medium at resonance (the arrow grows). Right:
transmitted intensity of an absorbing and of an amplifying slab near the resonance.}}
\label{fig:gain}
\end{figure}

Everything is in the phase of the small wave that each sheet adds to the light,
Eq.~\eqref{eq:sheet}, i.e.\ in the phase of the dipoles relative to the field
(Fig.~\ref{fig:gain}). Far from resonance the dipoles follow the field (or oppose it), and the
added wave is a quarter of a period behind: it turns the arrow of the light without changing its
length---refraction, Sec.~\ref{sec:index}. At resonance the dipoles lag the field by a quarter of
a period, and the added wave is in \emph{opposition} with the light: the arrow shortens, sheet after
sheet, and the light is absorbed, with the exponential decay of the Beer--Lambert law. Energy goes
from the field to the oscillators, at the rate of Eq.~\eqref{eq:absorbed}.

Now let the atoms be prepared in an excited state, with more atoms in the upper level of a
transition than in the lower one---a population inversion, obtained by pumping the medium with
another source of energy. As noted at the end of Sec.~\ref{sec:one-atom}, such atoms respond like
oscillators with a negative strength: their dipoles run a quarter of a period \emph{ahead} of the
field, and the wave added by each sheet is \emph{in phase} with the light. The arrow lengthens:
the light is amplified. Energy flows from the atoms to the field: this is stimulated emission.

The same coherence that makes refraction possible gives stimulated emission its remarkable
properties. The waves added by the excited atoms are emitted in the forward direction, at the same
frequency and with the same phase as the light that stimulates them: forward scattering is always
coherent (Sec.~\ref{sec:order}). An amplifier therefore reproduces faithfully, only stronger, the
wave that enters it. \rev{Put it between two mirrors, and it becomes a source: the laser.}

\begin{figure}[!htbp]
\centering
\includegraphics[width=\textwidth]{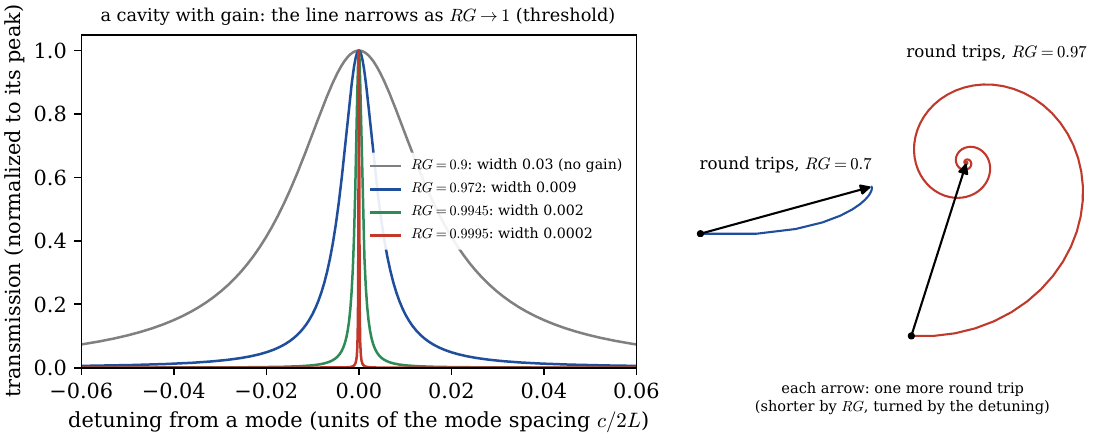}
\caption{\rev{A cavity with gain (mirror reflectance $R=0.9$). Left: transmission near a resonance, normalized
to its peak, as the round-trip factor $RG$ approaches one: the line narrows by orders of magnitude, while
the peak transmission grows from 1 to $3.7\times10^4$. Right: the arrows of the successive round trips, a
little off resonance; with $RG=0.7$ the spiral converges at once, with $RG=0.97$ it winds many times.}}
\label{fig:threshold}
\end{figure}

{\revon
\subsection{\texorpdfstring{\rev{The laser: light narrow in frequency, or narrow in time}}{The laser: light narrow in frequency, or narrow in time}}

\emph{From amplifier to oscillator.} Place the amplifying medium between two mirrors of reflectance $R$,
as in the plate of Sec.~\ref{sec:reflection}. Light that enters is reflected back and forth, and at the
output the arrows of all the round trips are added. In a passive plate each round trip makes the arrow
shorter, by the factor $R$, and turns it by the round-trip phase: the arrows form a converging spiral,
and the sum is the Airy formula. With gain, each pass also lengthens the arrow by the single-pass gain
$G$, and the ratio of the geometric series becomes $RG$. As $RG$ approaches one the spiral converges
more and more slowly, the transmission at resonance grows without limit, and the resonance becomes
narrower and narrower (Fig.~\ref{fig:threshold}). At $RG=1$, the \emph{threshold}, the series diverges:
no input is needed any more, and the light grows from the photons of spontaneous emission---the one
quantum ingredient that our classical oscillators cannot supply (Sec.~\ref{sec:one-atom}). The growth
stops when the strong field depletes the inversion and the gain saturates at the value $RG=1$ exactly:
the gain just compensates the losses.
\par}

\begin{figure}[!htbp]
\centering
\includegraphics[width=\textwidth]{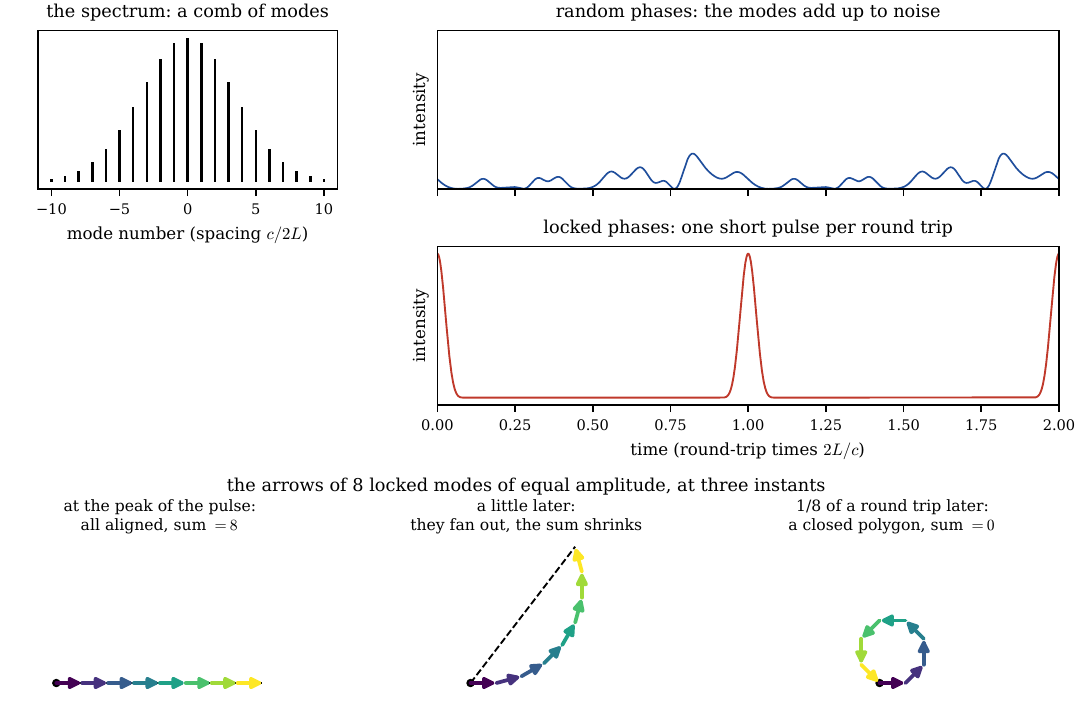}
\caption{\rev{Top: twenty-one modes of a laser cavity (the spectrum, left). With random phases their sum
fluctuates like noise; with locked phases it is one short pulse per round trip. Bottom: the arrows of eight
locked modes of equal amplitude (one colour per mode), each turning at its own frequency. At the peak of the
pulse they are all aligned and their sum (dashed) is eight times one arrow; a little later they fan out and
the sum shrinks; after one eighth of a round trip they close a polygon and the sum is zero: between two
pulses the laser is dark.}}
\label{fig:modelock}
\end{figure}

{\revon
\emph{Narrow in frequency.} The light that builds up is the one that comes back in step with itself
after a round trip: a mode of the cavity, at a frequency for which the round trip contains a whole number
of wavelengths. Above threshold its line is far narrower than the resonance of the atoms and than the
passive resonance of the cavity; Schawlow and Townes, in the paper that proposed the laser, estimated its
ultimate width~\cite{SchawlowTownes}. This purity is what LIGO relies upon (Sec.~\ref{sec:instruments}):
its frequency-stabilized neodymium laser, at 1064~nm, keeps the same phase over the long, folded paths of
the arms, so that a change of their difference by $10^{-18}$~m can be read from the interference. The
same principle, with the frequency locked to an atomic transition, makes the optical clocks, the most
precise instruments ever built.

\emph{Narrow in time.} A long cavity has many modes, separated by $\Delta\nu=c/2L$, and an amplifier with a
broad gain band can make many of them oscillate at once. Their sum depends on their phases
(Fig.~\ref{fig:modelock}). With random phases the arrows of the modes add up, at each instant, to a
random walk: the output fluctuates like noise. If the phases are \emph{locked}---by a modulator, or by a
nonlinear element that favours intense light, such as the Kerr lens described at the end of this section---the arrows
of all the $N$ modes point the same way once per round trip, and cancel in between: the laser emits a
train of pulses, each lasting about $1/(N\Delta\nu)$, the inverse of the whole bandwidth. Titanium-sapphire
lasers produce pulses of a few femtoseconds~\cite{SpenceKeanSibbett}, a few cycles of light. With them one
can watch a chemical bond break (Zewail's ``femtochemistry'', Nobel prize 1999~\cite{Zewail}) and follow,
step by step, the energy that a molecule of chlorophyll has absorbed from sunlight as it hops between the
pigments of the antenna towards the reaction centre, in times from a few hundred femtoseconds to a few
picoseconds (Sec.~\ref{sec:nature}).

\emph{Never both.} A line of width $\Delta\nu$ and a pulse of duration $\Delta t$ are Fourier partners,
$\Delta\nu\,\Delta t\gtrsim1$: a laser can be very narrow in frequency, or very short in time, but not both
at once. It is the same arithmetic of arrows that limits the focus of a lens (Sec.~\ref{sec:instruments}):
to make something narrow, one needs many components spread over a wide range.
\par}

\subsection{Red in, green out?}

Shine the infrared beam of a neodymium laser (1064~nm) into a small crystal of potassium
dihydrogen phosphate, and a green beam (532~nm) comes out: light of exactly twice the frequency,
which was not there before. It looks like magic; it is a spring that is not perfectly elastic.

\begin{figure}[!htbp]
\centering
\includegraphics[width=0.95\textwidth]{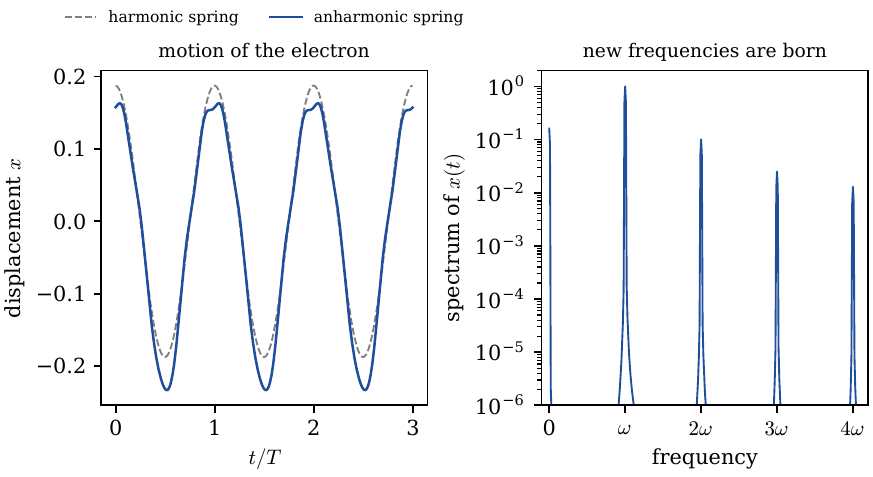}
\caption{The anharmonic oscillator of Eq.~\eqref{eq:anharmonic}, driven below resonance. Left:
the motion (solid) is distorted and shifted with respect to the harmonic one (dashed). Right: its
spectrum contains, besides the driving frequency $\omega$, a static part and the harmonics
$2\omega$, $3\omega$, \dots}
\label{fig:anharmonic}
\end{figure}

The restoring force on a real electron is linear only for small displacements. In a crystal
without a centre of symmetry, the next term of the force is quadratic, and the equation of motion
becomes
\begin{equation}
\ddot x+2\gamma\dot x+\omega_0^2x+\beta x^2=-\frac em\,E_0\cos\omega t .
\label{eq:anharmonic}
\end{equation}
For a weak field the motion is almost the linear one, $x_1=a\cos(\omega t-\varphi)$; but the term
$\beta x_1^2$ acts as an additional driving force, and since
$\cos^2\omega t=\tfrac12(1+\cos2\omega t)$, it contains a constant part and a part oscillating at
\emph{twice} the frequency. The electron acquires a small static displacement (optical
rectification) and a small oscillation at $2\omega$, and the dipole radiates at $2\omega$
(Fig.~\ref{fig:anharmonic}). At higher order, all the harmonics appear.

\begin{figure}[!htbp]
\centering
\includegraphics[width=0.7\textwidth]{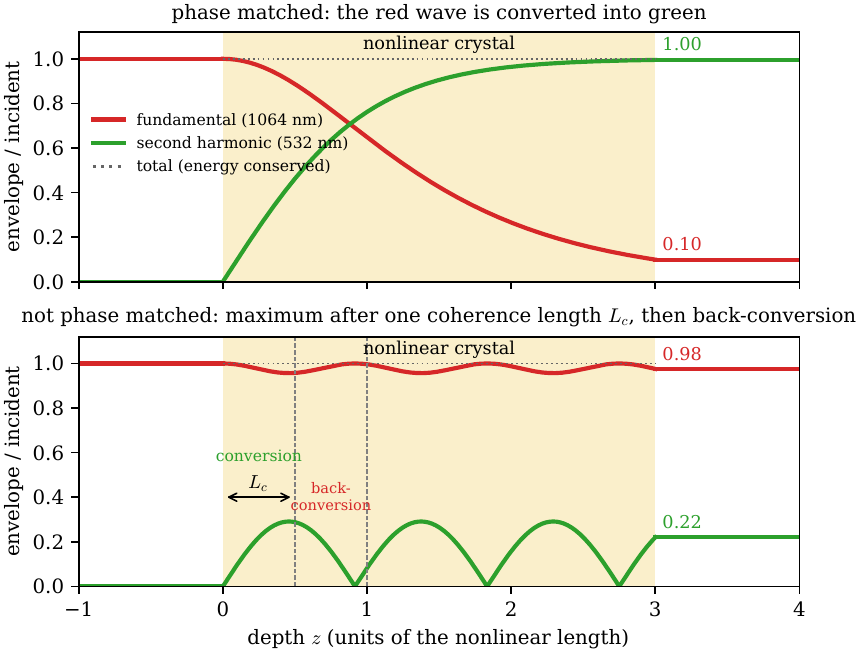}
\caption{\rev{Second-harmonic generation, 1064~nm $\to$ 532~nm: envelopes of the fundamental (red)
and of the second harmonic (green) through a nonlinear crystal (shaded), from the coupled-wave
equations including the depletion of the fundamental; depth in units of the nonlinear length
$L_{NL}$, over which conversion becomes substantial. Top: phase matched; the red wave is
progressively converted into green, and the total energy (dotted) is conserved. Bottom: not phase
matched ($L_c=L_{NL}/2$): the green wave reaches its maximum after one coherence length $L_c$ and is
converted back into red over the next one, and so on.}}
\label{fig:shg}
\end{figure}

Every atom of the crystal now radiates a little green light. Will the green wavelets add up? We
know the rule: they must add in phase in the forward direction. The atom at depth $z$ is driven by
the infrared wave, with phase $2k(\omega)z$ for its second harmonic, and its green wavelet travels
on with the wave number $k(2\omega)$. Because of dispersion, $n(2\omega)\neq n(\omega)$, and the
wavelets of successive sheets slip in phase by $\Delta k=2k(\omega)-k(2\omega)$ per unit length.
Their arrows run round a circle, exactly as in Fig.~\ref{fig:phasorsEO}: the green light grows over
a \emph{coherence length} $L_c=\pi/|\Delta k|$, a few tens of micrometres, and then decreases,
oscillating with the length of the crystal (Fig.~\ref{fig:shg}). \rev{Within one coherence length
the conversion reaches its maximum; over the next one the phase slip has reversed the sign of the
interaction, and the green light is converted back into red. Energy keeps flowing to and fro
between the two waves, and the crystal never delivers more green light than it does after a
single coherence length.} Phase matching---choosing the
direction of propagation in a birefringent crystal so that $n(2\omega)=n(\omega)$---straightens the
chain of arrows: the green light then grows steadily, \rev{at first} as the square of the length,
\rev{while the red wave is depleted, until almost all of it has been converted (Fig.~\ref{fig:shg},
top)}, and conversion efficiencies of tens of per cent become possible. The green laser pointer is an infrared laser
\rev{with such a crystal inside its cavity (panel ``Inside a green laser pointer'').} Historically, the first observation of second-harmonic generation, by
Franken and coworkers in 1961, used the red light of a ruby laser (694~nm), converted into the
ultraviolet (347~nm)~\cite{Franken}.

\begin{figure}[!tbp]
\colorbox{shadecolor}{\begin{minipage}{\dimexpr\linewidth-2\fboxsep\relax}
\small\noindent\textbf{\rev{Inside a green laser pointer}}\par\smallskip\noindent\ignorespaces{\revon
The green laser pointer, in everyday use in lecture halls and by amateur astronomers to point at
stars, gathers in a few centimetres the two ideas of this section: a laser, that is,
an amplifier between two mirrors, and second-harmonic generation (Fig.~\ref{fig:glp}). A laser diode
emits 808~nm light, which a lens focuses into a crystal of neodymium-doped yttrium vanadate,
Nd:YVO$_4$. The neodymium ions absorb it and amplify light at 1064~nm, in a cavity formed by two
coatings on the outer faces of a small bonded stack of crystals, the multiple crystal assembly
(MCA): both coatings reflect 1064~nm almost completely, so that the infrared light builds up
between them. The doubling crystal, potassium titanyl phosphate (KTP), sits inside the same
cavity, where the infrared intensity is highest: the green power grows as the square of that
intensity. The output coating lets the 532~nm light through, and two lenses expand and
collimate it into the beam we see.

\medskip
{\centering
\includegraphics[width=0.94\linewidth]{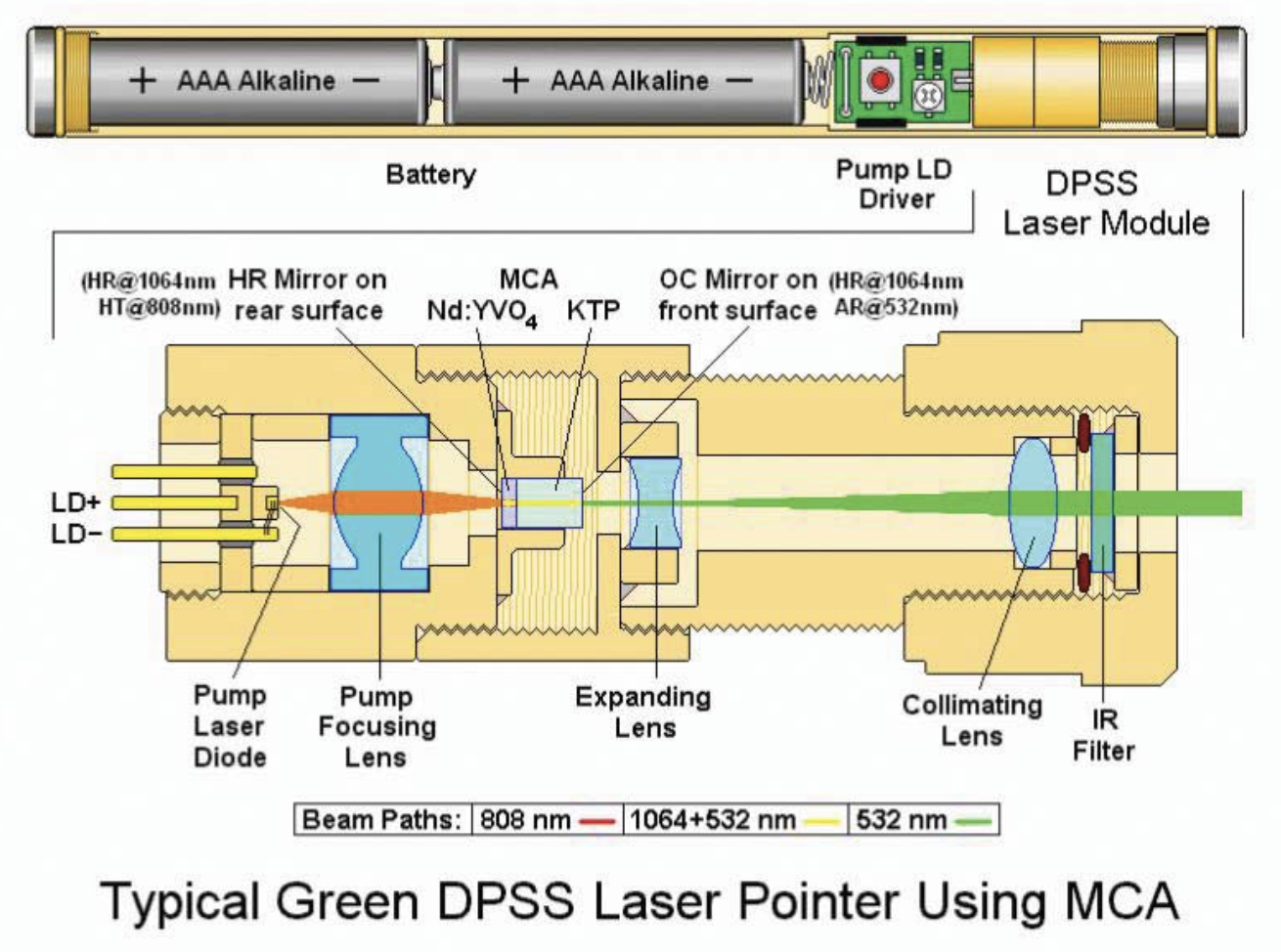}
\caption{Schematic of a green laser pointer (GLP) based on a multiple crystal assembly
(MCA). The familiar package holds two AAA batteries, a printed circuit board with the driver of
the pump laser diode (LD), and the diode-pumped solid-state (DPSS) laser module\rev{, which is shown expanded and exploded below}. The 808~nm pump
LD is focused into the Nd:YVO$_4$ crystal (violet section of the MCA), which emits 1064~nm light
into the KTP frequency-doubling crystal (light blue section); the cavity is closed by the HR coating
on the rear face and by the output coupler on the front face. The 532~nm light from the KTP crystal
is sent through an expanding and a collimating lens to produce a collimated output beam. An IR
filter stops the residual 808~nm and 1064~nm light; inexpensive pointers sometimes omit it, and
then emit invisible infrared light as well~\cite{Galang}. HR: high reflectivity; HT: high
transmissivity; AR: antireflective; OC: output coupler; IR: infrared. \rev{Figure credit: Samuel M.~Goldwasser~\cite{Goldwasser}.}}
\label{fig:glp}\par}
\par}
\end{minipage}}
\end{figure}

In materials with a centre of symmetry, such as glass, the quadratic term vanishes by symmetry,
and the first nonlinearity is cubic, $x^3$. It produces the third harmonic and, more importantly,
makes the refractive index depend on the intensity of the light itself, $n=n_0+n_2I$: the optical
Kerr effect. An intense pulse changes the index it travels in. It modulates its own phase, and its
spectrum broadens (self-phase modulation); an intense beam focuses itself (the Kerr lens). These
effects, balanced against dispersion and diffraction, are the basis of optical solitons in fibres
and of Kerr-lens mode locking, which produces the shortest laser pulses---a few cycles of light,
like the pulses of Sec.~\ref{sec:pulse}~\cite{SpenceKeanSibbett}.

%% file: ch9_nature.tex
\section{More natural phenomena: optics, vision and energy}
\label{sec:nature}

With the tools of the previous chapters we can now return to the spectacles of nature, the
starting point of the Prologue, and look at a few more of them. Each short section below names
the chapter that holds its key. Some of these phenomena are entirely explained by our classical
oscillators; others---vision and photosynthesis---lead us to the border where light has to be
treated as made of photons and the oscillators as quantum systems, and we shall say so.

\subsection{Optics of the sky}

\begin{figure}[!htbp]
\centering
\includegraphics[width=0.95\textwidth]{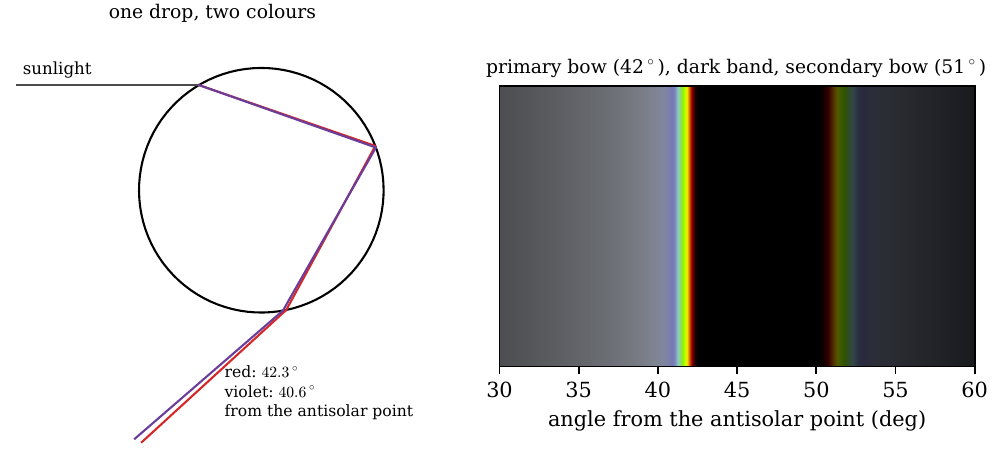}
\caption{Left: a ray of sunlight in a raindrop, refracted, reflected once and refracted again;
red and violet leave at slightly different angles. Right: the colours of the sky as a function of
the angle from the antisolar point, computed by tracing rays through spherical drops for all
visible wavelengths (with the Fresnel coefficients) and converting the spectra to colours: the
primary bow, the bright sky inside it, Alexander's dark band and the secondary bow.}
\label{fig:rainbow}
\end{figure}

\paragraph{The rainbow} (Secs.~3 and 6). Sunlight entering a raindrop is refracted, reflected once
inside and refracted again on the way out. Descartes, in 1637, traced rays through a drop by hand
and found that their deviation has a minimum: around $138^\circ$, i.e.\ $42^\circ$ from the
point opposite the sun, the rays bunch together, and that is where we see the bow
(Fig.~\ref{fig:rainbow}). Because of dispersion the angle is $42.3^\circ$ for red and
$40.6^\circ$ for violet: red outside, violet inside. Rays reflected twice inside the drop make the
fainter secondary bow at $51^\circ$, with the colours reversed; between the two, no ray can arrive,
and the sky is darker (Alexander's dark band). Two finer details need the wave picture. With small
drops, below half a millimetre or so, faint pastel arcs appear inside the primary bow: they are
interference fringes between rays that leave the drop in the same direction along different paths
(the supernumerary bows, explained by Young and by Airy). And the light of the rainbow is almost
completely polarized, tangentially to the bow: the internal reflection happens close to the
internal Brewster angle of water (Sec.~6), where the \emph{p} component is hardly reflected---as a
pair of polarizing sunglasses \rev{(panel in Sec.~\ref{sec:one-atom})} shows at once.

\paragraph{Halos.} Around the sun or the moon, seen through thin high clouds, a ring of
$22^\circ$ often appears: the minimum deviation of light crossing the $60^\circ$ prisms of
hexagonal ice crystals---the same physics as the rainbow, with ice prisms in place of drops.

\begin{figure}[!htbp]
\centering
\includegraphics[width=0.85\textwidth]{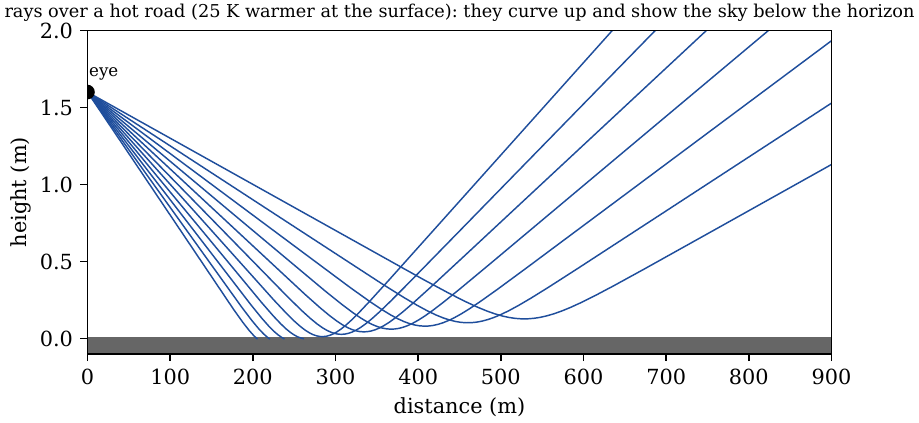}
\caption{Rays leaving the eye of an observer, traced through air whose temperature rises by 25~K
in the last few centimetres above a hot road. Steep rays reach the road; grazing rays curve back
up, so that the observer sees the sky in their direction.}
\label{fig:mirage}
\end{figure}

\paragraph{The mirage} (Sec.~3). Over a road heated by the sun, the air near the ground is
warmer and less dense, and its index is slightly lower: $n-1$ is proportional to the density, so
25~K make a difference of a few parts in $10^5$. Rays that skim the road at a grazing angle bend
upward, as in a very gentle total reflection, and the eye sees, below the horizon, an image of the
sky: the shimmering ``water'' on the road (Fig.~\ref{fig:mirage}). Over cold sea the gradient is
reversed, rays bend downward, and ships beyond the horizon appear suspended in the air, stretched
and multiplied: the Fata Morgana.

\paragraph{The green flash} (Sec.~3). Near the horizon the atmosphere bends the light of the sun
by about half a degree---the setting sun is already geometrically below the horizon when we see it
touch it. Since the index of air depends on wavelength, the blue and green images of the sun are
lifted slightly more than the red one, by about half a minute of arc. In the last instant of a
sunset, when only the upper rim of the sun is visible, the red image sets first, and for a second
or two the rim turns green (blue and violet having been removed by Rayleigh scattering along the
long path).

\paragraph{Twinkling.} Turbulence in the atmosphere makes the index fluctuate from place to place.
The light of a star, a point source, follows paths that wander and interfere, and its brightness
flickers; planets, which show a small disc, average the fluctuations over many paths and twinkle
much less.

\begin{figure}[!htbp]
\centering
\includegraphics[width=0.72\textwidth]{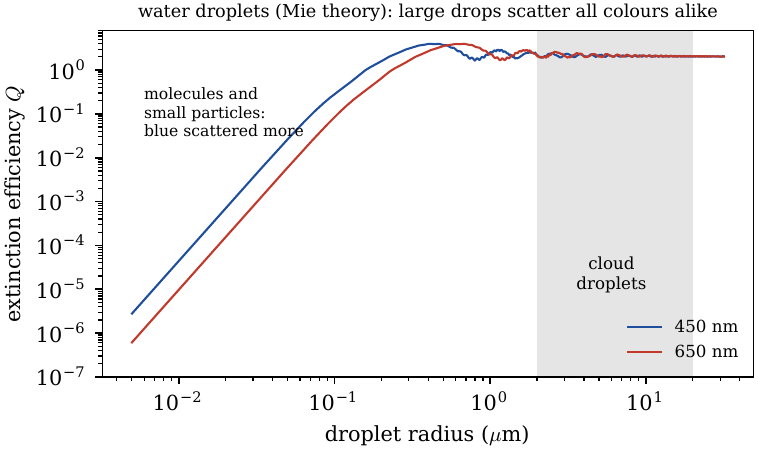}
\caption{Extinction efficiency (cross section divided by the geometric one) of water droplets
at 450 and 650~nm, from Mie's exact theory. Small particles are in the Rayleigh regime and favour
blue; cloud droplets (shaded) remove all colours equally, $Q\simeq2$.}
\label{fig:clouds}
\end{figure}

\paragraph{White clouds} (Secs.~1 and 3). Molecules and very small particles scatter blue light
more than red, and make the sky blue. Cloud droplets, around ten micrometres, are much larger than
the wavelength: each removes from the beam twice its geometric cross section, half by reflection and
refraction and half by diffraction around it (the ``extinction paradox'', a consequence of the
optical theorem of Sec.~3), whatever the colour (Fig.~\ref{fig:clouds}). Clouds are white, and the
haze of polluted air, made of particles of intermediate size, is greyish.

\begin{figure}[!htbp]
\centering
\includegraphics[height=4.3cm]{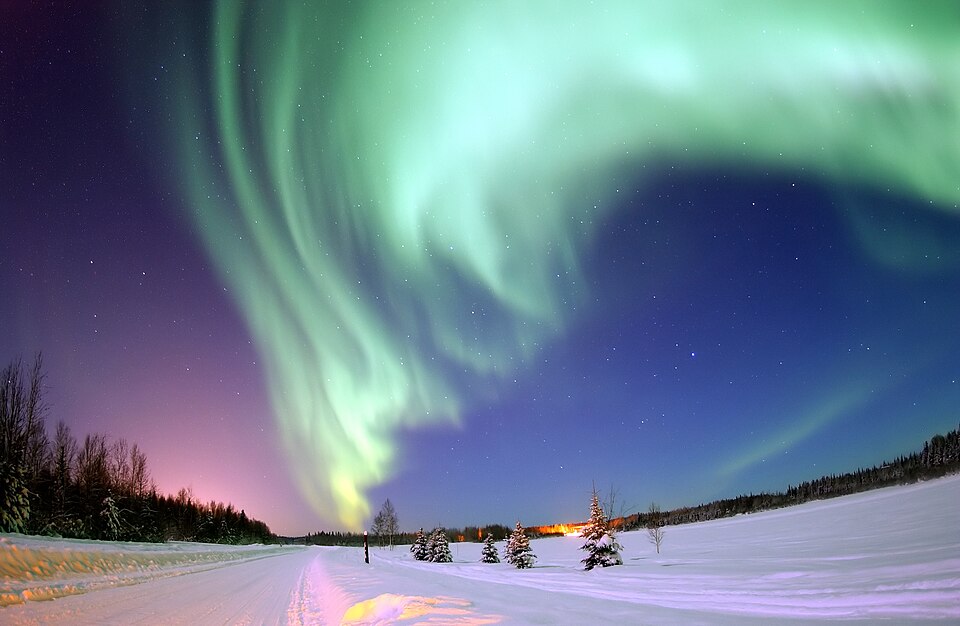}\hfill
\includegraphics[height=4.3cm]{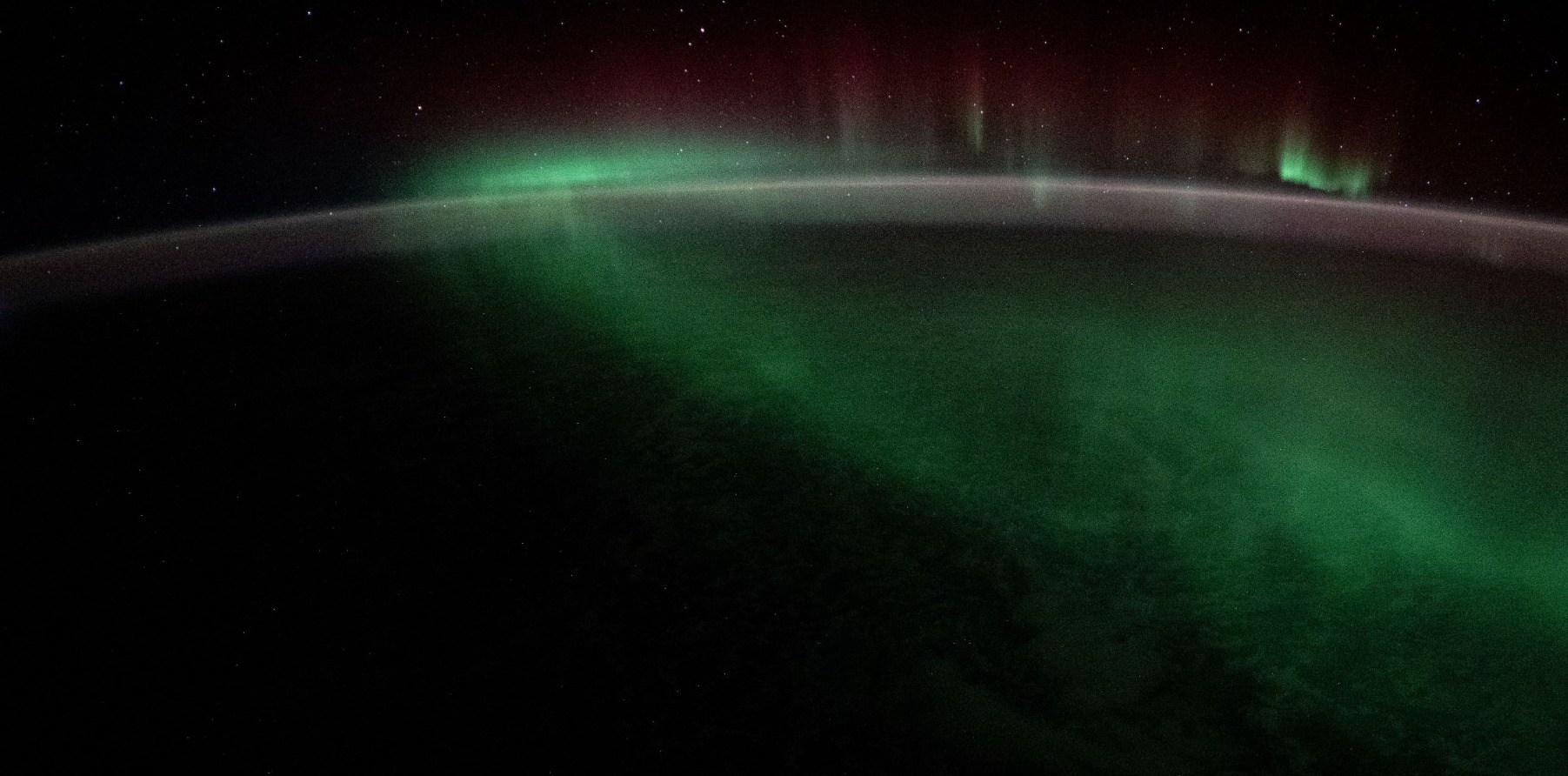}
\caption{\rev{The aurora seen from below and from above. Left: aurora borealis over Bear Lake,
Eielson Air Force Base, Alaska, January 2005; the curtain is green, the colour of the oxygen line at
557.7~nm (U.S.\ Air Force photo by Senior Airman Joshua Strang, public domain). Right: aurora australis
over the Indian Ocean, photographed from the International Space Station on 3 June 2024 (image
credit: NASA). Seen edge-on above the limb of the Earth, the green layer lies low and the red glow of
the 630.0~nm line rises above it, as in Fig.~\ref{fig:aurora}b.}}
\label{fig:aurora-photo}
\end{figure}

\begin{figure}[!htbp]
\centering
\includegraphics[width=\textwidth]{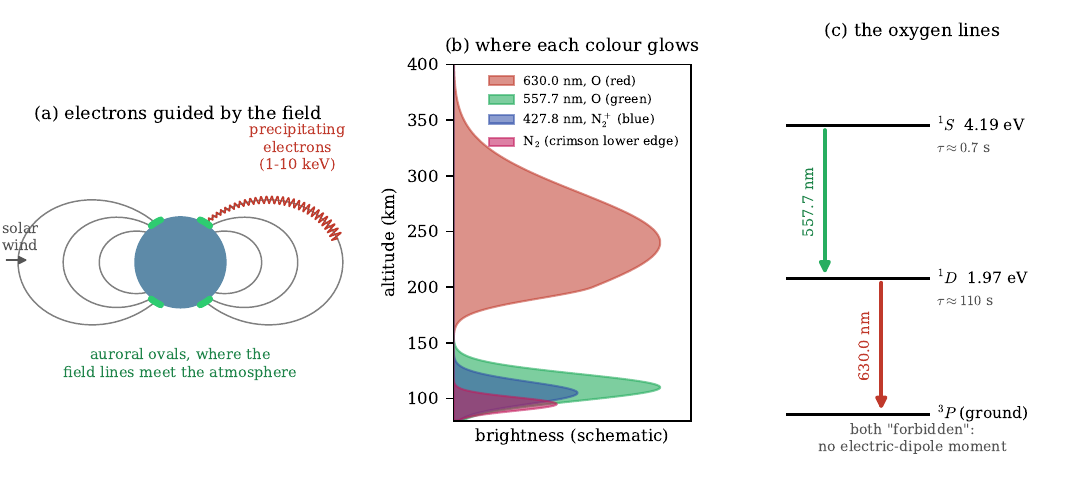}
\caption{\rev{(a) Electrons accelerated in the magnetosphere spiral along the field lines and reach
the atmosphere in two ovals around the magnetic poles. (b) Altitude of the main auroral emissions
(schematic): the red oxygen line above about 150--200~km, the green oxygen line around 100--150~km,
the blue band of N$_2^+$ and the crimson lower edge of N$_2$ lowest. (c) The levels of atomic
oxygen that give the green (557.7~nm) and red (630.0~nm) lines, with their radiative lifetimes;
both transitions are forbidden for electric-dipole radiation.}}
\label{fig:aurora}
\end{figure}

\paragraph{\rev{The aurora}} \rev{(Secs.~\ref{sec:one-atom} and~\ref{sec:beyond}).}
{\revon
Every phenomenon of this section so far is sunlight scattered, refracted or reflected by matter. The
polar aurora is different (Fig.~\ref{fig:aurora-photo}): its light is \emph{emitted} by the atoms of the upper
atmosphere,
and it brings us to the border where the classical oscillator stops.

The energy comes from the sun, but not as light. The solar wind, a stream of protons and
electrons, is held off by the Earth's magnetic field; in the tail of the magnetosphere electrons are
accelerated to energies of 1--10~keV and are guided down along the field lines. A charge moves
freely along a magnetic field but circles around it, so the electrons spiral along the lines and
reach the atmosphere only where the lines come down, in two ovals around the magnetic poles
(Fig.~\ref{fig:aurora}a). There, between 100 and 300~km of altitude, they collide with
atoms and molecules of oxygen and nitrogen and leave them excited.

An excited atom left alone emits light: \emph{spontaneous emission}, the first item in the list of
things the classical model cannot do (panel ``Why classical oscillators?'' in Sec.~\ref{sec:one-atom}). The
oscillator gives the frequency of the light and, through the oscillator strength, how fast it is
emitted; why an atom at rest radiates at all is a property of the quantized field. The aurora
shows this border in a spectacular way. Its two most famous colours, the green line of atomic oxygen
at 557.7~nm and the red line at 630.0~nm, come from transitions that are \emph{forbidden}: the
atom has no oscillating electric dipole between those states, its oscillator strength is almost
zero, and it radiates through much weaker magnetic-dipole and quadrupole terms. An allowed
transition takes some ten nanoseconds; the green line takes about 0.7~s, the red line almost two minutes
(Fig.~\ref{fig:aurora}c)~\cite{NISTASD}.

Waiting so long is possible only where collisions are rare. In the laboratory, or in the lower
atmosphere, a collision removes the energy long before the atom can radiate, and these lines are
never seen. In the aurora the colours sort themselves by altitude (Fig.~\ref{fig:aurora}b; the photograph
from space in Fig.~\ref{fig:aurora-photo} shows it directly): the
red glows above about 150--200~km, where the air is thin enough for an atom to wait two minutes; the
green, which needs to wait only a second, glows lower down, around 100--150~km, where the most
energetic electrons stop~\cite{Barthelemy2025}. Below it, the allowed bands of nitrogen, emitted at once, add the violet
and blue of the ionized molecule (391.4 and 427.8~nm) and the crimson fringe at the lower edge of
active curtains. That the aurora looks mostly green is also a matter of the eye, as for the blue of
the sky: the green line falls close to the peak of its sensitivity, while the red one, faint and
far less well seen, often shows up only in long-exposure photographs.
\par}

\subsection{Colours of water and of living things}

\begin{figure}[!htbp]
\centering
\includegraphics[width=0.95\textwidth]{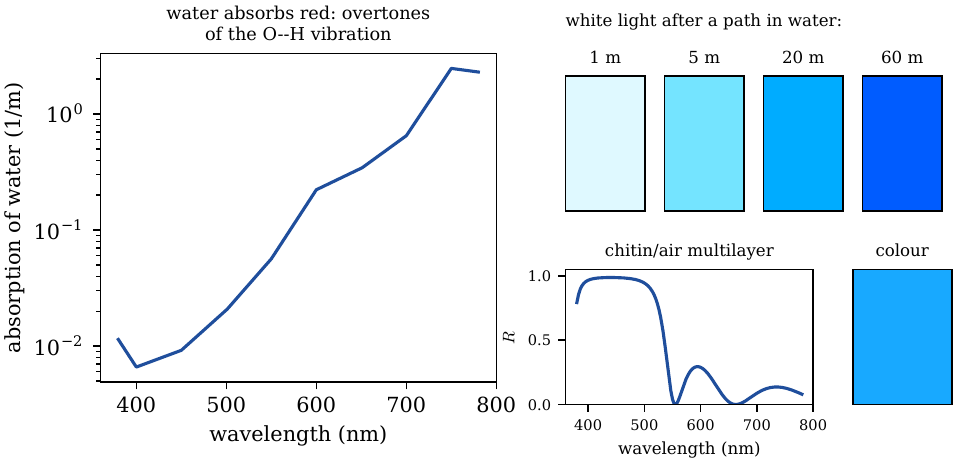}
\caption{Left: absorption coefficient of pure water in the visible (approximate values; up to
700~nm from Ref.~\cite{PopeFry}). Top right: colour of white light after 1, 5, 20 and 60~m of water. Bottom
right: reflectance of a stack of six chitin layers ($n=1.56$) separated by air, of the kind found in
the scales of \emph{Morpho} butterflies, and its colour.}
\label{fig:colours}
\end{figure}

\paragraph{Why water is blue} (Sec.~1). Water is the most transparent liquid we know, but not
perfectly: it absorbs red light weakly, a few tenths of a percent per centimetre. The cause is a
molecular oscillator of a different kind: the vibration of the hydrogen atoms against the oxygen,
whose fundamental resonance is in the infrared, at 2.9~$\mu$m, but whose weak overtones reach the
red end of the visible~\cite{BraunSmirnov}. A glass of water looks colourless; a swimming pool, a few
metres deep, is cyan; deep sea water is blue (Fig.~\ref{fig:colours}). Heavy water, whose heavier
deuterium atoms vibrate more slowly, pushes the overtones into the infrared and is almost colourless.

\paragraph{Colours without pigments} (Sec.~5). The brilliant blue of the \emph{Morpho} butterflies,
the metallic greens of beetles, the eyes of a peacock's feather and the iridescence of
mother-of-pearl contain no blue or green pigment. They are made by stacks of thin transparent
layers---chitin and air, keratin and melanin, aragonite and protein---that work as the multilayer
mirrors of Sec.~5: the arrows reflected at the interfaces add up for one band of colours and cancel
for the others (Fig.~\ref{fig:colours}). Change the angle of view, and the optical thicknesses and
the colour change: iridescence.

\subsection{Vision}

Why do we see between 400 and 700~nm? Because that is where the light is. The spectrum of the sun
peaks in the visible; the atmosphere is transparent there, between the ultraviolet absorption of
ozone and the infrared bands of water vapour; and liquid water---the substance of which the eye
and its ancestors in the sea are made---is transparent only in that same narrow window
(Fig.~\ref{fig:colours}). The eye evolved where the light arrives (Fig.~\ref{fig:life}). Colour
vision rests on three kinds of cones with overlapping sensitivities, peaking near 420, 535 and
565~nm; every colour we perceive is the ratio of their three signals, which is why so different
spectra---a laser line, a mixture of red and green, a broad yellowish spectrum---can look the same.
This is the principle of the colour-matching functions used in this text to compute the colours
of the sky, of soap films and of the rainbow.

The first step of vision, however, is beyond our oscillators. A photon absorbed by a molecule of
retinal in a rod or a cone changes its shape, from the bent \emph{11-cis} to the straight
\emph{all-trans} form, in less than a picosecond, and this change sets off the cascade that ends in
a nerve signal. A single photon suffices to trigger a rod. Here light is absorbed in quanta, and
the energy of one photon, not the amplitude of a wave, is what matters. Many insects add one more
sense: their photoreceptors are sensitive to polarization, and bees read the polarization pattern
of the blue sky (Sec.~6) to find the direction of the sun when it is hidden.

\subsection{Energy}

\begin{figure}[!htbp]
\centering
\includegraphics[width=0.8\textwidth]{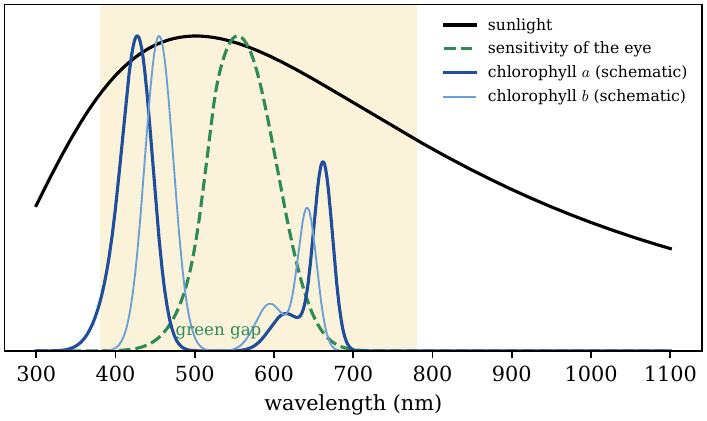}
\caption{The spectrum of sunlight (black-body at 5778~K), the sensitivity of the human eye (CIE
photopic curve) and the absorption of chlorophylls $a$ and $b$ (schematic, peaks at 430, 662 and
455, 642~nm). The visible band is shaded; between the two absorption peaks of chlorophyll lies the
``green gap''.}
\label{fig:life}
\end{figure}

\paragraph{Photosynthesis.} Plants harvest sunlight with chlorophyll, which absorbs strongly in the
blue and in the red but weakly in between (Fig.~\ref{fig:life}): leaves are green because green is
the light they do not use. Beyond 700~nm leaves become highly reflecting---the ``red edge'' used by
satellites to map vegetation. The absorption bands are, once more, molecular oscillators with their
strengths and widths. But what happens next is not classical: the energy of each absorbed photon is
passed, by quantum-mechanical resonance transfer, from pigment to pigment to a reaction centre, where
it separates an electron from its molecule and starts the chemistry of life. Nearly every absorbed
photon that reaches a reaction centre is used: a quantum efficiency that our best devices envy.

\begin{figure}[!htbp]
\centering
\includegraphics[width=0.8\textwidth]{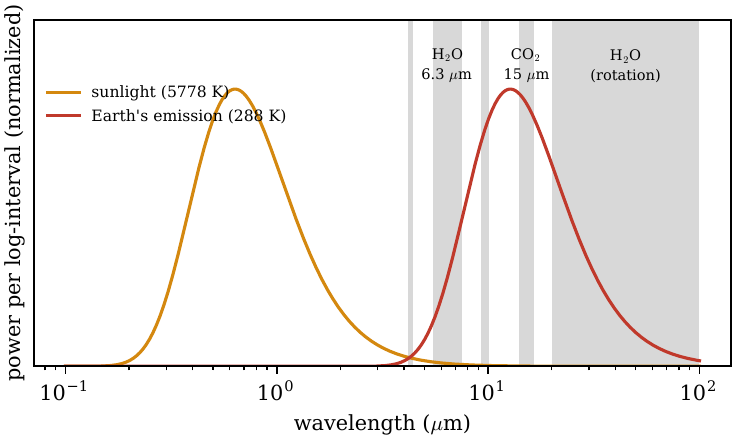}
\caption{Spectra of sunlight and of the thermal emission of the Earth (black bodies at 5778 and 288~K,
power per logarithmic interval of wavelength, each normalized). Shaded: the main infrared absorption
bands of water vapour, carbon dioxide (4.3 and 15~$\mu$m) and ozone (9.6~$\mu$m) (schematic).}
\label{fig:greenhouse}
\end{figure}

\paragraph{The greenhouse effect} (Sec.~1). The sun, at 5800~K, sends us mostly visible and
near-infrared light; the Earth, at 288~K, radiates back in the far infrared, around 10~$\mu$m
(Fig.~\ref{fig:greenhouse}). The main gases of the air, nitrogen and oxygen, are symmetric molecules:
their vibrations carry no oscillating dipole, and they are transparent to both. Water vapour, carbon
dioxide and ozone are different: they have vibrations that do make an oscillating dipole---for carbon
dioxide, the bending of the molecule at 15~$\mu$m---and they are Lorentz oscillators resonant in the
infrared, right where the Earth radiates. Transparent to sunlight, they absorb and re-emit the
Earth's infrared radiation, part of which returns to the ground. Without them the surface of the
Earth would be at about 255~K instead of 288~K. This is the same physics as Sec.~1, at frequencies
tens of times lower, with nuclei instead of electrons on the springs.

\paragraph{The push of light} (Sec.~1). Light carries momentum as well as energy: a beam of intensity
$I$ absorbed by a surface pushes it with the pressure $I/c$. The origin is visible in the oscillating
charge of Sec.~1: the electric field drives the electron, and the magnetic field of the same wave, acting
on the moving electron, pushes it along the direction of propagation; on average the push survives
whenever the electron takes energy from the wave. Sunlight at the Earth, 1361~W/m$^2$, exerts a pressure
of only 4.5 micronewtons per square metre---but in space, with no friction, this is enough to blow the
dust tails of comets away from the sun, and to propel solar sails such as that of the Japanese probe
IKAROS (2010).

%% file: ch10_epilogue.tex
\section*{\rev{Conclusions and epilogue}: two roads to the same field}
\addcontentsline{toc}{section}{Conclusions and epilogue: two roads to the same field}
\label{sec:epilogue}

We have told one story many times. A light wave sets the charges of matter oscillating; each
oscillating charge radiates a wave of its own, which travels at the speed of light and reaches the
other charges a little later; and what we see is the sum of all these waves, each with its proper
delay and its proper phase. From this single picture came the blue of the sky and the white of the
clouds, the transparency of glass and the slowing of light within it, the extinction of the
incident wave and the birth of the refracted one, the reflection of a window and the colours of a soap
bubble, \rev{the mirror of a metal and the colours of gold and of cadmium yellow,} the vanishing
reflection at Brewster's angle, \rev{the double image of calcite,} the front of a pulse travelling at
the speed of light while the pulse itself lags behind, \rev{the blue glow of a charge faster than light
in water,} the amplification of light in a laser and the doubling of its frequency in a crystal, the
rainbow and the mirage.

{\revon
\medskip\noindent\textbf{Three ingredients.} Looking back, every result has been obtained from three ingredients,
always the same.
\begin{enumerate}
\item \emph{The oscillator} (Sec.~\ref{sec:one-atom}). An electron bound by a spring responds to the
field with an amplitude and a phase fixed by its resonance: in step below it, in antiphase above it,
in quadrature at resonance, where it absorbs. Quantum mechanics confirms this response for weak
fields, with one oscillator for each transition.
\item \emph{The sheet} (Sec.~\ref{sec:sheet}). The wavelets of an ordered layer of oscillators add up
only in the forward and backward directions, and their sum is a quarter of a period out of step
with the dipoles that emit it. Disorder, instead, lets the sideways light escape.
\item \emph{Self-consistency} (Sec.~\ref{sec:extinction}). Every oscillator is driven not by the
incident wave alone but by the total field, which includes the waves of all the others, each emitted
at the earlier time required by the speed of light. Solving this condition gives the extinction of
the incident wave, the refracted wave travelling at $c/n$, and the Lorentz--Lorenz relation.
\end{enumerate}
Table~\ref{tab:map} shows which of them decides each phenomenon of the text. It can also be read as a
guide for the reader who enters the text from the middle.

\begin{table}[!htbp]
\centering\small
\caption{A map of the text: each phenomenon, the ingredient that decides it (O: the oscillator, its
amplitude and phase; S: the sheet, its quarter-period delay and the role of order; C: self-consistency
and retardation; Q: where the classical picture must be completed by quantum theory), and where it is
discussed.}
\label{tab:map}
\renewcommand{\arraystretch}{1.12}
\begin{tabular}{p{0.68\textwidth}cc}
\hline
Phenomenon & Ingredient & Section\\
\hline
Blue sky, red sunset, polarized skylight & O & \ref{sec:one-atom}\\
Polarizers and Malus's law; single photons through a polarizer & O, Q & \ref{sec:one-atom}\\
Transparency: why an ordered medium does not scatter sideways & S & \ref{sec:sheet}\\
Refractive index, dispersion, the prism; index and absorption & S, O & \ref{sec:index}\\
Extinction of the incident wave; local field & C & \ref{sec:extinction}\\
Reflection, Feynman's arrows, thin films, multilayer mirrors & S, C & \ref{sec:reflection}\\
Metals and semiconductors: mirrors, colours, transparent conductors & O, C (Q) & \ref{sec:metals}\\
Brewster's angle, external and internal & O, S & \ref{sec:brewster}\\
Birefringence, wave plates, liquid-crystal displays & O & \ref{sec:birefringence}\\
Mirrors, lenses, telescopes, microscopes, the eye, interferometers & S & \ref{sec:instruments}\\
A pulse in a slab: the front at $c$, the energy velocity & C & \ref{sec:pulse}\\
\v{C}erenkov radiation & S, C & \ref{sec:cerenkov}\\
Absorption, amplification, the laser; second-harmonic generation & O (Q) & \ref{sec:beyond}\\
Rainbow, mirage, green flash, clouds, colours of water and of insects & S, O & \ref{sec:nature}\\
Aurora, vision, photosynthesis & Q & \ref{sec:nature}\\
\hline
\end{tabular}
\end{table}

\medskip\noindent\textbf{What the picture adds, and where it stops.} The textbook results---Snell's law, Fresnel's
formulas, the dispersion formula---come out of this picture exactly, not
as approximations. What the picture adds is the answer to questions that the usual treatment does not
ask. Feynman's surface arrow, an empirical datum in \emph{QED}, is the radius of the circle described
by the arrows of the sheets of the medium (Sec.~\ref{sec:reflection}); in a metal the circle collapses,
and the arrows add up to one in a few tens of nanometres (Sec.~\ref{sec:metals}). The incident wave
is not stopped at the surface: it is cancelled, everywhere inside, by the waves of the medium, and
this cancellation takes a few optical periods to establish itself, behind a front that always
travels at $c$ (Sec.~\ref{sec:pulse}). The slowing of a pulse is the storage of energy in the motion
of the electrons, shared collectively, and not a sequence of absorptions and delayed re-emissions
(Secs.~\ref{sec:myth} and~\ref{sec:pulse}). The zero of the \emph{p} reflection is a property of the
dipole pattern, the same for external and internal reflection (Sec.~\ref{sec:brewster}). And the whole
process can be watched molecule by molecule in a simulation of coupled dipoles, which assumes nothing
but the polarizability and the speed of light (Sec.~\ref{sec:extinction}).

The picture stops where the light, or the atom, must be quantized. An excited atom left alone
radiates, and why it does so is a property of the quantized field (Sec.~\ref{sec:one-atom}); the
forbidden lines of the aurora show how slowly it can happen (Sec.~\ref{sec:nature}). The sharp
threshold of absorption at the gap of a semiconductor, the single photon that triggers a rod of the
retina, the photon that either passes a polarizer or does not---all these are counts of quanta, not
amplitudes of waves. Yet even there the classical oscillator keeps its value: it gives the
frequencies, the strengths and the phases with which the quantum processes happen.

Much has been left out: optical activity and magneto-optics, the scattering of light by particles
comparable with the wavelength, the near field and the surface waves of metals, and the genuinely
quantum states of light. Each of them can be approached in the same way, by adding up the waves of the
oscillators; but each would need a chapter of its own.
\par}

\medskip\noindent\rev{\textbf{Two roads.}} There is another way of telling the same story. Instead of following
each wave from its source,
one can describe matter as a continuous medium with a refractive index, and light as a solution of
Maxwell's equations in that medium, satisfying conditions at its boundaries. This is the way of most
textbooks, and it is enormously efficient: Snell's law and Fresnel's formulas follow in a few lines.
In it, nothing is said about retardation, about the extinction of the incident wave, about the waves
of the individual sheets of atoms: all of this is hidden in the local equations. The two descriptions
look so different that one might ask which of them is true. Papers~I and~II~\cite{RealiI,RealiII}
showed, analytically and numerically, that they give the same field---not approximately, but
exactly, to the last digit the computer can resolve. They are two roads to the same field.

Which road, then, \emph{explains}? Perhaps neither, in the strong sense of the word. Physical
theories describe how nature behaves, with the economy and the precision of mathematics; they do not
tell us what nature \emph{is}. The local description is more economical; the description by
retarded waves is closer to our intuition of cause and effect, and it answers questions---why is the
reflection white? where does the incident wave go? why does the front travel at the speed of
light?---that the local description answers only in its own language. Both are maps of the same
territory, drawn for different travellers. Behind both, at a deeper level, lies quantum
electrodynamics, Feynman's ``strange theory of light and matter'', in which light is made of photons
and every process is a sum of amplitudes---arrows---over all the ways in which it can happen; the
arrows of this text are their classical shadow.

\rev{A companion text, \emph{Light and living matter}~\cite{LivingMatter}, follows the same
oscillators into tissue, from the transparency of the cornea to laser surgery and to the light made
by a squid.}

\medskip\noindent\rev{\textbf{Wonder.}} And the wonder? Go back into the wood of the Prologue, on a misty morning,
and look again at the
shafts of light between the trees. Every droplet of the mist is sending a little of the sunlight
towards you, and that is why you see the beams. Every molecule of the air is doing the same, and
their wavelets, adding in step only in the forward direction, keep the beams straight and sharp
across the wood. The sun itself, low on the horizon, is a little lower than it seems. Nothing has been taken
away from the spectacle. Something has been added: the knowledge of the countless tiny oscillators,
working together in perfect step, that make it.